\documentclass[final,1p,times]{elsarticle}
\usepackage{amssymb}
\usepackage{lipsum}
\usepackage{amsmath}
\usepackage{xcolor}
\usepackage{color}
\usepackage{float} 
\usepackage{graphicx}
\usepackage{caption}
\usepackage{subcaption}
\usepackage{soul}
\usepackage[T1]{fontenc} 
\setcitestyle{square} 
\biboptions{comma}
\usepackage[colorlinks=true, linkcolor=red, citecolor=green, urlcolor=magenta]{hyperref}

\begin{document}

\begin{frontmatter}
\title{Joint Detection and Characterization of the Standing Accretion Shock Instability for Core-Collapse Supernovae with cWB XP}

\author[First]{Vicente Sierra}
\ead{vicente.sierra@ligo.org}

\author[Second,Sixth,Seventh]{Zidu Lin}
\ead{linzd9@mail.sysu.edu.cn}

\author[Third]{Michelle Zanolin}
\ead{zanolinm@erau.edu}

\author[First]{Claudia Moreno}
\ead{claudia.moreno@academicos.udg.mx}

\author[Fourth]{Javier M. Antelis}
\ead{mauricio.antelis@tec.mx}

\author[Fifth]{Marek J. Szczepa\'nczyk}
\ead{Marek.Szczepanczyk@fuw.edu.pl}

\affiliation[First]{organization={Departamento de F\'isica, CUCEI, Universidad de Guadalajara},
	addressline={}, city={Guadalajara},
	postcode={44430}, 
	state={Jalisco},
	country={M\'exico}}

\affiliation[Second]{organization={Sino-French Institute of Nuclear Engineering and Technology, Sun Yat-Sen University},
	addressline={}, city={Zhuhai},
	postcode={519082}, 
	state={},
	country={China}}

\affiliation[Third]{organization={Embry-Riddle Aeronautical University, Prescott campus},
	addressline={}, city={Prescott},
	postcode={86301}, 
	state={AZ},
	country={USA}}

\affiliation[Fourth]{organization={Tecnol\'ogico de Monterrey, Escuela de Ingeniería y Ciencias},
	addressline={}, city={Monterrey},
	postcode={64849}, 
	state={N.L.},
	country={M\'exico}}

\affiliation[Fifth]{organization={Faculty of Physics, University of Warsaw},
	addressline={Ludwika Pasteura 5}, city={Warsaw},
	postcode={02-093}, 
	state={},
	country={Poland}}

\affiliation[Sixth]{organization={Department of Physics and Astronomy, University of Tennessee Knoxville},
	addressline={}, city={Knoxville},
	postcode={37996}, 
	state={TN},
	country={USA}}

\affiliation[Seventh]{organization={Department of Physics \& Astronomy, University of New Hampshire},
	addressline={9 Library Way}, city={Durham},
	postcode={03824}, 
	state={NH},
	country={USA}}

\begin{abstract}
	
	The most sensitive to-date multimessenger detection of the standing accretion shock instability in real interferometric data is presented, which quantitatively identifies the presence of the SASI in core-collapse supernovae using neutrino and gravitational-wave (GW) signals. In the GW channel, the \textit{coherent WaveBurst} (cWB) software on its version \textit{XP} is implemented, among with real LIGO data from the O3 and O4 observing runs. With this, a more accurate estimation of parameters, such as the central frequency and signal duration, is obtained for both sets of data. The SASI identification probability versus false alarm rates is presented in the form of Receiver Operating Characteristic (ROC) curves. For O3, the new study for the combined GW and neutrino detection condition, labeled as $x + y$, shows an identification probability (previous best results from Lin et al.\cite{Zidu_SASImeter_Joint}) of 1 (1), 0.90 (0.70) and 0.37 (0.34) at 1, 5 and 10 kpc for a false identification probability of 0.10. On the other hand, using O4 shows that the GW channel by itself is sensitive enough to provide almost perfect identification probability scores, with identification probability values of 1, 0.99 and 0.97 for a false identification probability of 0.01 at 1, 5 and 10 kpc, respectively.
\end{abstract}
\end{frontmatter}

\section{Introduction} \label{Sect_Intro}

In the newly born age of multimessenger astrophysics, Core-Collapse Supernovae (CCSNe) are one of the most promising candidates to be analyzed through the three possible signal channels: neutrinos, electromagnetic radiation, and gravitational waves \cite{SN_Multimess}. A CCSN is the process through which a massive star, of $M \gtrsim 8 M_\odot$, dies \cite{Michele_Tony_CCSNe, CCSNe_Theory}, producing a shock front that, while traveling through the iron core and through the stellar layers above it, is referred to as a \textit{dynamical} shock. In doing so, the shock interacts with matter in the medium through mass dissociation and neutrino losses, losing energy in the process. When the shock has lost sufficient energy, it ``stalls", becoming instead a \textit{standing} accretion shock. This stagnation generates a global instability known as the Standing Accretion Shock Instability (SASI), which was first identified by Blondin, Mezzacappa $\&$ DeMarino \cite{Blondin_2003}. Hence, the SASI is a self-sustained hydrodynamic instability of the entire stalled accretion shock system that leads to large-scale oscillatory deformations of the shock front. It presents two oscillation modes known as the ``sloshing" and ``spiraling" modes, which reflect the dynamical effect they produce on fluids. 

The SASI plays an important role in several processes in a CCSNe; for example, it assists neutrino heating by increasing the size of the heating region, which in turn is responsible for a \textit{successful explosion} \cite{Michele_Tony_CCSNe, Scheck2008}. Furthermore, the SASI modulates the accretion flow onto the Proto-Neutron Star (PNS), which determines the core pulsations that generate high frequency GWs \cite{Michele_Tony_CCSNe}. Hence, the SASI features are crucial to understand processes in a CCSNe; for example, a long-lasting SASI can be a signature of a black hole formation. Also, since all of these processes can be highly energetic and asymmetric, both neutrinos and gravitational wave emissions can be modulated by the SASI \cite{MDrago_Multimess, Tamborra:2014hga}. Furthermore, it is broadly believed that these signals are actually related, specifically with respect to their characteristic central frequency.  While it is believed that there could be an exact relationship between the SASI frequency in the GW and neutrino channel as in the case of the orbital frequency with the GW frequency for a compact binary system, experimental evidence will play an essential role in settling the debate. For this reason, as it will be discussed further on, the multimessenger analysis here presented will estimate separately the frequency in the two channels.

If detected, both neutrinos and gravitational waves associated with the SASI can provide crucial information regarding the average radius of the PNS and the stalled shock front, the coupling mechanism between the shock wave and the PNS \cite{Blondin_2006, Walk_2020}, the EoS of nuclear matter in the PNS \cite{EoS} and the conditions under which the supernova undergoes a successful explosion \cite{Exp_Mechanism}. This information, which is crucial to understand the engine that drives a CCSN, can only be unveiled via neutrino and GW signals, since the dynamical media is opaque to electromagnetic signals.

The LIGO-Virgo-KAGRA (LVK) collaboration has published about 100 GW detections and over 200 candidates from the first observing run (O1) to the first segment of the fourth observing run (O4a) \cite{GWTC1, GWTC2, GWTC2.1, GWTC3, GWTC4}, but none have been associated with CCSNe \cite{Allsky_O1, Opt_O1_O2_SN_Search, Allsky_O2, Allsky_O3, Opt_O3_SN_Search, AllSky_O4a, SN2023ixf}. However, as neutrinos and GWs from a CCSN are likely to be detected in the next galactic event and for LIGO's O4 sensitivity, there is scientific interest in the development of methodologies that can quantitatively identify the presence and the features of the SASI. Indeed, Lin et al. (hereon referred to as Lin2020) \cite{Zidu_SASImeter_Neutrino} proposed a methodology, baptized as \texttt{SASImeter}, to determine the presence of a SASI signature by using a likelihood ratio as a test-statistics and to estimate the frequency and the amplitude of the SASI modulations in the neutrino channel. On a subsequent work, Lin et al. (hereon referred to as Lin2023) \cite{Zidu_SASImeter_Joint}, injected the simulated waveform into real LIGO interferometric O3 noise and reconstructed it by using the LVK coherent WaveBurst 2G (second-generation) \cite{cwb} detection software in order to create a sibling gravitational wave \texttt{SASImeter}. Then, the authors tested the joint and individual detection efficiency, whose results showed that a joint detection and parameter estimation performs better than an individual one, and that the neutrino channel performs better than gravitational waves at closer distances (1 kiloparsec), while the gravitational wave channel performs better than neutrinos at larger distances (5 and 10 kpc). 

There are also other proposals in the literature on how to extract SASI-features from GW signals. For example, there are authors that propose the implementation of the Hilbert-Huang transform, which is a method to decompose a signal into its intrinsic mode functions, which could allow to isolate a SASI-born signal and estimate its instantaneous frequency, which is important to note that not always corresponds to the signal's Fourier frequency. For example, in \cite{HuangT_Takeda}, authors apply this method into a 3D CCSN GW simulation and estimate its frequency and duration estimation with simulated Gaussian noise, while in \cite{HuangT_Veutro}, authors employ the transform in more 3D CCSN GW models and test the SASI identification in the Einstein Telescope, a third generation GW detector \cite{GW_Detect_Perspectives}. However, none of these implementations test the identification probability in real interferometric noise. Another and very recent work \cite{Freq_SASI_Analysis} proposes a dedicated-frequency framework; although not focused in the identification of the SASI imprint, this work evaluates the detection of overall low-frequency signals in order to constrain explosion mechanisms in CCSNe.

Regarding the GW channel, considerable progress has been made recently in the efforts for the search and detection of signals beyond the compact binary coalescences that are commonly detected by Advanced LIGO \cite{LIGO_Adv} and Advanced Virgo detectors \cite{Virgo_Adv}, such as those from CCSNe; one of these is a new version of cWB, labeled as XP. The previous version, cWB 2G, uses a multiresolution Wilson–Daubechies–Meyer (WDM) wavelet transform for its time–frequency analysis \cite{cWB_2G}, while the newer cWB XP (``Cross-power'') variant replaces the WDM basis with a multi-resolution \textit{WaveScan} transform based on Gabor wavelets \cite{wavescan}. This change dramatically reduces temporal and spectral leakage in the time–frequency plane, yielding much finer localization of power for short, broadband transients. Furthermore, cWB XP implements a new coherent Cross-power statistic (CRS), additional to the usual excess‐power statistic, in order to combine detector data and suppress incoherent noise glitches. For a comprehensive comparison between 2G and XP, and an application of the latter, refer to \cite{AllSky_O4a}. Furthermore, it is relevant to stress that the LVK collaboration has achieved an unprecedented sensitivity to gravitational waves \cite{O4_Performance} in the Advanced LIGO detectors \cite{LIGO_Adv} during the first part of its fourth observing run, O4a \cite{O4a_data}.

Given the potential of all of these improvements, it is of great interest to update the GW \texttt{SASImeter} by using cWB XP in O3 data \cite{O3_data} in order to compare it with its predecessor and with the neutrino \texttt{SASImeter}, as well as testing it with O4 data. Hence, this present work is a natural continuation of the two preceding \texttt{SASImeter} related papers, where the current goal is to follow closely the joint neutrino and gravitational wave analysis of Lin2023 \cite{Zidu_SASImeter_Joint} in order to provide an updated state of the art detection and parameter estimation of the SASI-born signals in the neutrino and gravitational wave channels. For comparison, the time-frequency region for the SASI established in Lin2023 \cite{Zidu_SASImeter_Joint}, based on 3D simulations \cite{Michele_Tony_CCSNe, Tamborra_2013, Lund_2010, Andresen2017, Mezzacappa2020, Kawahara2018}, is still employed. Recent works have pointed out that SASI, in particular spiral SASI, could also contain energy outside the defined frequency band \cite{Murphy2009, Powell2025, SASI_Spiral_Blondin, SASI_Spiral_Foglizzo, Vartanyan2023}. However, there is consensus that no energy from the high frequency features leaks in this chosen SASI region; in this regard, the implementation adopted here produces conservative SASI identification. Spiral SASI and possible further optimizations are left for future works. Accordingly, the choice of representative simulations are kept the same, also because they are based on 3D simulations and because both GW and neutrinos are available.

Furthermore, the detection efficiency analysis for the updated GW results will be shown for both O3 and O4, and a joint analysis (neutrino + GW), only for O3, will evaluate if the implementation of cWB XP shows improvements. Also, the central frequency and time duration estimations in the GW channel are updated, particularly presenting a new, more conservative estimation of the time duration as it is crucial not to overestimate this parameter due to its role on the physical information regarding a successful explosion and possibly in a black hole formation.

This work is organized as follows: in section \ref{Sect_SASImeter}, the overall workflow of the two \texttt{SASImeter}s is displayed, including detailed explanations on the GW and neutrino channels, focusing on the definition of the metrics that will aid on the parameter estimations and the identification probability. Furthermore, on the GW \texttt{SASImeter}, a brief explanation on the procedure of the Receiver Operating Characteristic analysis, the tool used to quantify the detection efficiency, is presented. Then, in the combined analysis, the procedure upon which both channels are fused to provide a joint detection efficiency is also shown. Then, section \ref{Sect_Results} contains the results of the individual and joint \texttt{SASImeter}s, with the distinction of the GW analysis containing separate results for O3 and O4. Furthermore, the individual channels include the parameter estimations and the ROC results, while the joint analysis contains only the latter. Lastly, section \ref{Sect_Conclusions} contains the conclusions and discussions of the results, underlying the improvements of this work with respect to Lin2023 and a brief picture of the future of the \texttt{SASImeter} is drawn.

\section{\texttt{SASImeter}} \label{Sect_SASImeter}

\subsection{Gravitational Waves} \label{Sect_SASImeter_GW}

The Gravitational Wave \texttt{SASImeter} (hereon referred to as GW-\texttt{SASImeter}) is a Python-based detection and parameter estimation pipeline that operates sequentially on a set of steps. A detailed explanation of each of these steps is now on place.

\begin{figure}[h!]
    \centering
    \includegraphics[scale=0.35]{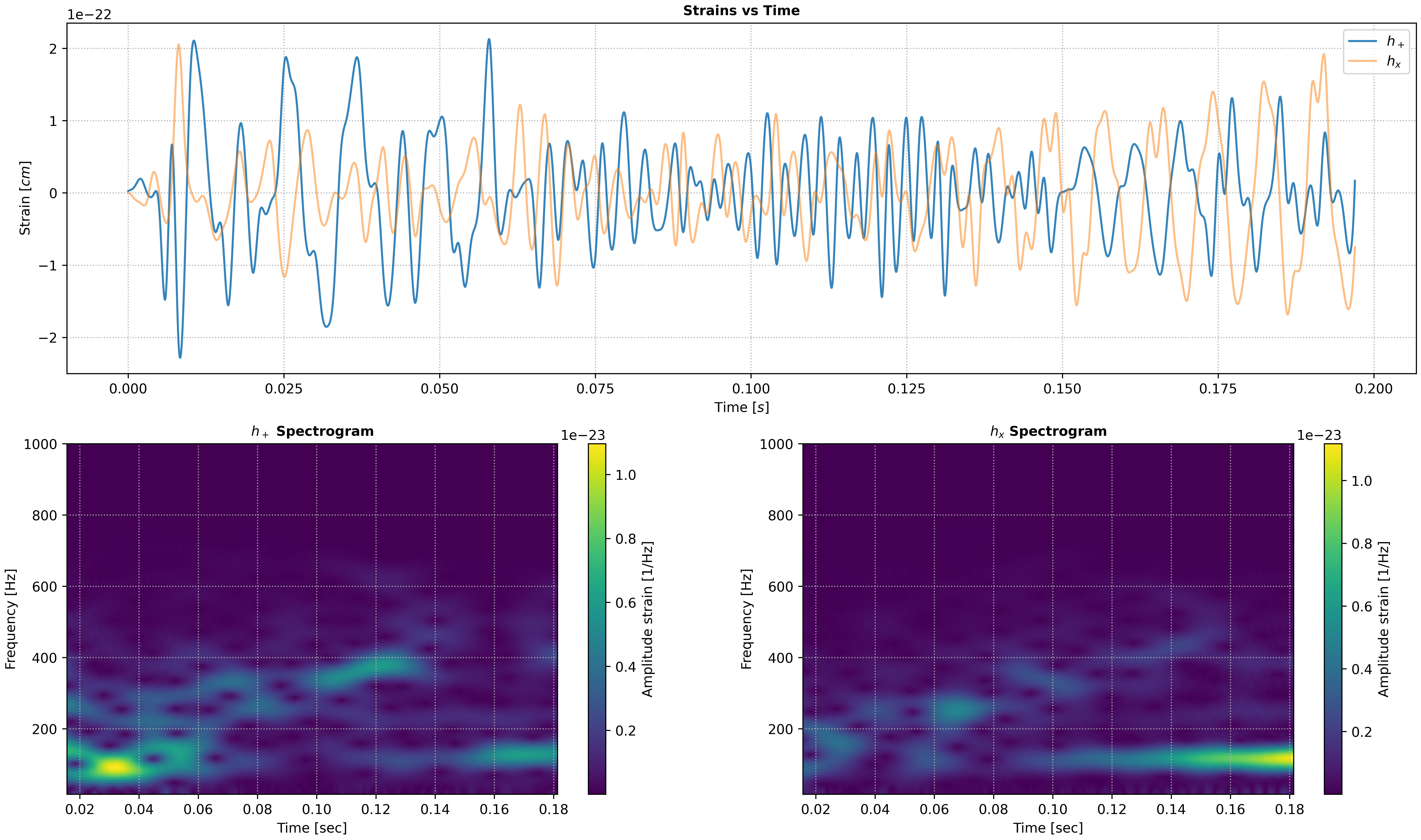}
    \caption{
    Kuroda2017 waveform (at 10 kpc), a 3D CCSNe numerical general relativistic simulation from a $15 M_{\odot}$, non-rotating progenitor with the SFHx EoS. This particular GW signal has both a strong SASI signal below $200 \; Hz$ and around after $t=0.12 \;s$ across both polarizations, and strong HFF at frequencies $>200 \; Hz$ and for the first couple of tens of milliseconds. \textit{Top row}: strain plot of both polarizations, $h_+$ and $h_\times$. \textit{Lower left corner}: spectrogram of the $h_+$ polarization. \textit{Lower right corner}: spectrogram of the $h_\times$ polarization.}
    \label{Fig_Kuroda2017}
\end{figure}

\textit{1. cWB XP GW detection and reconstruction}: This first step consists of the production of Time-Frequency (TF) maps via the coherent WaveBurst software from simulated detections. Since no real GW data from CCSNe has been detected yet, a simulated waveform is used for this task. As in Lin2023 \cite{Zidu_SASImeter_Joint}, the selected waveform is a 3D CCSN numerical general relativistic simulation by Kotake et al. \cite{Kuroda2017} (hereon referred to as Kuroda2017), from a $15 M_{\odot}$, non-rotating progenitor with the SFHx equation of state (EoS). Figure \ref{Fig_Kuroda2017} shows the strain plot and spectrograms of the $h_+$ and $h_\times$ polarizations at 10 kpc; spectrograms were obtained using the magnitude mode of the SciPy's \texttt{signal.spectrogram} module, prescribing a Hamming window of a length of 512 samples, an overlap of 507 and a zero-padded FFT with a length of 2048 samples. On this figure, a clear and strong SASI activity can be observed at frequencies below $200 \; Hz$ and around after $t=0.12 \;s$, while the high-frequency features (HFF) are also observed during the post-bounce stage at frequencies $>200 \; Hz$. 

Using real interferometric data from LIGO's Hanford (H1) and Livingston (L1) interferometers, cWB detects and reconstructs Kuroda2017 from injections at different instances in defined time windows. For O3, this is done under the same configuration prescribed in Lin2023 \cite{Zidu_SASImeter_Joint}, with the only change being the XP version. For O4, the injections are performed between the GPS times $1377561618$ and $1379061618$, which correspond to data from the first part of the LVK fourth observing run, O4a \cite{O4a_data}. This is done with a rate and jitter of $0.01 \; \text{Hz}$ and $10 \; \text{s}$, respectively, at a specific distance determined by the user, at a restricted frequency band from $32$ to $2048 \; \text{Hz}$, and with a source sky distribution defined by a right ascension of $169.59$\textdegree and a declination of $-32.83$\textdegree. The specific cWB parameters used for this work are similar to those used in \cite{SN2023ixf}, obtaining a comparable False Alarm Rate for the candidate events. However, it is important to stress that, in the context of the \texttt{SASImeter}, the True and False Identification Probabilities are obtained from the pipeline itself and not from the cWB reconstructions.

\begin{figure}[h!]
    \centering
    \includegraphics[scale=0.3]{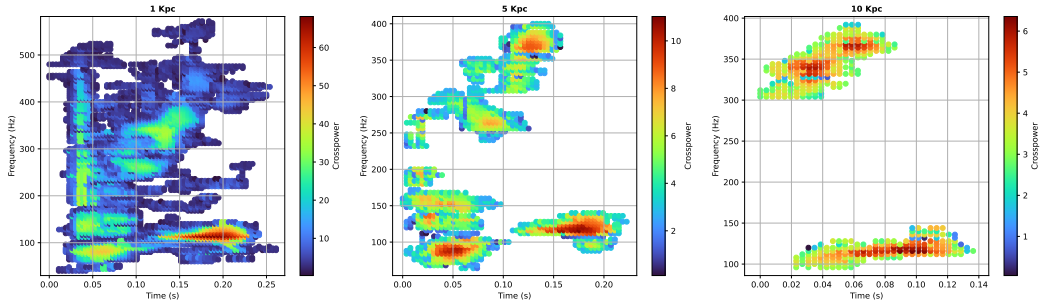}
    \caption{Examples of cWB XP reconstructions of Kuroda2017 injected in O4 noise at 1, 5, and 10 kpc. Since the distance injection affects the overall reconstruction, it is noted that the \textit{x}- and \textit{y}-axis' spans and the CRS scale are all different across the three events.}
    \label{Fig_Reconstructions}
\end{figure}

This first step, which can be regarded as a \textit{pre}-GW-\texttt{SASImeter} stage, is only performed six times for the whole analysis. It is repeated for three galactic distances ($1$, $5$, and $10$ kpc), and for each, it is run for two cases: one where SASI activity is present in the injected waveform and one where it is artificially removed; to perform the last, a Butterworth band-stop filter is implemented via SciPy's modules \textit{signal.butter} and \textit{signal.sosfilt}. The filter is created by the first module using the \texttt{bandstop} filter option of order 5 and with the \texttt{sos} output option, choosing the frequency band to filter out as $[0.1, \;200]$ and specifying the sampling frequency as $16,384 \; \text{Hz}$ (LIGO's standard). Then, the filter is applied via the second module directly on the strain data. This is done because both scenarios, when SASI is present and when it is not, are needed in order to evaluate the pipeline's detection efficiency (as it will be discussed in step 4). Each of these cWB XP implementations will generate numerous TF maps, some of which the GW-\texttt{SASImeter} will work on directly. These maps are also referred to as \textit{triggers} or \textit{pixel files} due to the fact that each data item contains three entries: time, frequency and CRS. To produce these files, a special cWB plugin is used to specifically dump them; an example of such reconstructed GW signals, for the case where SASI activity is present in the O4 injections, can be observed in figure \ref{Fig_Reconstructions}. In contrast, the following two steps are executed cyclically for a specific number of triggers for each of the six sets of TF maps.

\textit{2. High Frequency Features parameter estimation}: The GW HFF, which are produced by PNS pulsations during the CCSNe and which are mainly generated by the so-called oscillation modes \textit{f} and \textit{g}, are post–core-bounce signals whose frequencies begin at a few hundred Hertz and grow over time up to 1–2 kHz \cite{Michele_Tony_CCSNe}. Many physical properties from the CCSNe and its progenitor can be extracted from this particular period of GW emission. Consistently to the literature review performed in Lin2023 \cite{Zidu_SASImeter_Joint}, it is assumed that SASI activity appears not earlier than $50 \; \text{ms}$ after the HFF start. A sufficiently accurate estimation of the HFF starting time is achieved with a $\chi^2$ minimization that approximates the frequency growth as linear. The characterization of the HFF is, on its own, a topic of research, see, for example \cite{Casallas2023, Casallas2025, HFF_Slope_Daniel}, and \cite{Casallas_DCC} for an inclusion of the curvature. In the SASImeter module, the estimated slope of the HFF is not used.

Hence, step 2 consists of estimating the slope of the HFF and determining the time at which it intersects with the $200 \; Hz$ frequency band, which is where the SASI GW imprint is expected to be located. To do so, the first sub-step is to isolate the HFF region; for this, pixels with $f \leq 200 \; \text{Hz}$ are removed from the TF map and the remaining pixels, all in the high-frequency region ($> 200 \; \text{Hz}$), are passed through a CRS filter, retaining only those with a CRS greater than the CRS arithmetic mean across all pixels in this region. Then, a secondary CRS filter is applied: the remaining pixels are divided in time windows of 0.2 seconds and only those in the window with the largest total CRS sum are preserved. Throughout the GW-\texttt{SASImeter}, different CRS filters will be applied in order to ensure that the pixels used on the final parameter estimations are those corresponding to \textit{real} signal-related energy and not to interferometric noise. If in any of the above sub-steps there are fewer than two pixels left in the HFF region, then it is assumed that there are no \textit{usable} pixels in it, and the GW-\texttt{SASImeter} skips to step 3. Otherwise, pixels in the most energetic window are then used to compute an initial estimation of the HFF slope via a least square linear regression and, with it, the slope's intercept with the $200 \; \text{Hz}$ frequency band is obtained. Then, this first estimation is evaluated: if the slope does not lie in the interval $[500,5000] \; s^{-2}$, which was determined in Lin2023 \cite{Zidu_SASImeter_Joint} as the physically acceptable range for the HFF slope, then the event, i.e., this particular TF map, is discarded for the analysis and the GW-\texttt{SASImeter} restarts for another one.

In the case where the slope is within the physically acceptable range, then this first estimation and the remaining pixels in the high-frequency band are used for a secondary and more robust slope estimation; for the GW-\texttt{SASImeter} goals, a $\chi^2$ minimization process is sufficient enough to provide a more confident final estimation. This process takes the initial HFF slope ($m$) and intercept ($c$) estimations and vary their values $33\%$ to generate 200 new samples and, with them, create a $200 \times 200$ grid in  the $m \times c$ parameter space. Then, the \textit{i}-th element of the $\chi^2$ surface is computed as:

\begin{equation}
    \chi^2_i = \sum_j \rho_j (w_j)^2 \Xi_j \left[ t_j - \frac{f_j - c_i}{m_i} \right]^2, \label{eq_chi_sqr}
\end{equation}

\noindent where $\rho_j$, $t_j$, and $f_j$ are the CRS, time and frequency of the \textit{j}-th pixel in the final HFF region, respectively; $(w_j)^2$ is a weight function used to compensate for the detectors' noise curves, $\Xi_j$ is a measure of the density of pixels around the \textit{j}-th pixel, and $m_i$ and $c_i$ are elements of the $m \times c$ grid. Then, the final slope and intercept estimations are the $m_i$ and $c_i$ corresponding to the minimum $\chi^2_i$ value.

\textit{3. SASI parameter estimation}: This next step starts by isolating the SASI region. This is done by, from the original TF map, defining such region as that of pixels with a frequency $\leq 200 \; \text{Hz}$ and a time greater or equal than $t_i^\text{SASI} = t_i^{\text{HFF}} + 50\; \text{ms}$, where $t_i^{\text{HFF}}$ is the HFF initial time estimated through the slope and its intercept obtained from the previous step, i.e., $t_i^{\text{HFF}} = \frac{200 - c_i}{m_i}$; or set as zero if the HFF parameter estimation stage was skipped due to a lack of usable pixels in the region. After isolating pixels in the SASI region, these are passed through a CRS filter, retaining only those with a CRS greater than half the maximum CRS in the region.

With the SASI region defined and filtered, the next step is to perform the cWB XP central frequency and time duration estimations. The first one is computed as the weighted average of the frequency with respect to the CRS of all pixels in the region:

\begin{equation}
    f_\text{GW}^c = \frac{\sum_i \rho_i^\text{SASI} f_i^\text{SASI}}{\sum_i \rho_i^\text{SASI}}, \label{eq_f_c_sasi}
\end{equation}

\noindent where $\rho_i^\text{SASI}$ and $f_i^\text{SASI}$ are, respectively, the CRS and frequency of the \textit{i}-th pixel in the SASI region. 

Unlike previous implementations, in this one the time duration is computed in two different ways. The first one, also used in Lin2023 \cite{Zidu_SASImeter_Joint}, is obtained as:

\begin{equation}
    \begin{aligned}
        \tau_\text{GW}^r & = \left( t_f^\text{SASI} + \frac{\delta t_f^\text{SASI}}{2} \right) - \left( t_0^\text{SASI} + \frac{\delta t_0^\text{SASI}}{2} \right), \\
        & = t_f^\text{SASI} - t_0^\text{SASI},
    \end{aligned} \label{eq_time_dur_f1}
\end{equation}

\noindent where $t_f^\text{SASI}$ and $t_0^\text{SASI}$ are the time of the edge pixels in the SASI region and $\delta t_f^\text{SASI}$ and $\delta t_0^\text{SASI}$ are their associated time resolutions; one of the first noticeable differences between cWB XP and 2G is that, for the first, these two are identical and so are canceled, something that does not occur for the latter. It is noticed that this way of computing the time duration of the SASI is clearly a very relaxed one, considering all of the remaining pixels in the SASI region as truly SASI-associated pixels. 

In contrast, the second and new time duration estimation is more conservative. It starts by computing the central time of the SASI region in the same fashion as for the central frequency:

\begin{equation}
    t_\text{GW}^c = \frac{\sum_i \rho_i^\text{SASI} t_i^\text{SASI}}{\sum_i \rho_i^\text{SASI}}, \label{eq_t_c_sasi}
\end{equation}

\noindent where now $t_i^\text{SASI}$ is the time of the \textit{i}-th pixel in the SASI region. Then, with this central time, the standard deviation is obtained as:

\begin{equation}
    \sigma_{\text{GW}_t} = \sqrt{\frac{\sum_i \left( t_i^\text{SASI} - t_\text{GW}^c \right)^2}{N-1}}, \label{eq_sigma_t_sasi}
\end{equation}

\noindent where $N$ is the amount of pixels in the SASI region. Finally, the time duration is defined as:

\begin{equation}
    \begin{aligned}
        \tau_\text{GW}^\sigma & = t_{f_\sigma}^\text{SASI} - t_{0_\sigma}^\text{SASI}, \\
        & = (t_\text{GW}^c + \sigma_{\text{GW}_t}) - (t_\text{GW}^c - \sigma_{\text{GW}_t}), \\ 
        & = 2 \sigma_{\text{GW}_t}.
    \end{aligned} \label{eq_time_dur_f2}
\end{equation}

\noindent In general, $\tau_\text{GW}^\sigma < \tau_\text{GW}^r$, which confirms that, indeed, the second estimation is more conservative as it underestimates the time duration with respect to the first. Without further observational information on the time duration of the SASI, both of these estimations are considered as valid.

The final sub-step of this stage requires the definition of a metric that sets the first step into evaluating the GW-\texttt{SASImeter} detection and identification performance. This, defined in Lin2023 \cite{Zidu_SASImeter_Joint}, is proposed to be the normalized likelihood of the SASI pixels with respect to the whole trigger, defined as:

\begin{equation}
    \rho_\text{GW} = \frac{\sum_ i \rho_i^\text{SASI}}{\sum_j \rho_j^\text{All}}, \label{eq_GW_likelihood}
\end{equation}

\noindent which leverages on the relative importance of the detected energy in the time-frequency region of the SASI with respect to the overall energy of the event.

\textit{4. Receiver Operating Characteristic Analysis}: This is the final stage of the GW-\texttt{SASImeter}, and its goal is to evaluate the algorithm's performance at identifying the SASI signal from noise. Hence, this step is performed once for every pair of SASI (containing SASI activity) and No-SASI (not containing SASI activity) analysis (i.e., once per distance), for which the normalized likelihood, equation \eqref{eq_GW_likelihood}, plays a central role since it measures the ratio between the energy content of the SASI region and that of the rest of the whole event. Particularly, it is noticed that the absence of SASI activity on the injected waveform does not mean that the likelihood will be exactly zero; in fact, triggers with $\rho_\text{GW}=0$ are discarded from the ROC analysis since, even in this case, low energy pixels in the SASI region are expected from interferometric noise.

Hence, in the context of the ROC analysis, the fraction of events above a given threshold in $\rho_\text{GW}$ are considered as an estimate of the probability that the normalized likelihood is indeed above such threshold. Given this, the identification probability, $P^\text{GW}_\text{D}$, is obtained from the probability density function (PDF) of $\rho_\text{SASI}$, which corresponds to triggers with SASI activity; on the other hand, the false identification probability, $P^\text{GW}_\text{FI}$, is calculated from the PDF of $\rho_\text{No-SASI}$, which now corresponds to triggers with no SASI activity. Both of these probabilities are computed as the cumulative area under the PDF of each case for $\rho_\text{GW} > \Lambda_\text{GW}$, being $\Lambda_\text{GW}$ the threshold:

\begin{equation}
    P^\text{GW}_\text{D} = \int_{\Lambda_\text{GW}}^{\infty} \text{PDF} (\rho_\text{SASI}) \mathrm{d}\rho, \label{eq_PD_GW}
\end{equation}

\begin{equation}
    P^\text{GW}_\text{FI} = \int_{\Lambda_\text{GW}}^{\infty} \text{PDF} (\rho_\text{No-SASI}) \mathrm{d}\rho. \label{eq_PFI_GW}
\end{equation}

\noindent Finally, the ROC curves are defined as the plot of $P^\text{GW}_\text{D}$ as a function of $P^\text{GW}_\text{FI}$ for all possible thresholds $\Lambda_\text{GW}$, with the case $P^\text{GW}_\text{D}=P^\text{GW}_\text{FI}$ as the $50/50$ classification-detection scenario, i.e., the coin-toss case.

\subsection{Neutrino Channel} \label{Sect_SASImeter_Neutrino}

\begin{figure}[h!]
    \centering
    \includegraphics[scale=0.45]{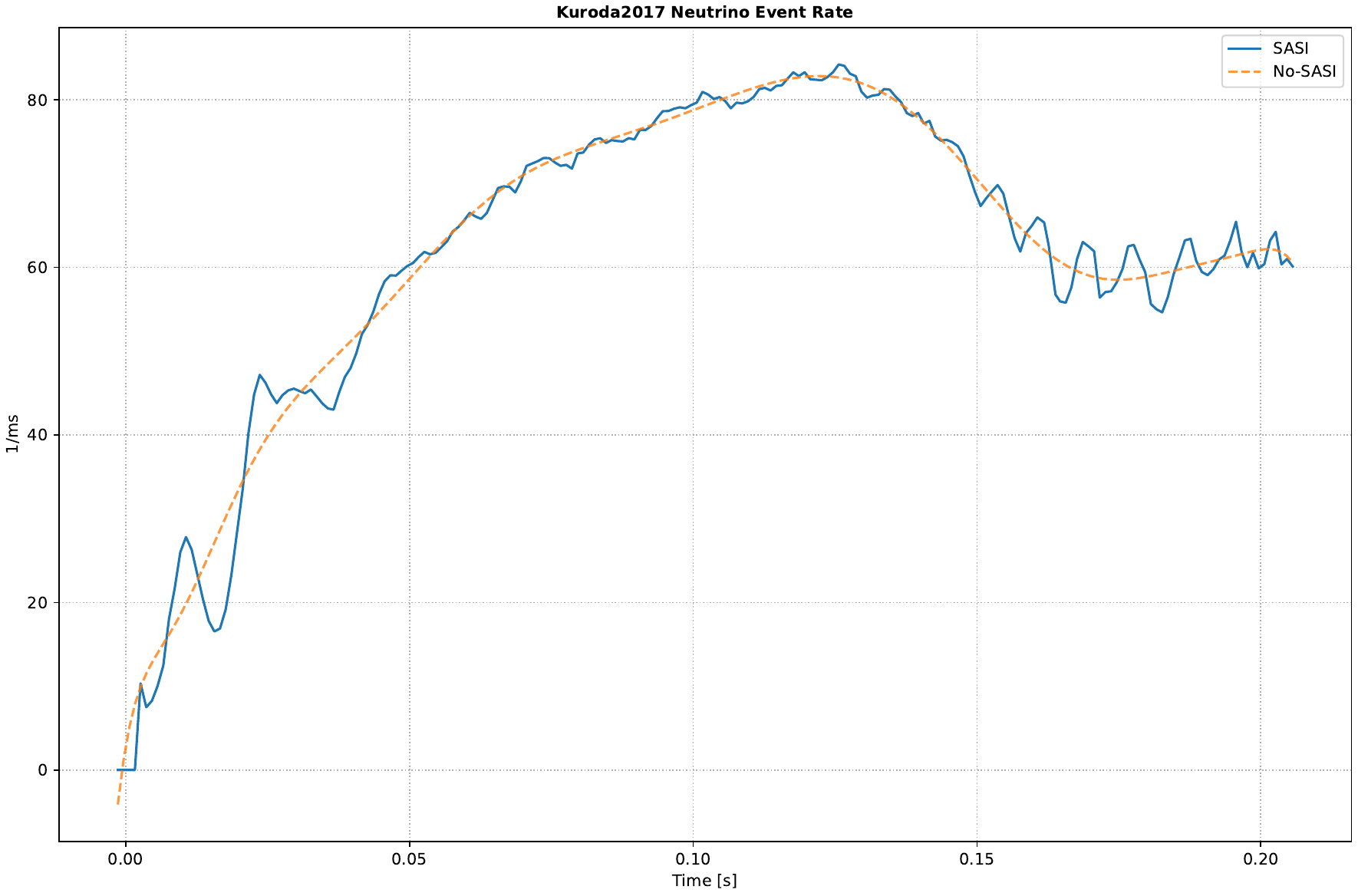}
    \caption{Kuroda2017 neutrino rate event, for both cases when SASI activity is present (continuous, blue line) and when it is removed artificially (dashed, orange line) by smoothing out the signal. The SASI activity is appreciated approximately after $t=0.145 \;s$.}
    \label{Fig_Kuroda2017_Nu_Rate}
\end{figure}

The neutrino \texttt{SASImeter} (hereon referred to as the $\nu$-\texttt{SASImeter}) is designed to be a hypothesis test for the presence of SASI in a neutrino signal from a CCSNe on a specific detector, HyperK in this case. As stated before, the neutrino signal in this analysis corresponds to the same Kuroda2017 s15.0 model \cite{Kuroda2017}; this signal can be observed in figure \ref{Fig_Kuroda2017_Nu_Rate}, where the cases where SASI activity is present and removed are shown. As prescribed in Lin2020 \cite{Zidu_SASImeter_Neutrino}, the “No-SASI” case is obtained by smoothing out the SASI oscillations from the signal; this is done by dividing the entire signal into time bins of length of 1 millisecond and averaging the event rate inside each, and then performing a polynomial interpolation onto the averaged rates.

A brief overview on how the $\nu$-\texttt{SASImeter} operates is now in place; for an in-depth explanation, refer to Lin2020 \cite{Zidu_SASImeter_Neutrino} and Lin2023 \cite{Zidu_SASImeter_Joint}. As for its GW sibling, this \texttt{SASImeter} yields a likelihood ratio sensitive to the quasi-periodic neutrino events induced by the SASI in the frequency domain:

\begin{equation}
    \mathcal{L}(\mathcal{\tilde P})=\frac{\mathrm{Max}[L(\mathcal{\tilde P}|P_\Omega)]}{\mathrm{Max}[L(\mathcal{\tilde P}|P_{\Omega_0})]}~, \label{eq:likeR}
\end{equation}

\noindent where $L(\mathcal{\tilde P}|P_{\Omega})$ and $L(\mathcal{\tilde P}|P_{\Omega_0})$ are the likelihoods that the observed neutrino power spectrum $\tilde P$ can be described by the hypothetical model of power spectrum P, which is based on the templates $\Omega$ and $\Omega_0$, the first corresponding to the case where SASI activity is present on the signal and the second where it is not, and each template being dependent on a corresponding set of parameters. First, the hypothetical neutrino events are defined, in the time domain, as:

\begin{equation}
    R_\Omega(t)=(A-n) \left[1 + a \sin(2 \pi  f_\nu t) \right] + n~, \label{eq:mod2}
\end{equation}

\begin{equation}
    R_{\Omega_0}(t)=A~, \label{eq:mod0}
\end{equation}

\noindent where $A$ is the time-averaged event rate (the ``DC component") in the detector including instrumental noise (after possible experimental cuts); $a$ is the relative SASI amplitude, $n$ is the mean value of the background rate ($n=0$ for HyperK),  and $f_\nu$ is the SASI nominal frequency. Given this, then a discrete Fourier transformation is performed as prescribed in Lin2020 \cite{Zidu_SASImeter_Neutrino}. Considering statistical fluctuations of neutrino events on a detector, the analytical probability density function of observed power $\tilde P_k$ at frequency $k$, based on template $P_k(\Omega/\Omega_0)$, is: 

\begin{equation}
    \begin{split}
        \mathrm{PDF}(\tilde{P_k}|P_k(\Omega/\Omega_0))&=\frac{N_{bins}^2}{4\sigma^2} \exp{ \left[ -\frac{N_{bins}^2}{4\sigma^2} \left(\tilde{P}_k + P_k(\Omega/\Omega_0) \right)\right]}\\
        &\times I_0\left( \frac{N_{bins}^2}{2\sigma^2} \sqrt{ \tilde{P_k}P_k(\Omega/\Omega_0)} \right)~, \label{eq:prob}
    \end{split}
\end{equation}

\noindent where $I_0$ is the modified Bessel function of the first kind, $N_{bins}$ is the number of neutrino event bins involved in the discrete Fourier transformation, and 

    \begin{equation}
        \begin{split}
        \sigma^2=\frac{N_{ev}}{2}~,
        \end{split} \label{eq:sigma2}
\end{equation}

\noindent where $N_{ev}$ is the total neutrino events. Hence, the likelihood $L(\tilde P|P_{\Omega/\Omega_0})$ is given by:

\begin{equation}
    L(\mathcal{\tilde P}|P_{\Omega/\Omega_0})=\prod_{k}\mathrm{PDF}(\mathcal{\tilde P}_k|P_k(\Omega/\Omega_0))~. \label{eq:likeli}
\end{equation}

\noindent Then, the maximum likelihood $\mathrm{Max}[L(\mathcal{\tilde P}|P_{\Omega/\Omega_0})]$ is obtained by surveying the parameter space of the template $\Omega/\Omega_0$; in turn, the selection of such maximum likelihood for each of the two possible templates determines the set of estimated parameters, which are $\{a, f_\nu \}$ and $\{Null \}$ respectively.

As in Lin2020 \cite{Zidu_SASImeter_Neutrino} and Lin2023 \cite{Zidu_SASImeter_Joint}, and as for the GW-\texttt{SASImeter}, Kuroda2017 \cite{Kuroda2017} is used for the simulated detected neutrino signal; particularly, the observed neutrino power spectrum $\tilde P$ is assumed to follow the prediction of this simulation. Due to the statistically fluctuating neutrino events, $\tilde P$ is probabilistic in nature rather than deterministic, so a simple Monte Carlo simulation is performed to generate an ensemble of $\tilde P$ from a sample of such. Correspondingly, two probability density functions of $ \ln(\mathcal{L})$ are obtained, namely $\mathrm{PDF}(\ln(\mathcal{L}_\mathrm{SASI}))$ and $\mathrm{PDF}(\ln(\mathcal{L}_\mathrm{No-SASI}))$. The former (latter) uses $\tilde P$ based on the Kuroda2017 model with (without) SASI.

Finally, the identification probability and false identification probability of the neutrino SASI-born signal is defined as:

\begin{equation}
    P_D^\nu=\int_{\ln (\mathcal{L})>\Lambda}\mathrm{PDF}(\ln (\mathcal{L}_\mathrm{SASI})) \ln (\mathrm{d} \mathcal{L}), \label{eq:pd}
\end{equation}

\begin{equation}
    P_{FI}^\nu=\int_{\ln (\mathcal{L})>\Lambda}\mathrm{PDF}(\ln (\mathcal{L}_\mathrm{No-SASI})) \ln (\mathrm{d} \mathcal{L}). \label{eq:pf}
\end{equation}

\noindent Again, the ROC curves are obtained as the plot of $P^\nu_\text{D}$ as a function of $P^\nu_\text{FI}$ for all possible thresholds. However, for the $\nu$-\texttt{SASImeter}, these are sensitive to the choice of a \textit{testing time window} since the SASI-born signal might be embedded at different times for different detections. In turn, this allows for a method to estimate the signal's starting time and duration: by determining a time window of a specific length, the neutrino signal can be divided into different blocks, each with a different starting time. Then, each time-windowed block will have a $P^\nu_\text{D}$ associated to it; by selecting a threshold for the identification probability, it is possible to discriminate among all time windows and select that (or those) for which there is a considerable probability for SASI activity with an estimated starting time and duration.

\subsection{Combined Analysis} \label{Sect_SASImeter_Joint}

The goal of the combined analysis is to evaluate the detection efficiency of SASI activity by combining the individual identification probability of each of the channels. For the individual signal channels, the identification probability, $P_\text{D}$, and the false identification probability, $P_\text{FI}$, are defined in terms of the probability density function of the each of the defined likelihood metrics for both the cases where the signal contains SASI activity and where it doesn't. Hence, in order to construct a joint identification probability, 2-dimensional probability density functions are defined as:

\begin{equation}
    \text{PDF}_\text{SASI} \left( \rho_\text{SASI}, \ln(\mathcal{L}_\text{SASI}) \right) = \text{PDF} \left (\rho_\text{SASI} \right) \times \text{PDF} \left(\ln(\mathcal{L}_\text{SASI}) \right), \label{eq_PDF_Joint_SASI}
\end{equation}

\begin{equation}
    \text{PDF}_\text{No-SASI} \left( \rho_\text{No-SASI}, \ln(\mathcal{L}_\text{No-SASI}) \right) = \text{PDF} \left(\rho_\text{No-SASI} \right) \times \text{PDF} \left(\ln(\mathcal{L}_\text{No-SASI}) \right). \label{eq_PDF_Joint_No_SASI}
\end{equation}

Then, integrating over a threshold on these PDFs yields the detection probabilities. However, in the joint analysis, there are infinite ways to stablish the thresholds. Following the procedure of Lin2023 \cite{Zidu_SASImeter_Joint}, three ways of combining the threshold are explored. The first one employs a \textit{logical and} structure, meaning that the SASI identification is passed when both the individual channels satisfy $\rho_\text{GW} > \Lambda_\text{GW}$ and $\ln (\mathcal{L}) > \Lambda_\nu$. Hence, the combined identification probability and the combined false identification probability are:

\begin{equation}
    \begin{aligned}
        P^\text{and}_\text{D} & = P^\text{GW}_\text{D} \times P^\nu_\text{D}, \\ 
        & = \int_{\Lambda_\text{GW}}^{\infty} \int_{\Lambda_\nu}^{\infty} \text{PDF} (\rho_\text{SASI}) \times \text{PDF} (\ln(\mathcal{L}_\text{SASI})) \ln (\mathrm{d} \mathcal{L}) \mathrm{d}\rho,
    \end{aligned} \label{eq_PD_Joint_And}
\end{equation}

\begin{equation}
    \begin{aligned}
        P^\text{and}_\text{FI} & = P^\text{GW}_\text{FI} \times P^\nu_\text{FI}, \\ 
        & = \int_{\Lambda_\text{GW}}^{\infty} \int_{\Lambda_\nu}^{\infty} \text{PDF} (\rho_\text{No-SASI}) \times \text{PDF} (\ln(\mathcal{L}_\text{No-SASI})) \ln (\mathrm{d} \mathcal{L}) \mathrm{d}\rho.
    \end{aligned} \label{eq_PFI_Joint_And}
\end{equation}

The next threshold combination uses a \textit{logical or} structure, meaning that the SASI identification is performed when either  $\rho_\text{GW} > \Lambda_\text{GW}$ or $\ln (\mathcal{L}) > \Lambda_\nu$ are satisfied. With this, the combined identification probability and the combined false identification probability are:

\begin{equation}
    \begin{aligned}
        P^\text{Or}_\text{D} & = 1 - \left( 1 - P^\text{GW}_\text{D} \right) \times \left( 1 - P^\nu_\text{D} \right), \\ 
        & = 1 - \int^{\Lambda_\text{GW}}_0 \int^{\Lambda_\nu}_0 \text{PDF} (\rho_\text{SASI}) \times \text{PDF} (\ln(\mathcal{L}_\text{SASI})) \ln (\mathrm{d} \mathcal{L}) \mathrm{d}\rho,
    \end{aligned} \label{eq_PD_Joint_Or}
\end{equation}

\begin{equation}
    \begin{aligned}
        P^\text{Or}_\text{FI} & = 1 - \left( 1 - P^\text{GW}_\text{FI} \right) \times \left( 1 - P^\nu_\text{FI} \right), \\ 
        & = 1 - \int^{\Lambda_\text{GW}}_0 \int^{\Lambda_\nu}_0 \text{PDF} (\rho_\text{No-SASI}) \times \text{PDF} (\ln(\mathcal{L}_\text{No-SASI})) \ln (\mathrm{d} \mathcal{L}) \mathrm{d}\rho.
    \end{aligned} \label{eq_PFI_Joint_Or}
\end{equation}

Finally the third threshold combination is labeled as a \textit{mixed} one, meaning that the SASI identification occurs for $f(\rho, \; \mathcal{L}) > \Lambda$, where $f(\rho, \; \mathcal{L})$ is an arbitrary function of the likelihood metrics, $\rho_\text{GW}$ and $\mathcal{L}$. Again, as in Lin2023 \cite{Zidu_SASImeter_Joint}, two scenarios are taken for this combination: $\rho \times \mathcal{L} > \Lambda$ and $\rho + \mathcal{L} > \Lambda$. The detection probabilities of each of these cases are:

\begin{equation}
    P^{\rho \times \mathcal{L}}_\text{D} = \int_{\rho \times \mathcal{L}> \Lambda }^{\infty} \text{PDF} (\rho_\text{SASI}) \times \text{PDF} (\ln(\mathcal{L}_\text{SASI})) \ln (\mathrm{d} \mathcal{L}) \mathrm{d}\rho,\label{eq_PD_Joint_Times}
\end{equation}

\begin{equation}
    P^{\rho \times \mathcal{L}}_\text{FI} = \int_{\rho \times \mathcal{L}> \Lambda }^{\infty} \text{PDF} (\rho_\text{No-SASI}) \times \text{PDF} (\ln(\mathcal{L}_\text{No-SASI})) \ln (\mathrm{d} \mathcal{L}) \mathrm{d}\rho. \label{eq_PFI_Joint_Times}
\end{equation}

\begin{equation}
    P^{\rho + \mathcal{L}}_\text{D} = \int_{\rho + \mathcal{L}> \Lambda }^{\infty} \text{PDF} (\rho_\text{SASI}) \times \text{PDF} (\ln(\mathcal{L}_\text{SASI})) \ln (\mathrm{d} \mathcal{L}) \mathrm{d}\rho,\label{eq_PD_Joint_Sum}
\end{equation}

\begin{equation}
    P^{\rho + \mathcal{L}}_\text{FI} = \int_{\rho + \mathcal{L}> \Lambda }^{\infty} \text{PDF} (\rho_\text{No-SASI}) \times \text{PDF} (\ln(\mathcal{L}_\text{No-SASI})) \ln (\mathrm{d} \mathcal{L}) \mathrm{d}\rho. \label{eq_PFI_Joint_Sum}
\end{equation}

\section{Results} \label{Sect_Results}

\subsection{Gravitational Waves Channel - O3} \label{Sect_Results_GW_O3} 

\begin{figure}[h!]
    \begin{center}
    	 \includegraphics[width=0.4\textwidth]{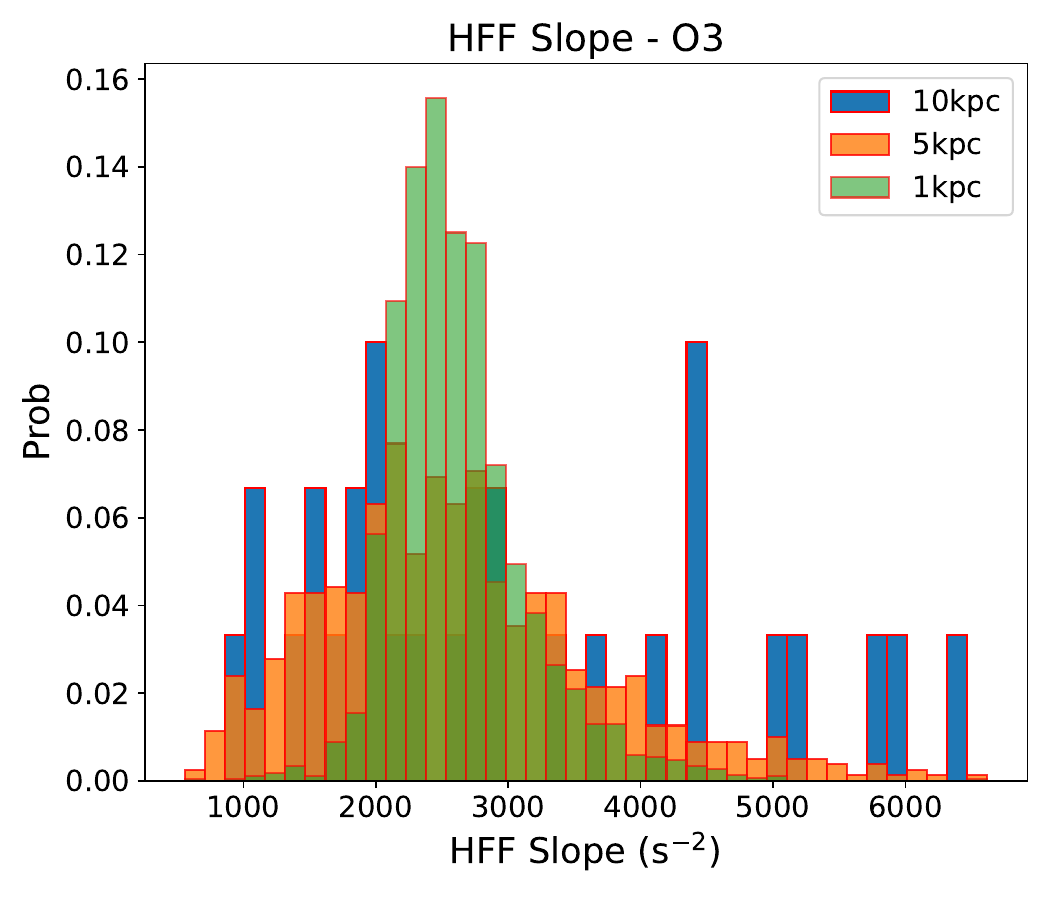}
        \includegraphics[width=0.4\textwidth]{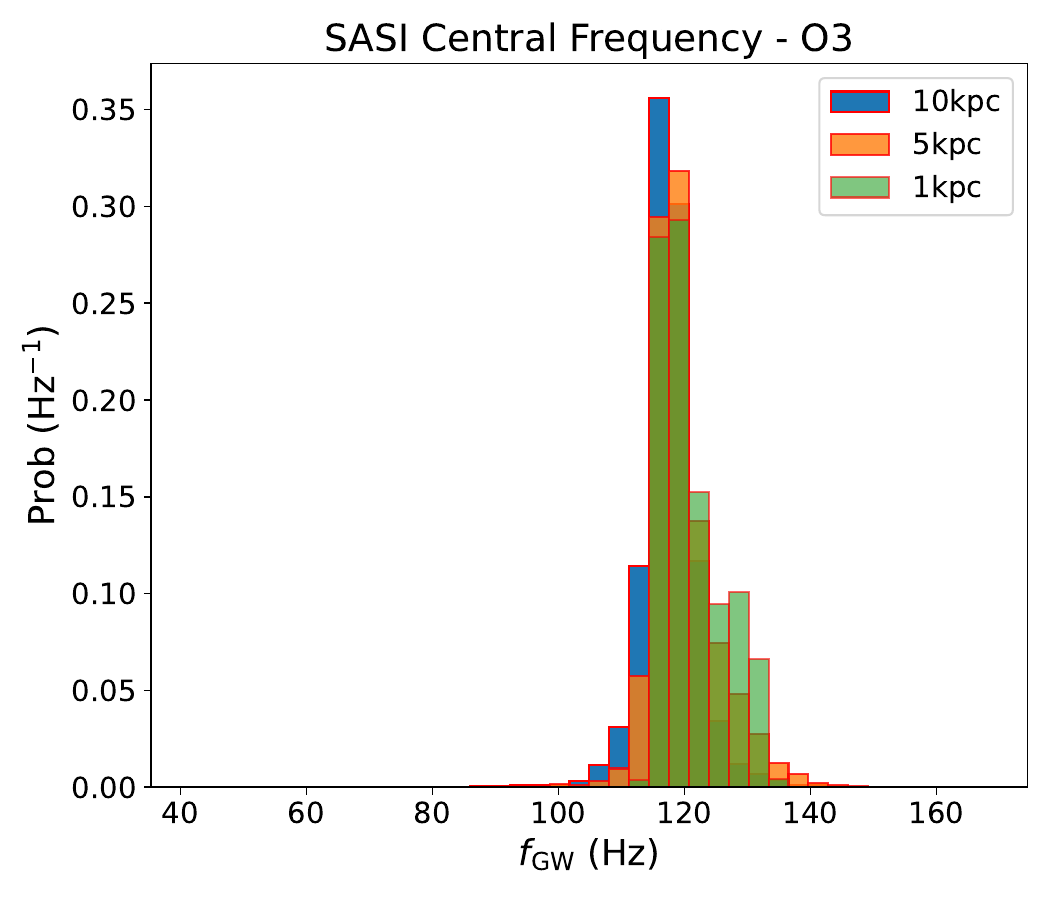}
        \includegraphics[width=0.4\textwidth]{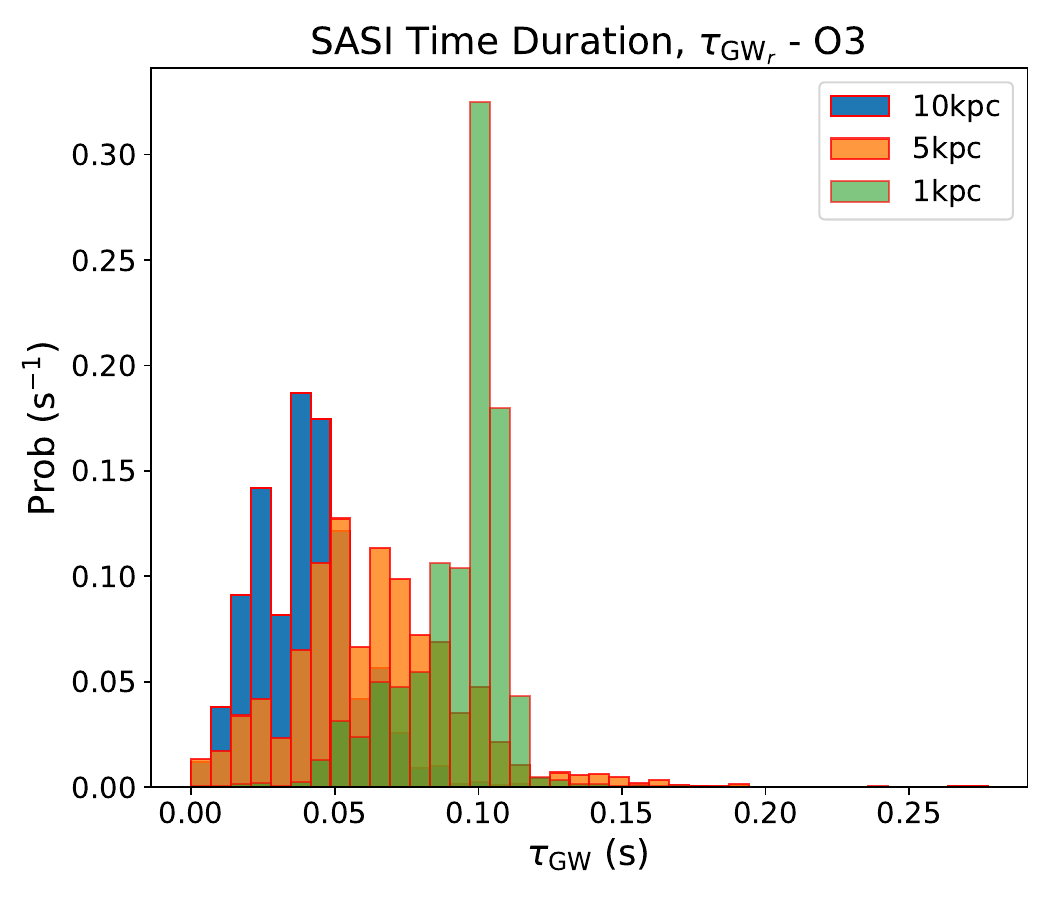}
        \includegraphics[width=0.4\textwidth]{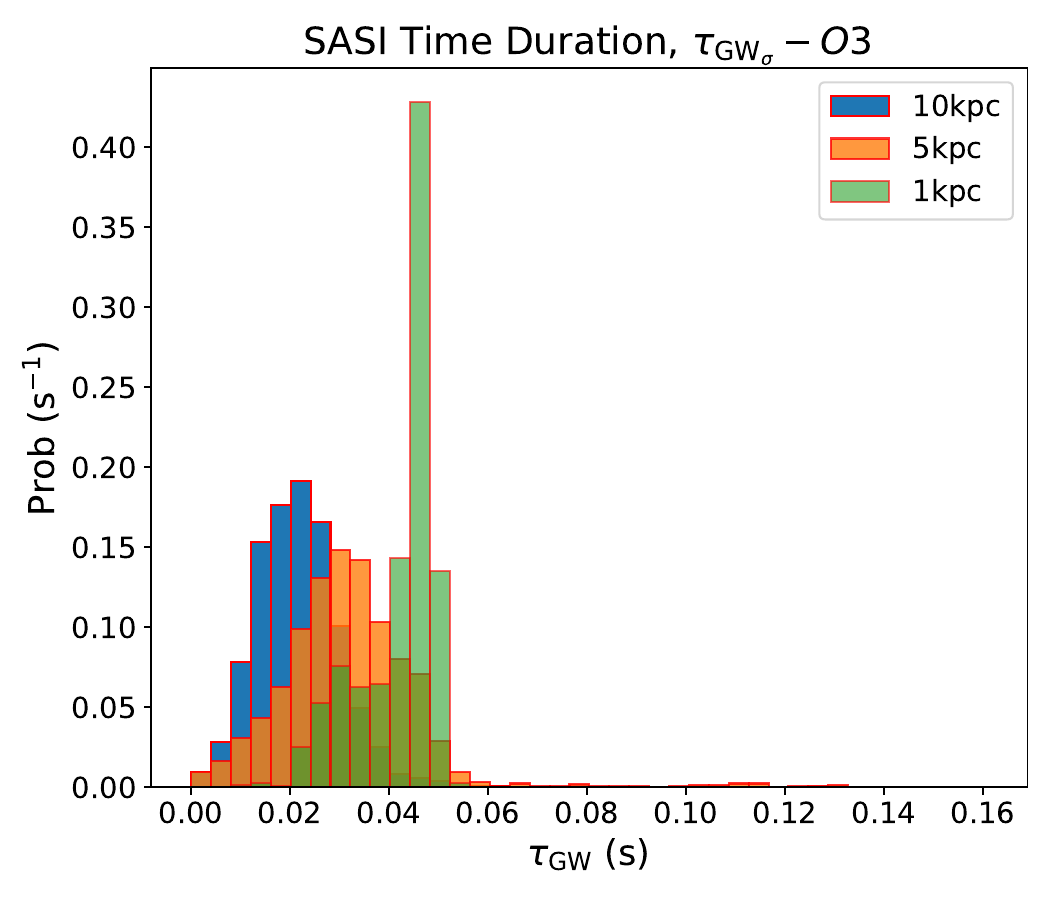}
    \end{center}
    \caption{Probability distributions from injections in O3 at 1, 5 and 10 kpc of the HFF slope estimations and SASI estimations of the central frequency and time durations computed from equation \eqref{eq_time_dur_f1}, $\tau_\text{GW}^r$, and from equation \eqref{eq_time_dur_f2}, $\tau_\text{GW}^\sigma$. }
    \label{Fig_PE_distributions_O3}
\end{figure}

For the GW-\texttt{SASImeter}, there are two sets of results. The first is the parameter estimations for the HFF slope and the SASI central frequency and time duration; the mean values of these estimations across all samples, for each distance and in O3 injections, are displayed in table \ref{Table_GW_PE_O3}. For the analysis where the signal included SASI activity, at 1 kpc 3097 triggers were used, out of which 78 had no pixels in the HFF region, 92 were discarded due to the HFF slope being out of the physically acceptable range, and 3000 where used for computing the $\rho_\text{GW}$ metric, with only 5 of these (which represent $0.16\%$ of the total number of the $\rho_\text{GW}$ computations) having no pixels in the SASI region. At 5 kpc, 3688 triggers were used on the analysis, from which 2274 had no pixels in the HFF region, 621 were discarded and 3000 were used for computing the SASI pixel likelihood, out of which only 67 ($2.23\%$) had $\rho_\text{GW}=0$. At 10 kpc, 3108 triggers were used, with only 105 of them having pixels in the HFF region and 75 out of these being discarded due to the slope laying outside the physically acceptable range; 3000 triggers were used to compute the $\rho_\text{GW}$ metric, with only 33 ($1.1 \%$) having no pixels in the SASI region. Also, in figure \ref{Fig_PE_distributions_O3}, the probability distributions for the estimated parameters at different distances are displayed, noticing that the closer the source, the less disperse the distribution is.

\renewcommand{\arraystretch}{1.3}
\setlength{\tabcolsep}{8pt}
\begin{table}[h!]
\centering
\resizebox{0.85\textwidth}{!}{%
\begin{tabular}{lllll}
& & \textbf{O3}\\
\hline
HFF Slope                    & SASI                                  & 1 kpc     & 5 kpc     & 10 kpc    \\ \hline
                                & $f_\text{GW}^c$ (Hz)                & 121.1966  & 119.7670  & 117.5006  \\ \cline{2-5} 
                                & $\delta f_\text{GW}^c$ (Hz)         & 4.8289    & 5.5790    & 5.0454    \\ \cline{2-5} 
                                & $\tau_\text{GW}^r$ (ms)             & 92.4960    & 63.5624   & 40.0481   \\ \cline{2-5} 
                                & $\delta \tau_\text{GW}^r$ (ms)      & 17.7414   & 28.9618   & 18.8267   \\ \cline{2-5} 
                                & $\tau_\text{GW}^\sigma$ (ms)        & 42.1964   & 32.0393   & 21.7379   \\ \cline{2-5} 
                                & $\delta \tau_\text{GW}^\sigma$ (ms) & 8.8476    & 15.0436   & 9.2095    \\ \hline
$m_\text{GW}$ ($s^{-2}$)        &                                       & 2622.7613 & 2614.1424 & 3013.4402 \\ \hline
$\delta m_\text{GW}$ ($s^{-2}$) &                                       & 517.0200  & 1037.6926  & 1575.4167 \\ \hline
\end{tabular}%
}
\caption{Mean values of the estimated SASI central frequency ($f_\text{GW}^c$), time duration ($\tau_\text{GW}^r, \; \tau_\text{GW}^\sigma$) and HFF slope ($m_\text{GW}$) across 1, 5, and 10 kpc, and their respective standard deviations ($\delta$), in O3 data.}
\label{Table_GW_PE_O3}
\end{table}

Comparison of the parameter estimation results with those from Lin2023 \cite{Zidu_SASImeter_Joint} show the first key differences between the two implementations. First, for the central frequency estimations, it is noted that both sets of results show a coincidental tendency of mean values around $f^c_\text{GW} \sim 120 \; \text{Hz}$; however, it is the standard deviations associated to these estimates that have changed. In Lin2023, the largest standard deviation corresponds to $18.65  \; \text{Hz}$ (10 kpc) and the smallest one to $5.48 \; \text{Hz}$ (1 kpc), while for the cWB XP O3 results, the largest standard deviation is $5.57 \; \text{Hz}$ (5 kpc) and the smallest is $4.82 \; \text{Hz}$ (1 kpc). This overall decrease in the standard deviations is translated into less disperse distributions of estimations, as observed in the histograms for the central frequency in figure \ref{Fig_PE_distributions_O3} and in Lin2023; this, in turn, also translates into a better estimation confidence for the new results in O3. Furthermore, these also show that the standard deviation at 10 kpc is smaller than at 5 kpc, which wasn't the case in Lin2023. The explanation for this is the amount of reconstructed pixels due to cWB XP and 2G; since the newest version uses the CRS as an additional metric for correlated energy across detectors, less selected pixels are expected at larger distances, translating into a narrower distribution.

Second, with respect to the time duration estimations, the only direct comparison that can be made between Lin2023 and these results are those for equation \eqref{eq_time_dur_f1}. In the first, these estimations ranged from about $150 \; \text{ms}$ to up to almost $500 \; \text{ms}$, with the largest standard deviation being $552 \; \text{ms}$ (5 kpc) and the smallest $261  \; \text{ms}$ (1 kpc), all of these larger than their respective estimations themselves. In contrast, the implementation of cWB XP in O3 show estimations of less than $100  \; \text{ms}$ and with associated standard deviations being less than half of the estimated values for 5 and 10 kpc, and even smaller for 1 kpc; particularly, the largest standard deviation is $28.96\; \text{ms}$ (5 kpc) and the smallest is $17.74 \; \text{ms}$ (1 kpc). This, again, shows that the introduction of cWB XP turns into narrower distributions of the estimated parameters, as clearly seen from the histogram for $\tau_\text{GW}^r$ in figure \ref{Fig_PE_distributions_O3}. Furthermore, the time estimations provided by equation \eqref{eq_time_dur_f2} reflect its conservative spirit, providing smaller values for both the estimations and their standard deviations with respect to the first time duration estimations.

\begin{figure}[h!]
	\begin{center}
		\includegraphics[width=0.3\textwidth]{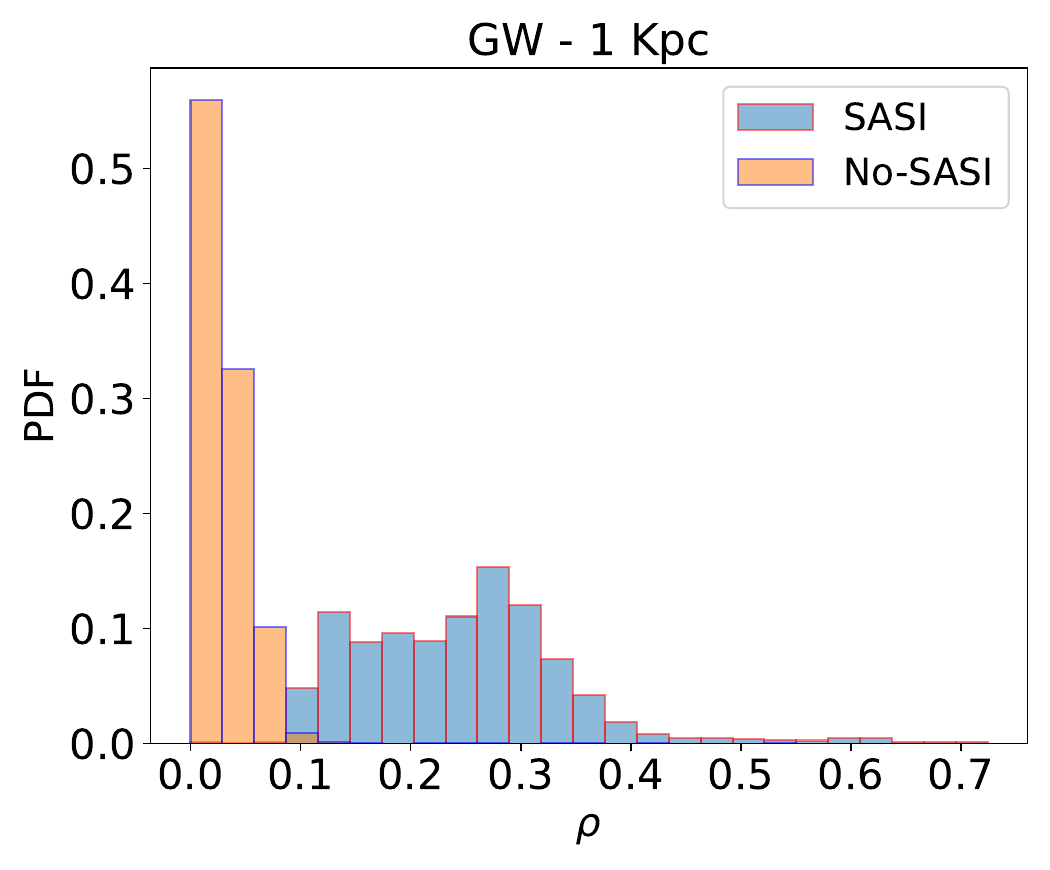}
		\includegraphics[width=0.3\textwidth]{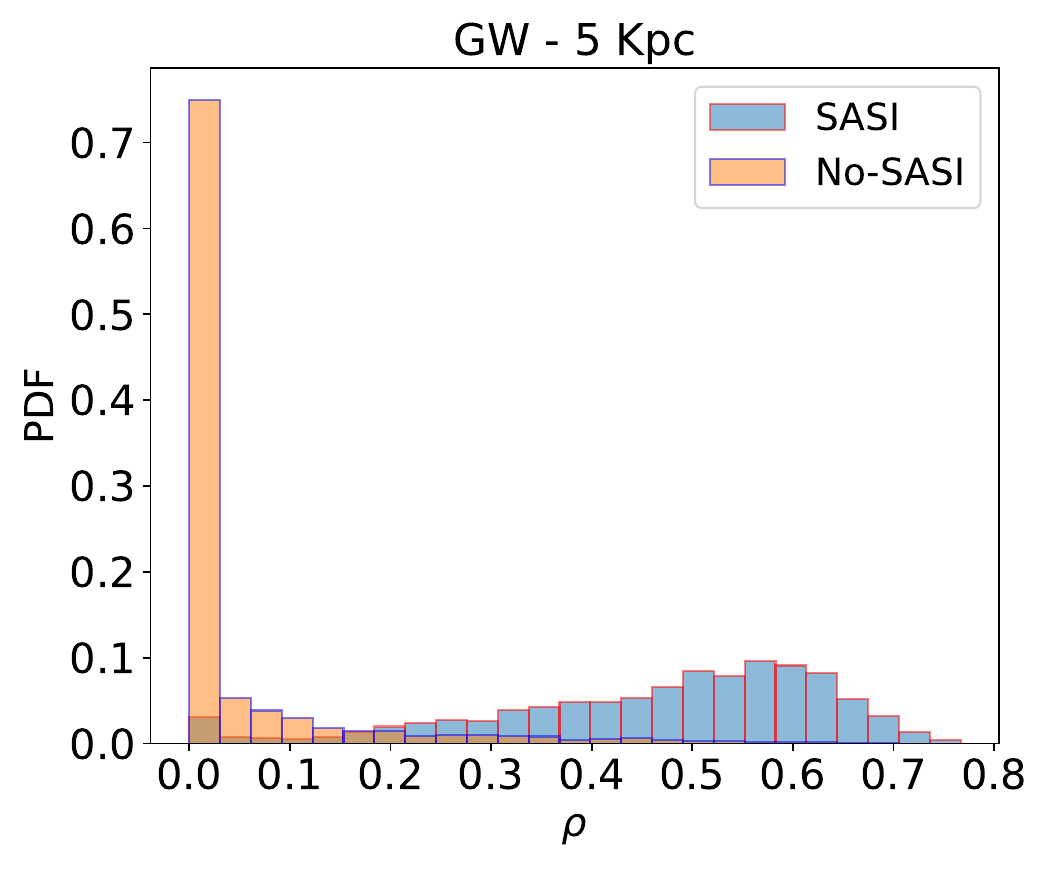}
		\includegraphics[width=0.3\textwidth]{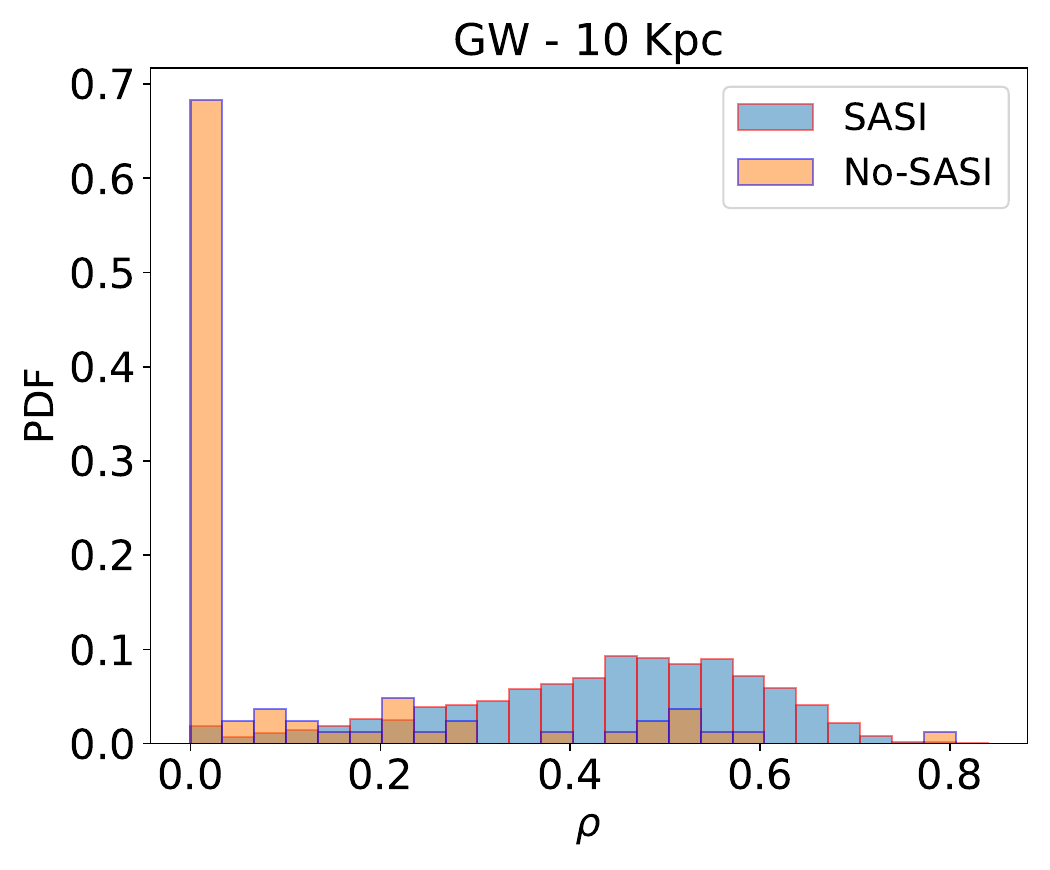}
        \includegraphics[width=0.3\textwidth]{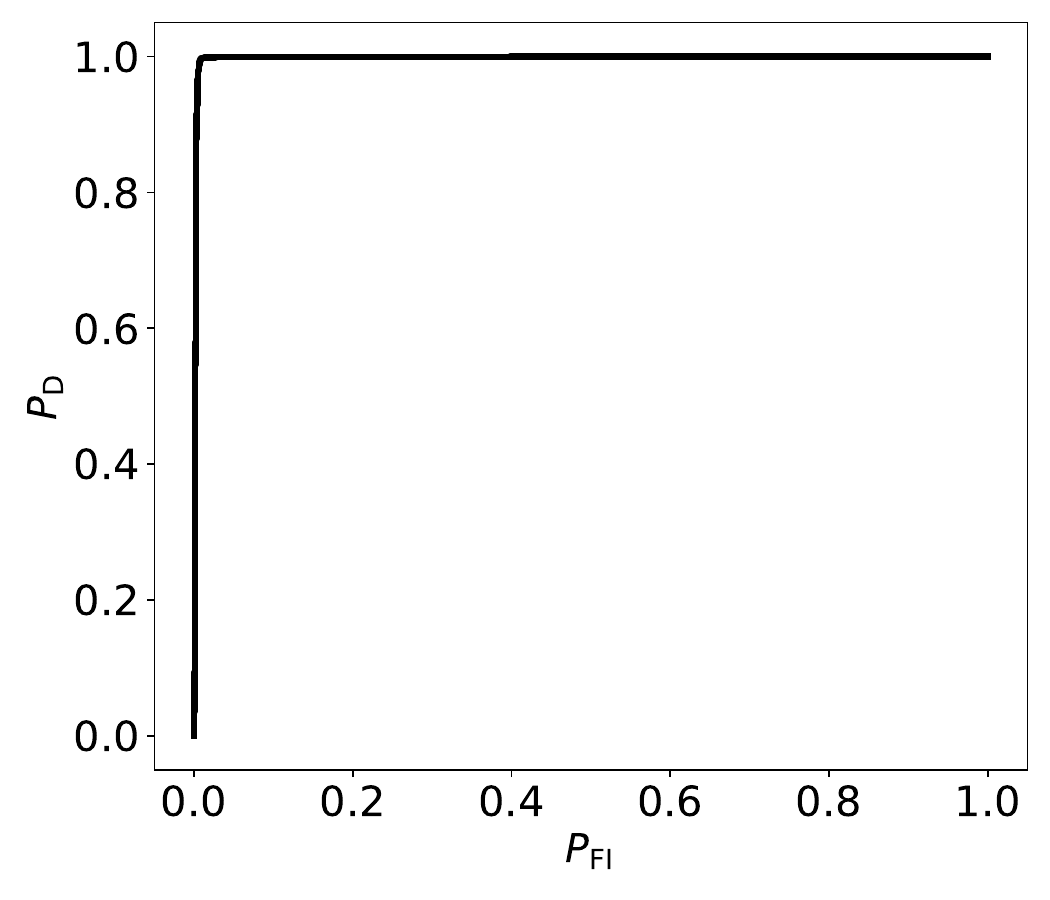}
        \includegraphics[width=0.3\textwidth]{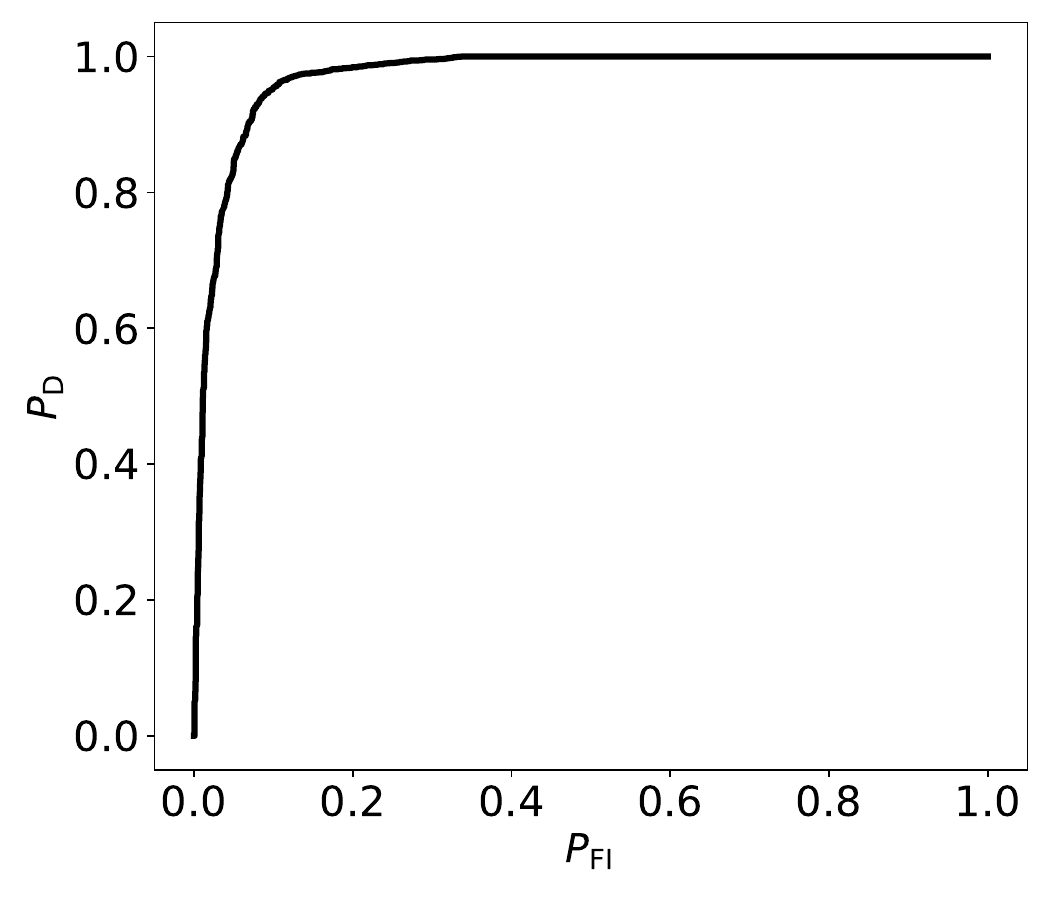}
        \includegraphics[width=0.3\textwidth]{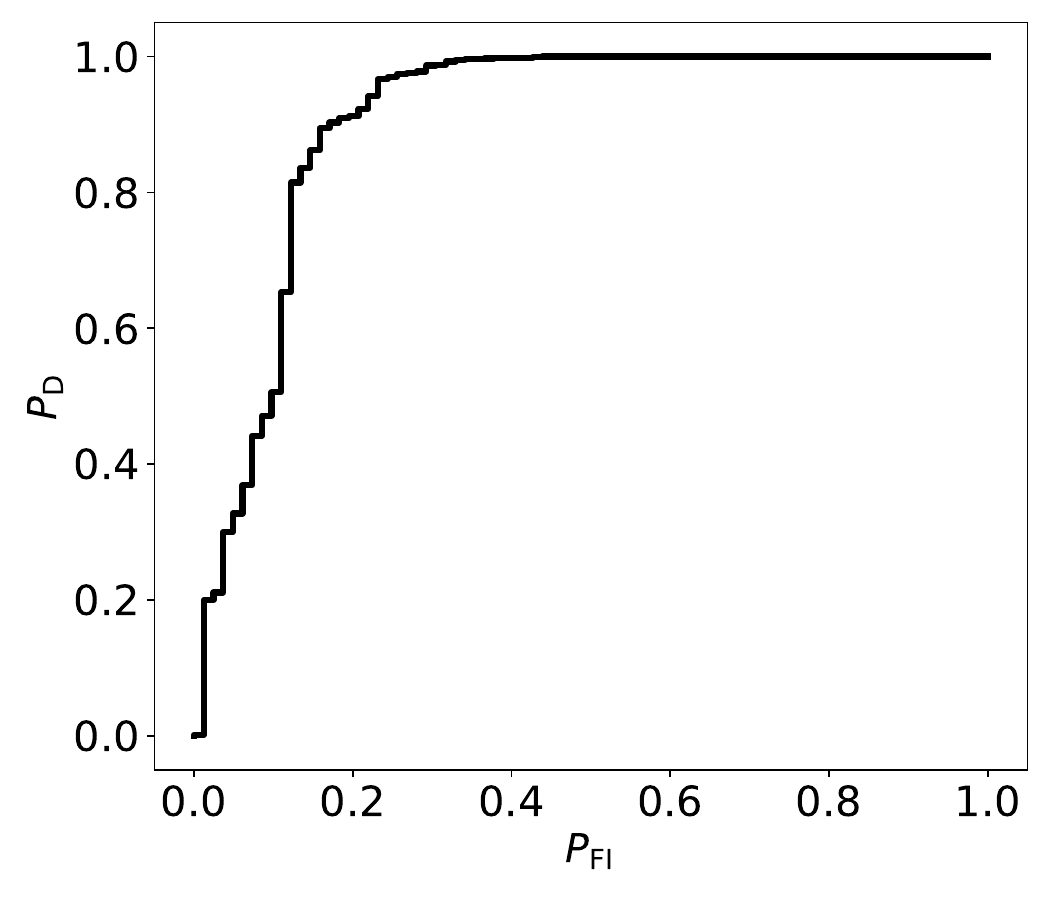}
	\end{center}
	\caption{\textit{Top row:} probability distributions of the $\rho_\text{GW}$ metric at 1, 5 and 10 kpc for the cases where SASI activity is present and removed from the injected waveform in O3. \textit{Bottom row:} ROC curves for the probability distributions.}
	\label{Fig_ROC_GWs_O3}
\end{figure}

The next set of results is that from the ROC analysis, which can be observed in figure \ref{Fig_ROC_GWs_O3}. The top row displays the probability density function of the $\rho_{\text{GW}}$ metric, equation \eqref{eq_GW_likelihood}, for both the cases when SASI activity is and is not present on the injected signal. For these histograms, the ``No-SASI'' case is expected to exhibit a denser PDF around close-to-zero $\rho_{\text{GW}}$, meaning that the GW-\texttt{SASImeter} assigns a low likelihood to these events of having SASI-related pixels in the SASI region. On the other hand, the ``SASI'' case is expected to have a broader distribution far from zero $\rho_{\text{GW}}$, meaning that the GW-\texttt{SASImeter} assigns a higher likelihood (relative to a zero-likelihood) that events will have pixels in the SASI region. Furthermore, on an ideal scenario where the identification of SASI activity is perfect, these two probability distributions are expected to be well differentiated from one another (i.e., no overlap), meaning that the GW-\texttt{SASImeter} is \textit{perfectly} identifying both cases. 

This \textit{perfect} classification scenario is almost fulfilled at 1 kpc, where confusion, i.e., significant histogram overlap, occurs mostly at the edge bins and for a low $\rho_{\text{GW}}$ value. However, it is noticed that for 5 and 10 kpc, the confusion increases: the ``No-SASI'' case records higher PDF values for higher likelihood values, while in the ``SASI'' case the opposite occurs. Nevertheless, these behaviors are expected since, at larger distances, cWB's detection and reconstruction efficiency diminishes, meaning that the SASI region is more polluted with interferometric noise; in turn, this implies that the ``No-SASI'' events will exhibit non-SASI-related activity in the region, increasing the probability of higher likelihoods. At the same time, less pixels are expected to be obtained from the cWB reconstructions, which in turn means that ``SASI'' events can have low to null activity in the SASI region even if SASI activity is expected in it, as observed in figure \ref{Fig_Reconstructions}, where an example of cWB XP reconstructions in O4 is displayed for the three considered galactic distances.

The bottom row of figure \ref{Fig_ROC_GWs_O3} includes the ROC curves obtained from $\rho_\text{GW}$'s PDF via equations \eqref{eq_PD_GW} and \eqref{eq_PFI_GW}. ROC curves are generally used to evaluate binary classifiers, and that is exactly what the GW-\texttt{SASImeter} does in the context of the PDF of $\rho_\text{GW}$: a methodological quantification of how good it is at classifying activity in the SASI region as true SASI-born signals or not. Hence, at 1 kpc, it is observed that the ROC curve behaves as an almost perfect classifier, growing almost vertically on detection efficiency ($P_D$) at low false identification probability ($P_\text{FI}$) values. For 5 and 10 kpc, the confusion observed in the PDF histograms is reflected onto the ROC curves, whose behavior stray away from the perfect classifier and approach the coin-toss (50/50) classifier, i.e., a diagonal line, with detection efficiency decreasing for increasing distance.

\renewcommand{\arraystretch}{1.3}
\setlength{\tabcolsep}{8pt}
\begin{table}[h!]
\centering
\resizebox{0.85\textwidth}{!}{%
\begin{tabular}{ccccccc}
\multicolumn{7}{c}{\textbf{O3}}                                                                                                                                                         \\ \hline
$P_\text{FI}$             & \multicolumn{6}{c}{$P_\text{D}$}                                                                                                                   \\
\multicolumn{1}{c|}{}     & \multicolumn{2}{c|}{1 kpc}                    & \multicolumn{2}{c|}{5 kpc}                    & \multicolumn{2}{c}{10 kpc}                         \\ \cline{2-7} 
\multicolumn{1}{c|}{}     & \textbf{XP O3} & \multicolumn{1}{c|}{Lin2023} & \textbf{XP O3} & \multicolumn{1}{c|}{Lin2023} & \textbf{XP O3}               & Lin2023             \\ \hline
\multicolumn{1}{c|}{0.01} & \textbf{0.99}  & \multicolumn{1}{c|}{0.63}    & \textbf{0.43}  & \multicolumn{1}{c|}{0.21}    & $\mathbf{1.6\times 10^{-3}}$ & $1.1\times 10^{-3}$ \\ \hline
\multicolumn{1}{c|}{0.05} & \textbf{1.00}  & \multicolumn{1}{c|}{0.84}    & \textbf{0.84}  & \multicolumn{1}{c|}{0.39}    & \textbf{0.32}                & 0.11                \\ \hline
\multicolumn{1}{c|}{0.10} & \textbf{1.00}  & \multicolumn{1}{c|}{0.90}    & \textbf{0.95}  & \multicolumn{1}{c|}{0.52}    & \textbf{0.47}                & 0.24                \\ \hline
\multicolumn{1}{c|}{0.25} & \textbf{1.00}  & \multicolumn{1}{c|}{1.00}    & \textbf{0.99}  & \multicolumn{1}{c|}{0.75}    & \textbf{0.97}                & 0.72                \\ \hline
\multicolumn{1}{c|}{0.50} & \textbf{1.00}  & \multicolumn{1}{c|}{1.00}    & \textbf{1.00}  & \multicolumn{1}{c|}{0.95}    & \textbf{0.99}                & 0.91                \\ \hline
\end{tabular}
}
\caption{Identification probability ($P_D$) of the GW-\texttt{SASImeter} for different false identification probabilities ($P_\text{FI}$) at 1, 5, and 10 kpc, comparing the cWB XP O3 results with those from Lin2023.}
\label{Table_GW_ROC_Thresh_O3}
\end{table}

These ROC curves reflect the most important update in the GW channel for O3. Comparing this figure with figure 8 in Lin2023 \cite{Zidu_SASImeter_Joint}, it can be observed that all of the ROC curves obtained via cWB XP in O3 show an improvement with respect to those using the 2G version in the same injection time window. For preciseness, table \ref{Table_GW_ROC_Thresh_O3} shows a numerical comparison between different $P_\text{D}$ values for some relevant $P_\text{FI}$ values. At 1 kpc, it is immediately observed an improvement with respect to the previous analysis as the probability detection reaches a value of 0.90 for a $P_\text{FI}$ of 0.1, while this new implementation exhibits an even greater $P_D$ value (0.99) for a $P_\text{FI}$ ten times smaller (0.01). On the other hand, the performance at 5 and 10 kpc varies from being very similar to even better than the previous results: in Lin2023, a $P_\text{D}$ of 0.52 takes place at $P_\text{FI}$ of 0.1, while a similar $P_\text{D}$ (0.43) is obtained for $P_\text{FI}=0.01$ on the XP O3 analysis; at 10 kpc, in Lin2023, to a $P_\text{FI}$ of 0.25 corresponds a $P_\text{D}$ of 0.72, while this same false identification probability has a $P_\text{D}$ of 0.97 in this analysis. Hence, this shows, quantitatively, that the use of cWB XP for the GW-\texttt{SASImeter} in the same data improves the overall detection efficiency of the pipeline across all distances.

It is relevant to highlight that the 1 kpc ROC curve is the most important update in the GW channel in O3. Comparing again figure \ref{Fig_ROC_GWs_O3} with figure 8 in Lin2023 \cite{Zidu_SASImeter_Joint}, it can be observed that it is indeed the 1 kpc case the one that has improved the growing rate of the identification probability more drastically, exhibiting now an almost perfect-classifier behavior. As it will be further discussed in section \ref{Sect_Results_Joint}, this updates Lin2023 \cite{Zidu_SASImeter_Joint} conclusions on which channel in O3, GW or neutrinos, is more reliable at which distances.

\subsection{Gravitational Waves Channel - O4} \label{Sect_Results_GW_O4}

\begin{figure}[h!]
    \begin{center}
    	 \includegraphics[width=0.4\textwidth]{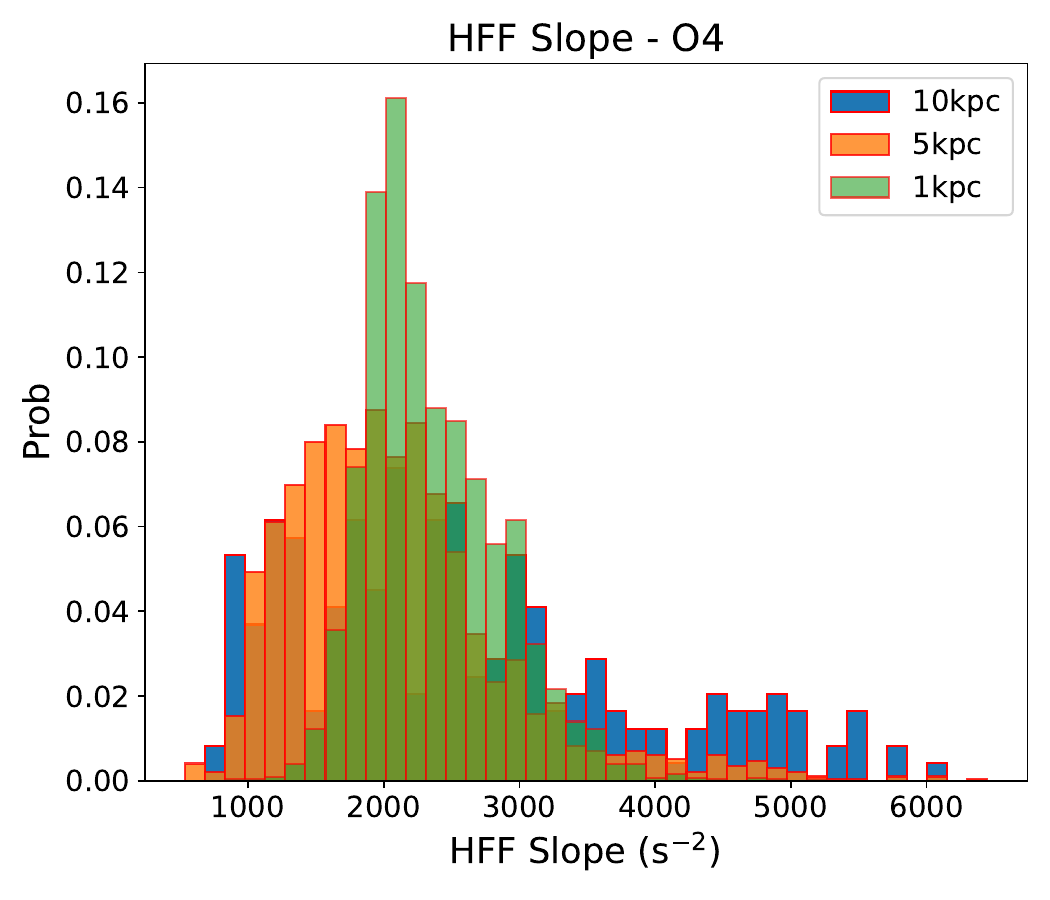}
        \includegraphics[width=0.4\textwidth]{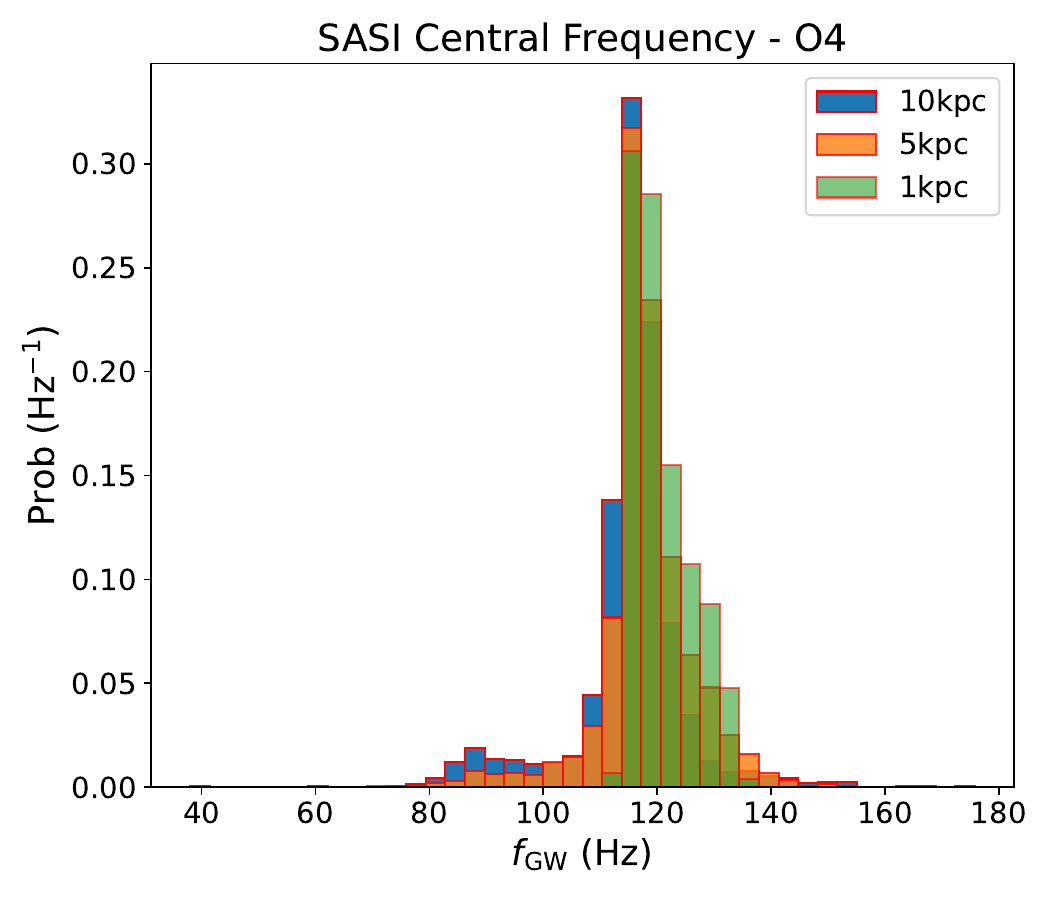}
        \includegraphics[width=0.4\textwidth]{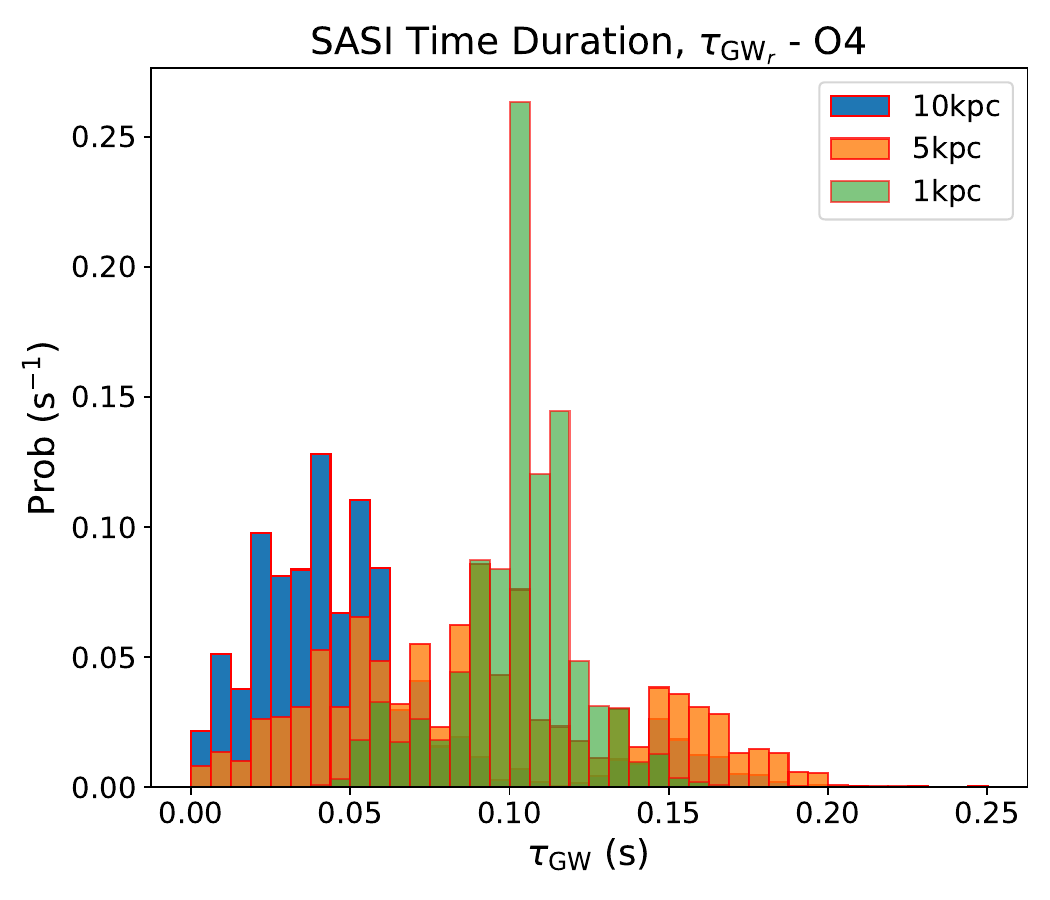}
        \includegraphics[width=0.4\textwidth]{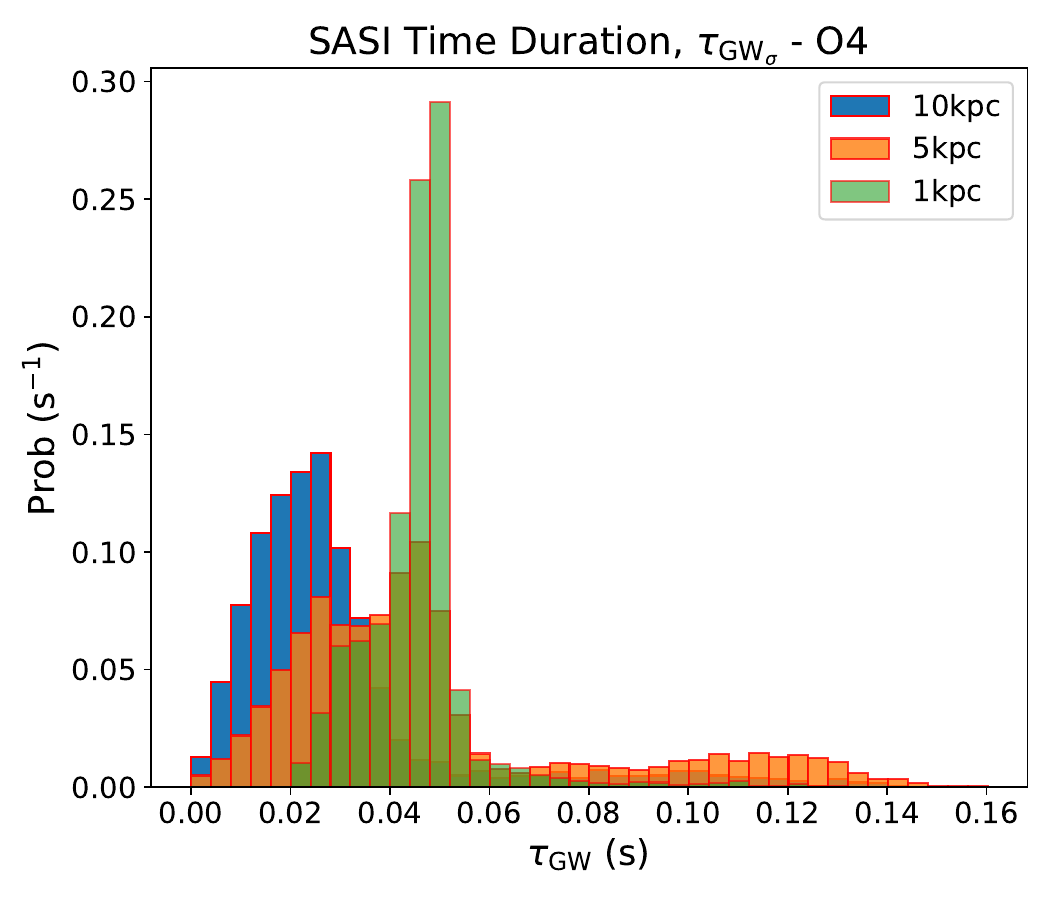}
    \end{center}
    \caption{Probability distributions from injections in O4 at 1, 5 and 10 kpc of the HFF slope estimations and SASI estimations of the central frequency and time durations computed from equation \eqref{eq_time_dur_f1}, $\tau_\text{GW}^r$, and from equation \eqref{eq_time_dur_f2}, $\tau_\text{GW}^\sigma$. }
    \label{Fig_PE_distributions_O4}
\end{figure}

The mean values of the estimated parameters across all reconstructed samples in O4, for each distance, are displayed in table \ref{Table_GW_PE_O4}. For the analysis where the signal included SASI activity, at 1 kpc 3018 triggers were used, all having pixels in the HFF region, 16 being discarded due to the HFF slope being out of the physically acceptable range, and 3000 being used for computing the $\rho_\text{GW}$ metric, with only 2 of these (which represent $0.06\%$ of the total number of the $\rho_\text{GW}$ computations) having no pixels in the SASI region. At 5 kpc, 3725 triggers were used on the analysis, from which 1157 had no pixels in the HFF region, 603 were discarded and 3000 were used for computing the SASI pixel likelihood, out of which only 122 ($4.06\%$) had $\rho_\text{GW}=0$. At 10 kpc, 3430 triggers were used, with only 521 of them having pixels in the HFF region and 277 out of these being discarded due to the slope laying outside the physically acceptable range; 3000 triggers were used to compute the $\rho_\text{GW}$ metric, with only 153 ($5.1 \%$) having no pixels in the SASI region. Likewise, figure \ref{Fig_PE_distributions_O4} shows the probability distributions for the estimated parameters at different distances, where the trend of dispersion growing with distance is still observed.

\renewcommand{\arraystretch}{1.3}
\setlength{\tabcolsep}{8pt}
\begin{table}[h!]
\centering
\resizebox{0.85\textwidth}{!}{%
\begin{tabular}{lllll}
& & \textbf{O4}\\
\hline
HFF Slope                    & SASI                                  & 1 kpc      & 5 kpc     & 10 kpc    \\ \hline
                                & $f_\text{GW}^c$ (Hz)                & 120.8190  & 117.9053  & 115.0791  \\ \cline{2-5} 
                                & $\delta f_\text{GW}^c$ (Hz)         & 5.0996    & 8.4522    & 9.8101    \\ \cline{2-5} 
                                & $\tau_\text{GW}^r$ (ms)             & 102.9413  & 91.4372   & 54.3622   \\ \cline{2-5} 
                                & $\delta \tau_\text{GW}^r$ (ms)      & 18.9740   & 45.7635   & 40.1186   \\ \cline{2-5} 
                                & $\tau_\text{GW}^\sigma$ (ms)        & 45.6599   & 48.3120   & 29.3868   \\ \cline{2-5} 
                                & $\delta \tau_\text{GW}^\sigma$ (ms) & 11.4121   & 31.5636   & 23.3937    \\ \hline
$m_\text{GW}$ ($s^{-2}$)        &                                     & 2358.1734 & 2078.9537 & 2565.5719 \\ \hline
$\delta m_\text{GW}$ ($s^{-2}$) &                                     & 496.7621  & 803.2093  & 1247.3102 \\ \hline
\end{tabular}%
}
\caption{Mean values of the estimated SASI central frequency ($f_\text{GW}^c$), time duration ($\tau_\text{GW}^r, \; \tau_\text{GW}^\sigma$) and HFF slope ($m_\text{GW}$) across 1, 5, and 10 kpc, and their respective standard deviations ($\delta$), in O4 data.}
\label{Table_GW_PE_O4}
\end{table}

Results in table \ref{Table_GW_PE_O4} show, again, an accordance with previous results in the estimated central frequency of around $f^c_\text{GW} \sim 120 \; \text{Hz}$. Comparison with Lin2023 \cite{Zidu_SASImeter_Joint} shows a considerable reduction in the standard deviation of the estimations, which is in accordance with the O3 results. However, it is noted that the O4 standard deviations for the central frequency are greater than those reported in O3, specially at 5 and 10 kpc. This can be explained by the fact that the implementation of O4 translates into TF maps containing more pixels; indeed, it was observed that O4 triggers contained, in average and with respect of O3, $59.54\%$, $118.95\%$, and $31.16\%$ more pixels per trigger at 1, 5, and 10 kpc respectively, which naturally means a larger dispersion of data across almost all distances. This is also reflected in both time estimations, which are all greater than the ones reported in table \ref{Table_GW_PE_O3} for O3; this is again explained by the fact that O4 allows the retention of more pixels per trigger, particularly exhibiting $3.65\%$, $17.6\%$, and $16.34\%$ more pixels per trigger in the filtered SASI region at 1, 5, and 10 kpc respectively. It is also noted that all standard deviations associated to the time estimations are greater than in O3 too. Particularly, it is observed that for 1 kpc, the standard deviations for both time estimations represent around $22 \%$ of the mean estimated value, while for 5 and 10 kpc they represent, respectively, $\sim 57 \%$ and $\sim 76 \%$ of the means, showcasing the increase in data dispersion with growing distance, and an increase with respect to the $\sim 50 \%$ observed in O3.

\begin{figure}[h!]
	\begin{center}
		\includegraphics[width=0.3\textwidth]{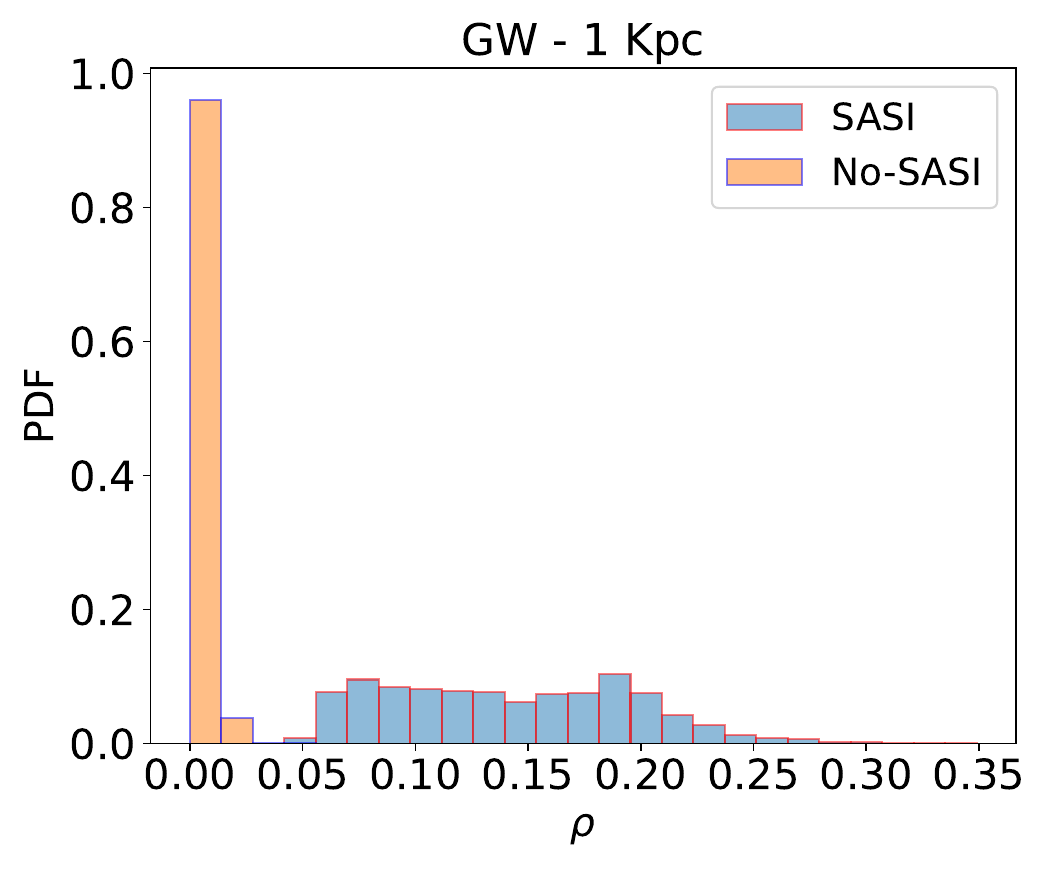}
		\includegraphics[width=0.3\textwidth]{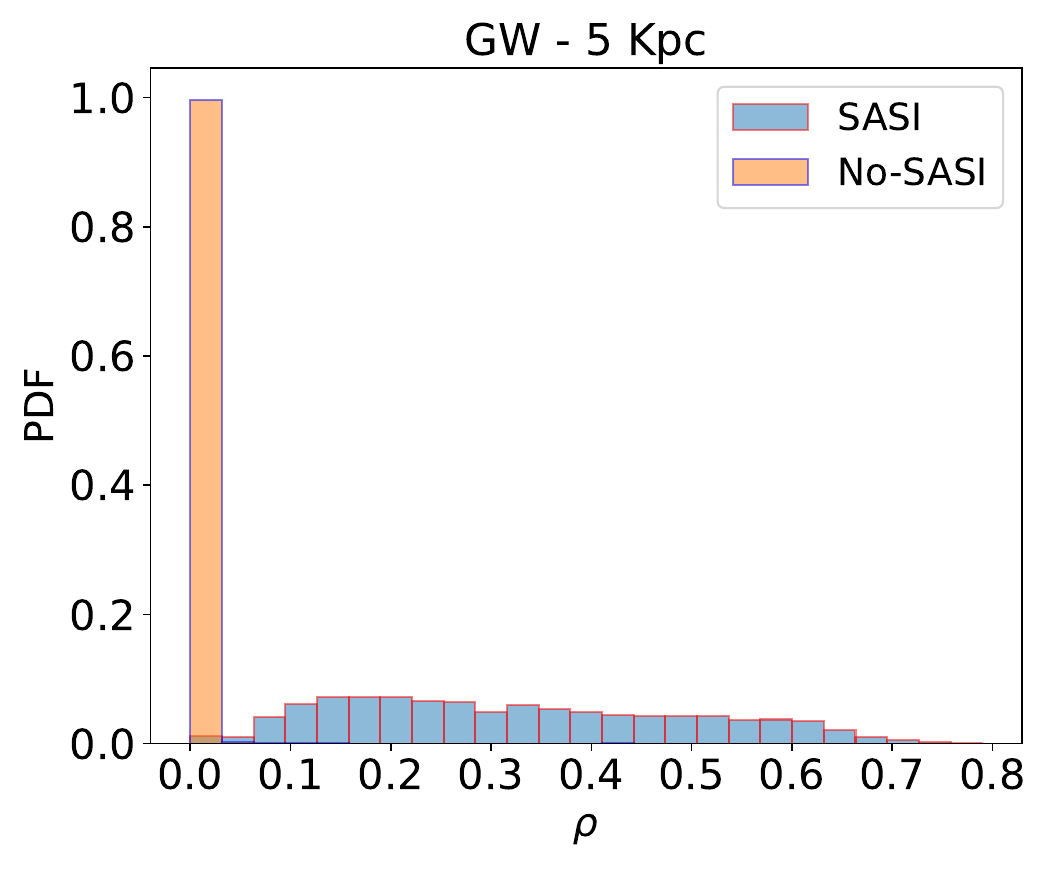}
		\includegraphics[width=0.3\textwidth]{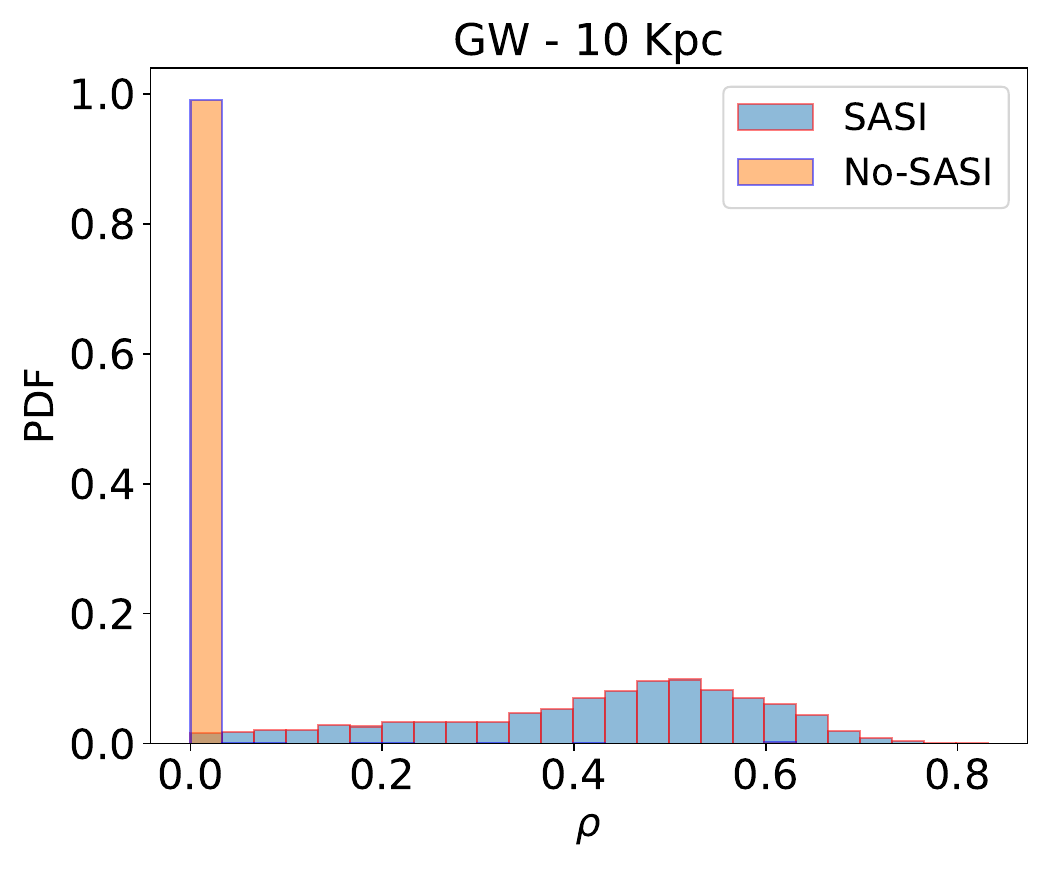}
        \includegraphics[width=0.3\textwidth]{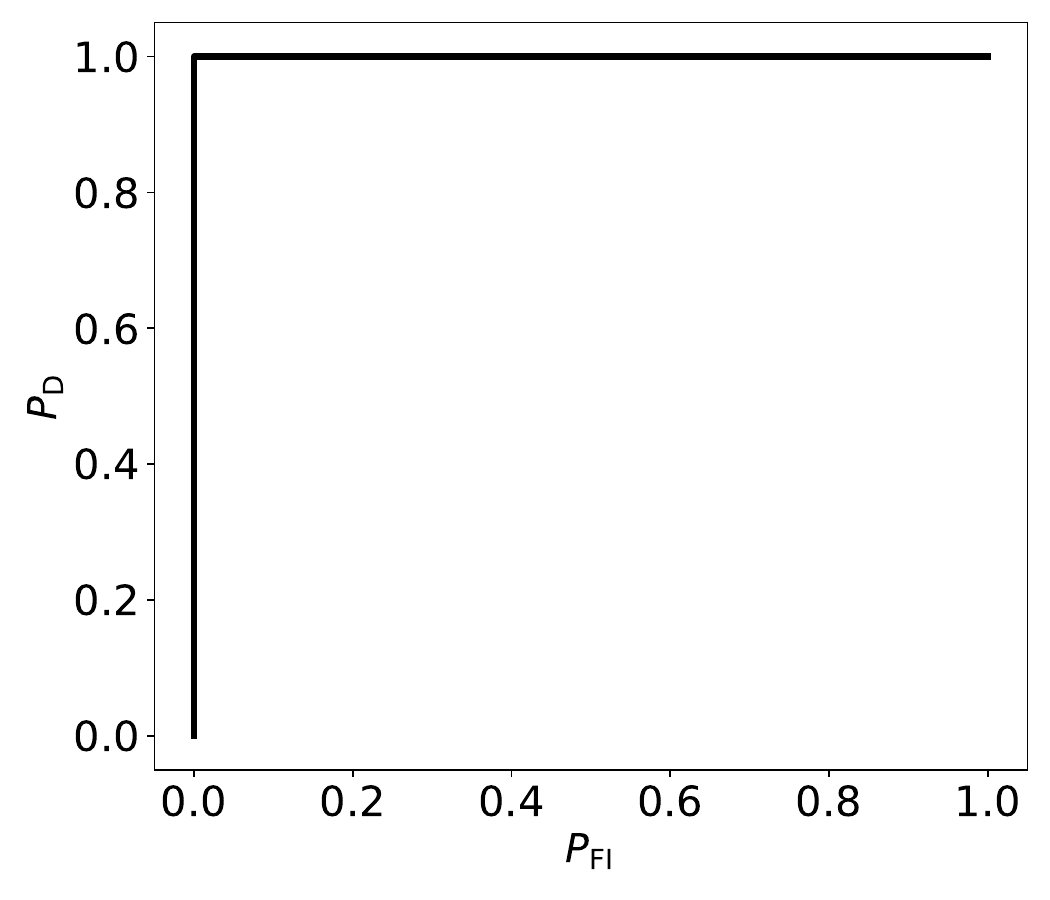}
        \includegraphics[width=0.3\textwidth]{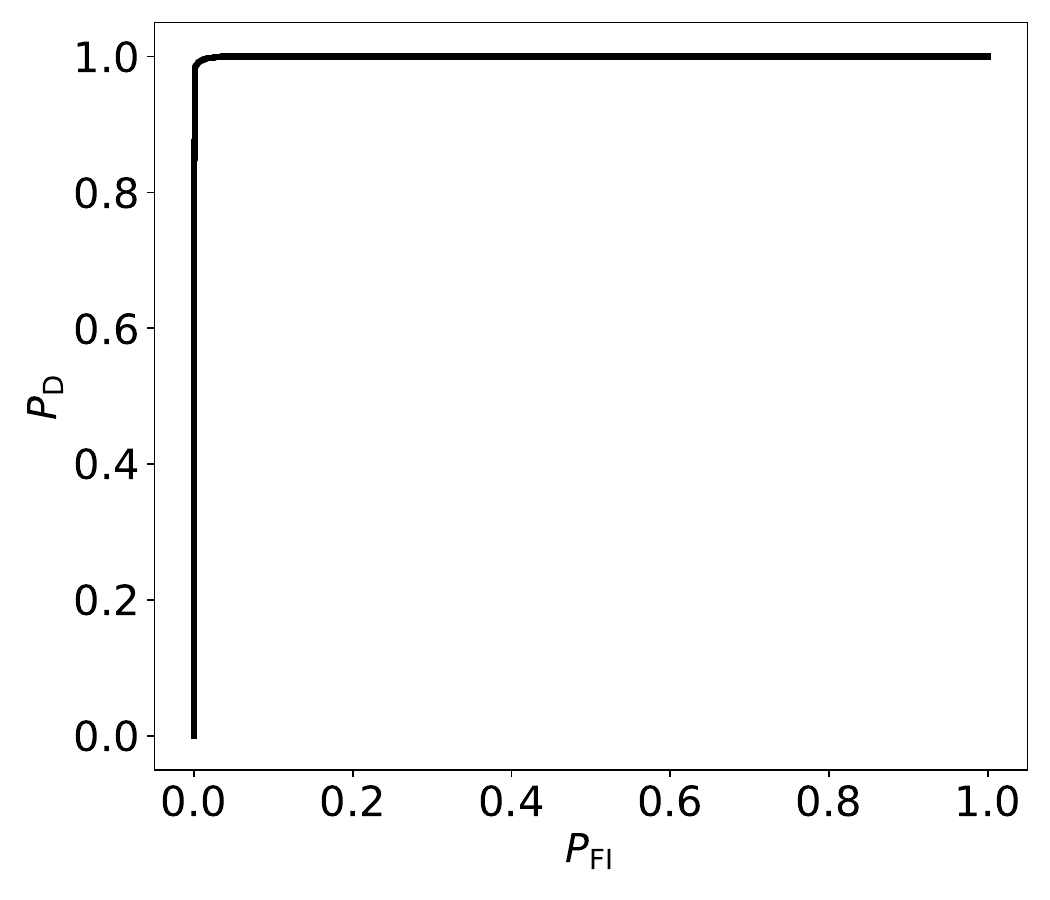}
        \includegraphics[width=0.3\textwidth]{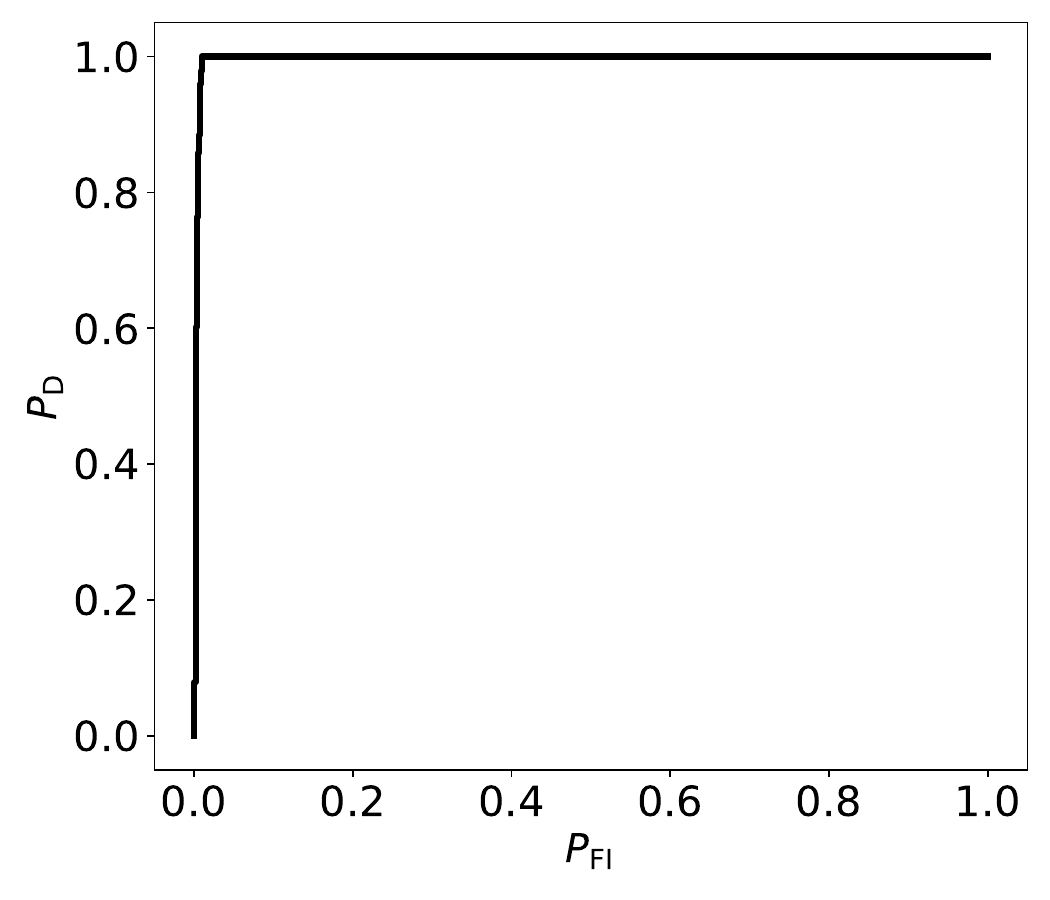}
	\end{center}
	\caption{\textit{Top row:} probability distributions of the $\rho_\text{GW}$ metric at 1, 5 and 10 kpc for the cases where SASI activity is present and removed from the injected waveform in O4. \textit{Bottom row:} ROC curves for the probability distributions.}
	\label{Fig_ROC_GWs_O4}
\end{figure}

Figure \ref{Fig_ROC_GWs_O4} shows the histograms for $\rho_\text{GW}$ and the ROC curves for the O4 implementation. For the first, it is observed that there is a drastic difference when comparing them with the O3 results, figure \ref{Fig_ROC_GWs_O3}, and it is that the \texttt{SASImeter}'s confusion, i.e., distributions overlap, is not visible at 1 kpc and is minimal for the rest of the distances. Particularly, for 1 kpc, it is noted that the ``No-SASI'' distribution is very densely concentrated around $\rho_\text{GW}=0$, even more than in O3, and that the SASI distribution is now more densely concentrated at lower values of the likelihood metric, showing a clear separation from the ``No-SASI'' distribution. These results at 1 kpc translate into a technically perfect ROC curve, even better than its O3 counterpart as clearly noted in table \ref{Table_GW_ROC_Thresh_O4}, where it is shown that, for a $P_\textbf{FI}$ of $0.01$, O4 increases the $P_D$ to $1$ in comparison to the $0.99$ in O3. All of this is attributed to the better noise floor and data quality in O4, which translates into better reconstructions: the ``No-SASI'' case overall contains less pixels in the SASI region due to better identification of instrumental noise, which translates into lower maximums for $\rho_\text{GW}$; for the SASI case, as more pixels are reconstructed across all TF maps, the portion of CRS contained in the SASI region is smaller, which explains the low $\rho_\text{GW}$ values, while the narrower distribution of likelihood values also shows a better convergence of estimations.

Although the O4 advancements at 1 kpc are remarkable, 5 and 10 kpc show the greatest improvement of all. This is firstly evident from the ROC curves which, at first sight, emulate closely the behavior of a perfect classifier, with the 10 kpc curve deviating more from such than the 5 kpc one. This improvement is quantitatively evident in table \ref{Table_GW_ROC_Thresh_O4}, where the identification probability values at 5 kpc are as good as those at 1 kpc in O3, and the O4 10 kpc being slightly worse than at 5 kpc for a false identification probability of $0.01$. These remarkable improvements are attributed to the combined advantages of cWB XP and O4 noise and data quality, which translate into the ``No-SASI'' case, as at 1 kpc, having a denser concentration of likelihood values around 0 in comparison with the O3 results; although the ``SASI'' distributions seem very similar to those in O3, further inspection shows that the mean and median values for $\rho_\text{GW}$ are reduced in O4, confirming the 1 kpc trend of the TF maps containing overall more pixels and so the SASI region containing a reduced fraction of the total CRS.

\renewcommand{\arraystretch}{1.3}
\setlength{\tabcolsep}{8pt}
\begin{table}[h!]
\centering
\resizebox{1\textwidth}{!}{%
\begin{tabular}{cccccccccc}
\multicolumn{10}{c}{\textbf{O4}}                                                                                                                                                                       \\ \hline
$P_\text{FI}$             & \multicolumn{9}{c}{$P_\text{D}$}                                                                                                                                           \\
\multicolumn{1}{c|}{}     & \multicolumn{3}{c|}{1 kpc}                            & \multicolumn{3}{c|}{5 kpc}                            & \multicolumn{3}{c}{10 kpc}                                 \\ \cline{2-10} 
\multicolumn{1}{c|}{}     & \textbf{XP O4} & XP O3 & \multicolumn{1}{c|}{Lin2023} & \textbf{XP O4} & XP O3 & \multicolumn{1}{c|}{Lin2023} & \textbf{XP O4} & XP O3               & Lin2023             \\ \hline
\multicolumn{1}{c|}{0.01} & \textbf{1.00}  & 0.99  & \multicolumn{1}{c|}{0.63}    & \textbf{0.99}  & 0.43  & \multicolumn{1}{c|}{0.21}    & \textbf{0.97}  & $1.6\times 10^{-3}$ & $1.1\times 10^{-3}$ \\ \hline
\multicolumn{1}{c|}{0.05} & \textbf{1.00}  & 1.00  & \multicolumn{1}{c|}{0.84}    & \textbf{1.00}  & 0.84  & \multicolumn{1}{c|}{0.39}    & \textbf{1.00}  & 0.32                & 0.11                \\ \hline
\multicolumn{1}{c|}{0.10} & \textbf{1.00}  & 1.00  & \multicolumn{1}{c|}{0.90}    & \textbf{1.00}  & 0.95  & \multicolumn{1}{c|}{0.52}    & \textbf{1.00}  & 0.47                & 0.24                \\ \hline
\multicolumn{1}{c|}{0.25} & \textbf{1.00}  & 1.00  & \multicolumn{1}{c|}{1.00}    & \textbf{1.00}  & 0.99  & \multicolumn{1}{c|}{0.75}    & \textbf{1.00}  & 0.97                & 0.72                \\ \hline
\multicolumn{1}{c|}{0.50} & \textbf{1.00}  & 1.00  & \multicolumn{1}{c|}{1.00}    & \textbf{1.00}  & 1.00  & \multicolumn{1}{c|}{0.95}    & \textbf{1.00}  & 0.99                & 0.91                \\ \hline
\end{tabular}
}
\caption{Identification probability ($P_D$) of the GW-\texttt{SASImeter} for different false identification probabilities ($P_\text{FI}$) at 1, 5, and 10 kpc, comparing the cWB XP O4 results with those in O3 and those from Lin2023.}
\label{Table_GW_ROC_Thresh_O4}
\end{table}

In general, the most important conclusion of the O4 results is that the \texttt{SASImeter}, in conjunction with the cWB XP applied to the most sensitive LIGO data, allows an almost perfect identification of the SASI in the GW channel at all galactic distances, overperforming all previous implementations and the neutrino channel itself, as it will be shown in the following subsection; this, in turn, also updates the conclusion of Lin2023 \cite{Zidu_SASImeter_Joint} on the importance of using both data channels to successfully identify SASI activity on an hypothetical real detection, as it will be discussed in section \ref{Sect_Conclusions}.

\subsection{Neutrino Channel} \label{Sect_Results_Neutrino}

\begin{figure}[h!]
	\begin{center}
		\includegraphics[width=0.3\textwidth]{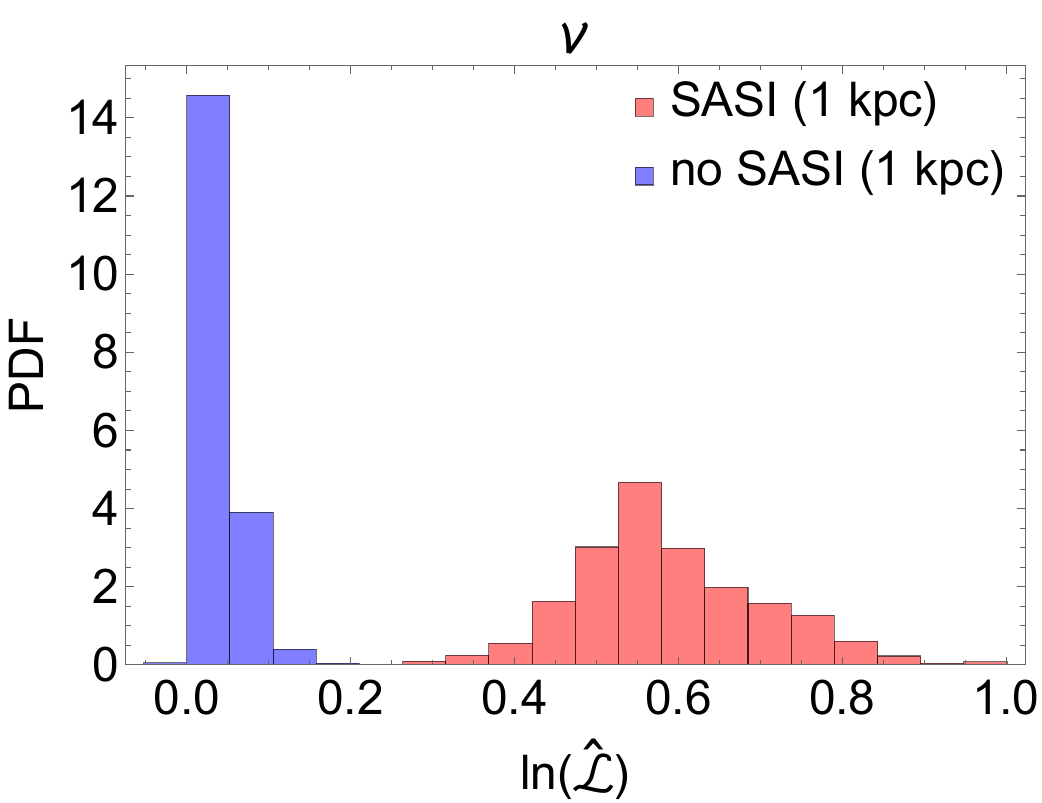}
		\includegraphics[width=0.3\textwidth]{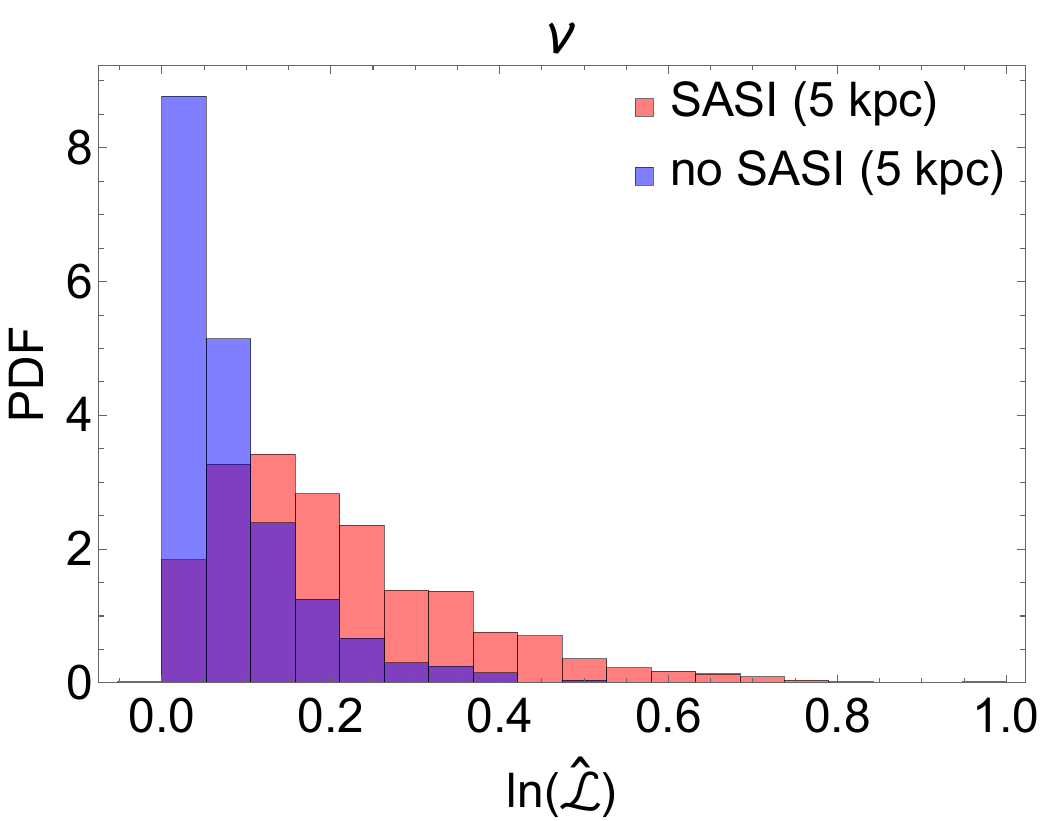}
		\includegraphics[width=0.3\textwidth]{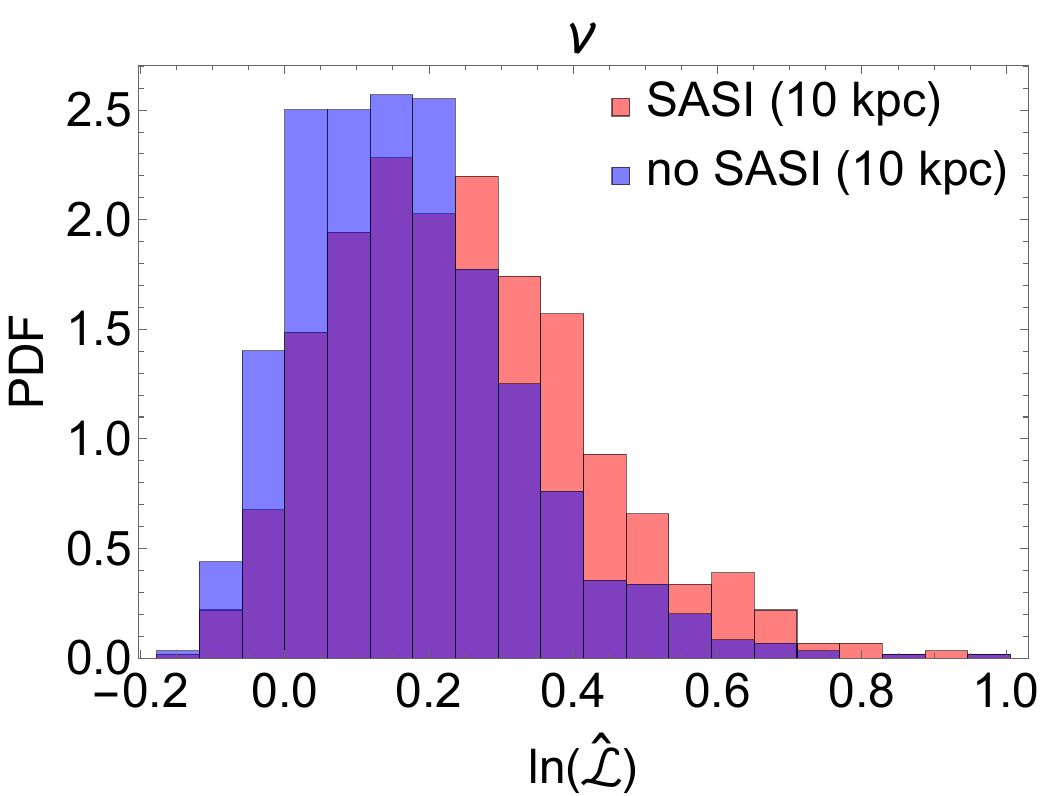}
        \includegraphics[width=0.3\textwidth]{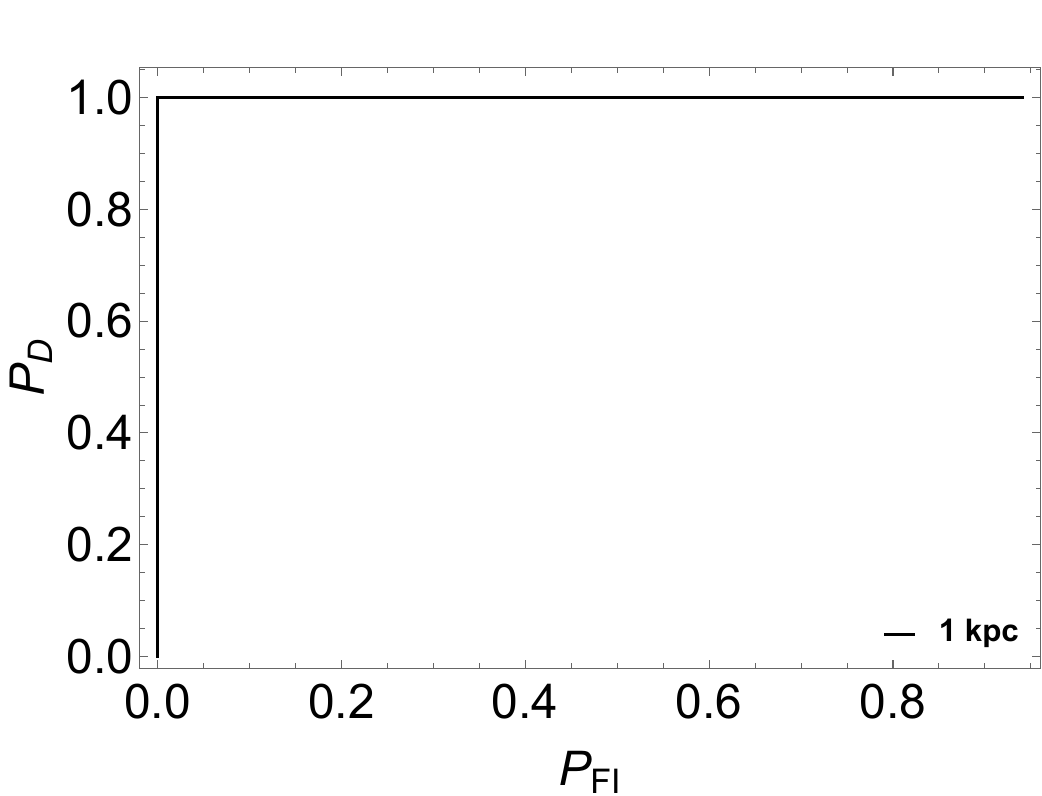}
        \includegraphics[width=0.3\textwidth]{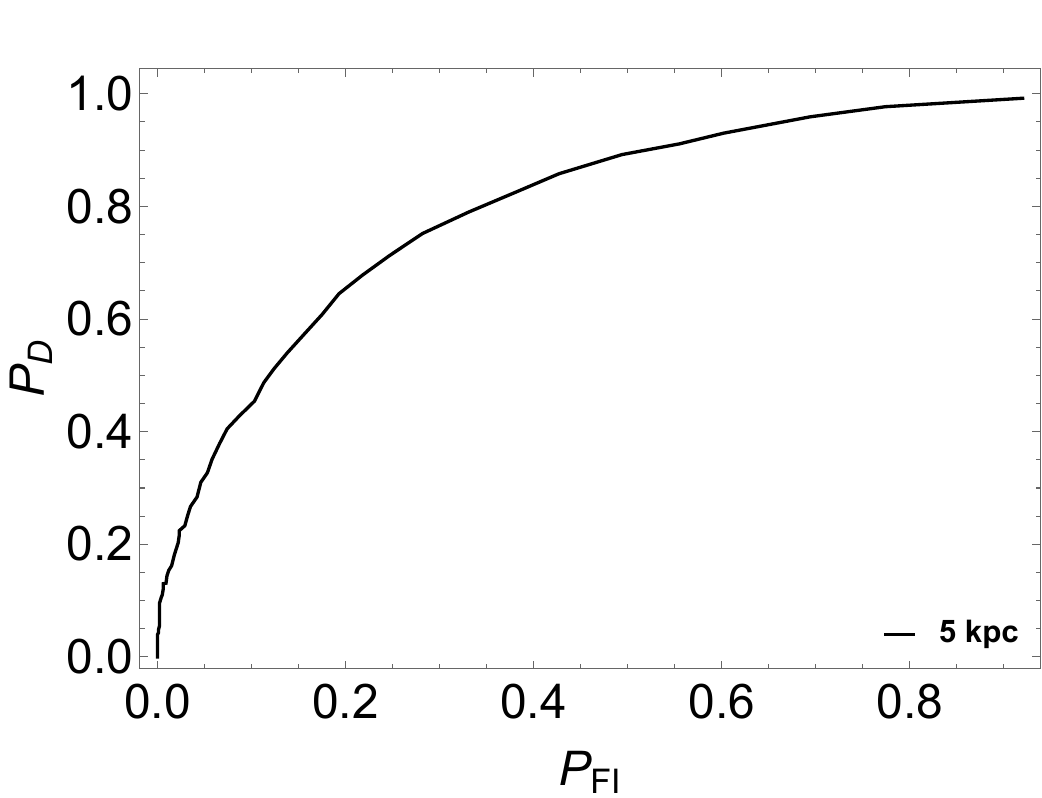}
        \includegraphics[width=0.3\textwidth]{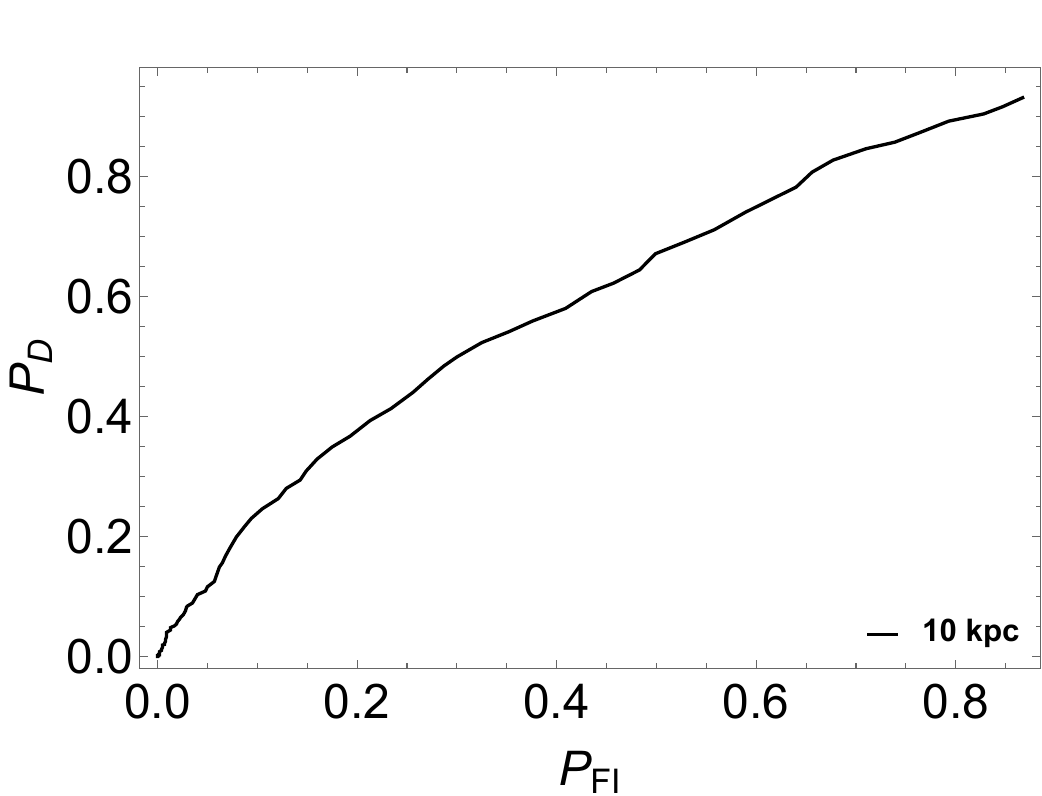}
	\end{center}
	\caption{\textit{Top row:} probability distributions of the $\ln(\mathcal{L})_\mathrm{SASI/no-SASI}$ neutrino likelihood metric at 1, 5 and 10 kpc. \textit{Bottom row:} ROC curves for the probability distributions.}
    \label{fig:nuroc}
\end{figure}

Figure \ref{fig:nuroc} displays the PDF of $\ln(\mathcal{L})$ and the ROC curves at 1, 5 and 10 kpc, all of which were computed for a time window that starts at $150 \; \text{ms}$ and lasts for $50 \; \text{ms}$ on Kuroda2017 neutrino events. In this figure, it can be observed that the 1 kpc case exhibits very well differentiated histograms; furthermore, the ROC curve behaves as a technically perfect classifier. These two plots show a very similar performance to their GW counterparts for 1 kpc, although further inspection shows a slightly better detection efficiency in the neutrino channel than in XP O3, as observed from the ROC curves for $P_{\text{FI}} \lesssim  0.05$, but a technically identical performance than in XP O4. In contrast, this is not the case for 5 and 10 kpc, where the performance of the neutrino channel is clearly worse than both GW results for all false identification probability values, as it is observed from the ROC curves.

\begin{figure}[h!]
	\begin{center}
		\includegraphics[width=0.3\textwidth]{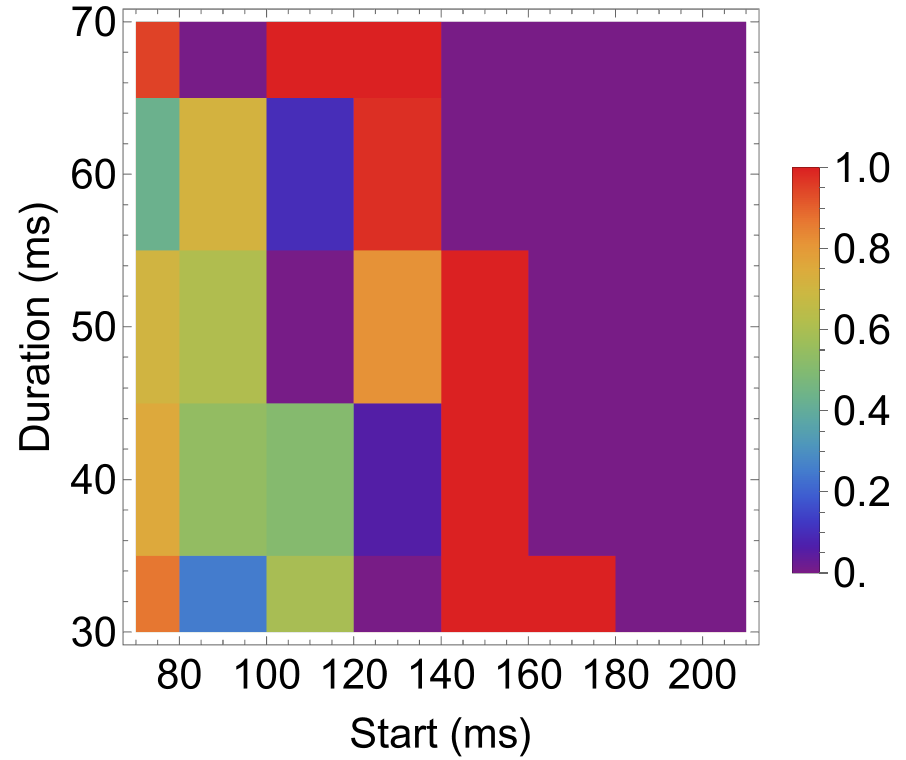}
		\includegraphics[width=0.3\textwidth]{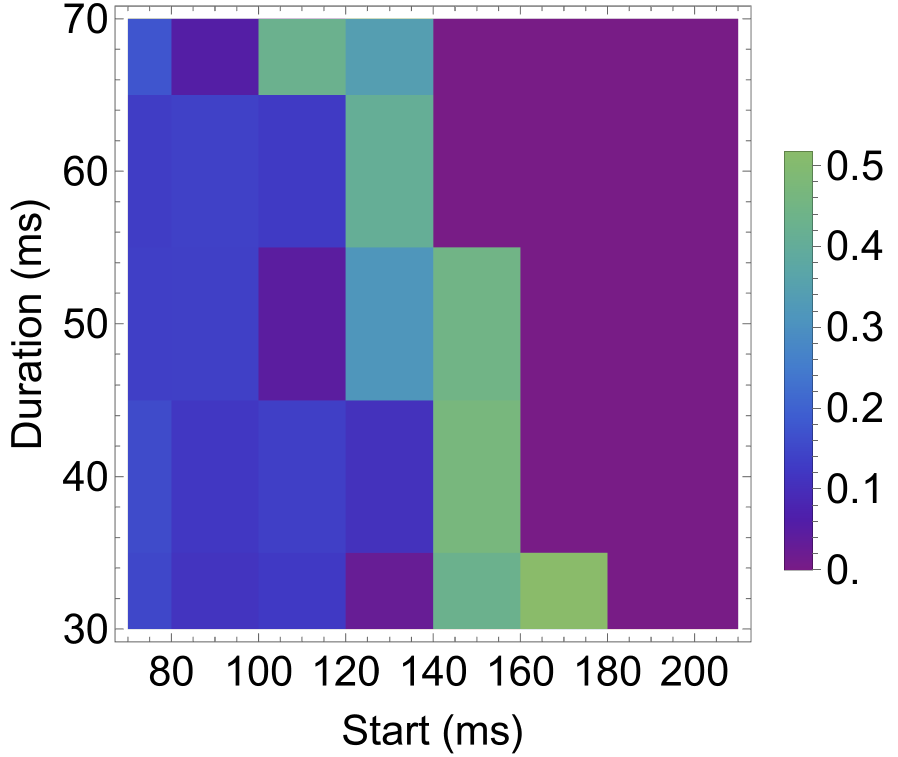}
		\includegraphics[width=0.3\textwidth]{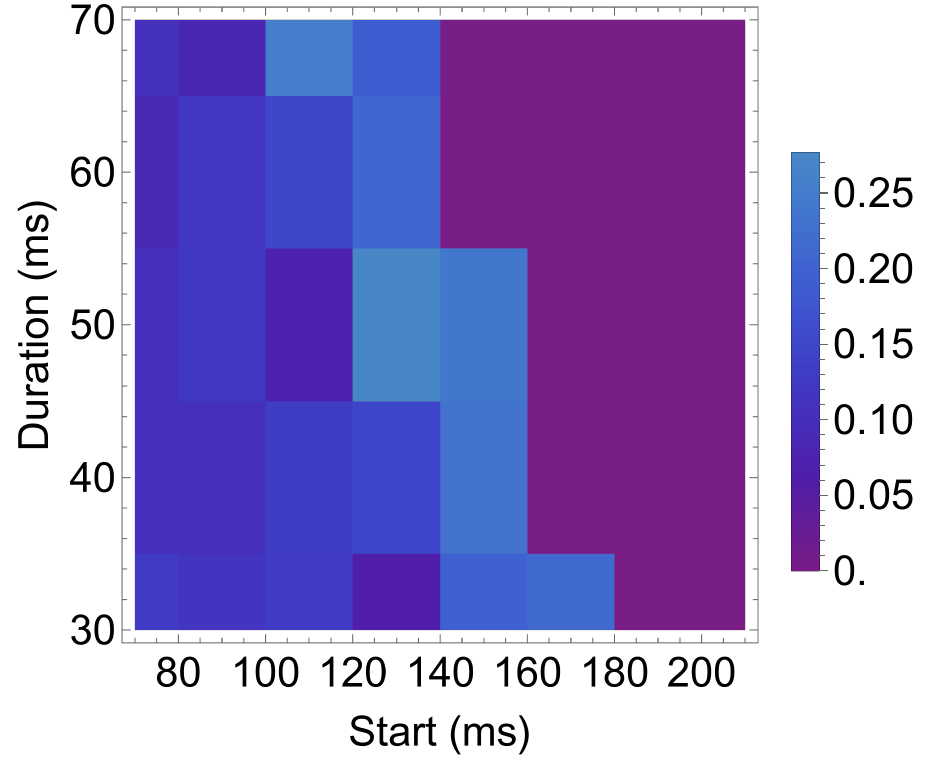}

	\end{center}
	\caption{The distribution of $P_D$ (color map) with varying starting time and duration of the window of SASImeter. The left, middle and right panels are at 1, 5 and 10 kpc. All the $P_D$ are calculated at $P_{FA}=10\%$.}
    \label{fig:nustartdur}
\end{figure}

On the other hand, figure \ref{fig:nustartdur} displays the possible time windows of the SASI activity in the neutrino signal, for all of which there is a $P_D^\nu$ assigned (via the color map). These values are obviously higher for windows that partly/fully cover the SASI activity, which approximately starts at around 150 ms and lasts for 50 ms. For example, at 1 kpc, for windows starting at 110 ms, only the time window that lasts for 70 ms has $P_D^\nu>0.8$; for windows starting at 130 ms, those with durations larger than 60 ms have $P_D^\nu>0.8$. This suggests that the SASI activity may be most obvious at around 180 ms, and that the SASI does not start before 150 ms since, for windows that end before 150 ms, their $P_D^\nu$s are low. In addition, the SASI does not end before 170 ms since the window starting here also has high $P_D^\nu$. Furthermore, and evidently, the identification probability is 0 when the tail of the window exceeds the ending point of Kuroda2017 neutrino data.

Lastly, table \ref{Table_Neutrino_PE} contains the numerical values of the estimated parameters via the $\nu$-\texttt{SASImeter}. The SASI frequency and amplitude, $f_\nu$ and $a_\nu$, are directly obtained from the mean value of all of the sets of parameters from templates with and without SASI activity, $\Omega / \Omega_0$, that maximize the likelihood ratio. The time duration estimations are obtained from the aforementioned method, although only as an approximated lower bound; for 10 kpc, the standard deviation is too large to provide this value. For further information on this, refer to Appendix A of Lin2023 \cite{Zidu_SASImeter_Joint}.

\renewcommand{\arraystretch}{1.3}
\setlength{\tabcolsep}{8pt}
\begin{table}[h!]
\centering
\resizebox{0.5\textwidth}{!}{%
\begin{tabular}{llll}
\hline
SASI                & 1 kpc            & 5 kpc            & 10 kpc \\ \hline
$f_\nu$ (Hz)        & 119.85           & 111.03           & 113.38 \\ \hline
$\delta f_\nu$ (Hz) & 1.22             & 22.6             & 113.38 \\ \hline
$a_\nu$             & 0.044            & 0.047            & 0.063  \\ \hline
$\delta a_\nu$      & 0.005            & 0.013            & 0.022  \\ \hline
$\tau_\nu$ (ms)     & \textgreater{}50 & \textgreater{}50 & N/A    \\ \hline
\end{tabular}
}
\caption{SASI parameter estimations via the $\nu$-\texttt{SASImeter} for the frequency, amplitude and signal time duration, and their respective standard deviations ($\delta$).}
\label{Table_Neutrino_PE}
\end{table}

\subsection{Combined Analysis} \label{Sect_Results_Joint}

\begin{figure}[h!]
	\begin{center}
		\includegraphics[width=0.35\textwidth]{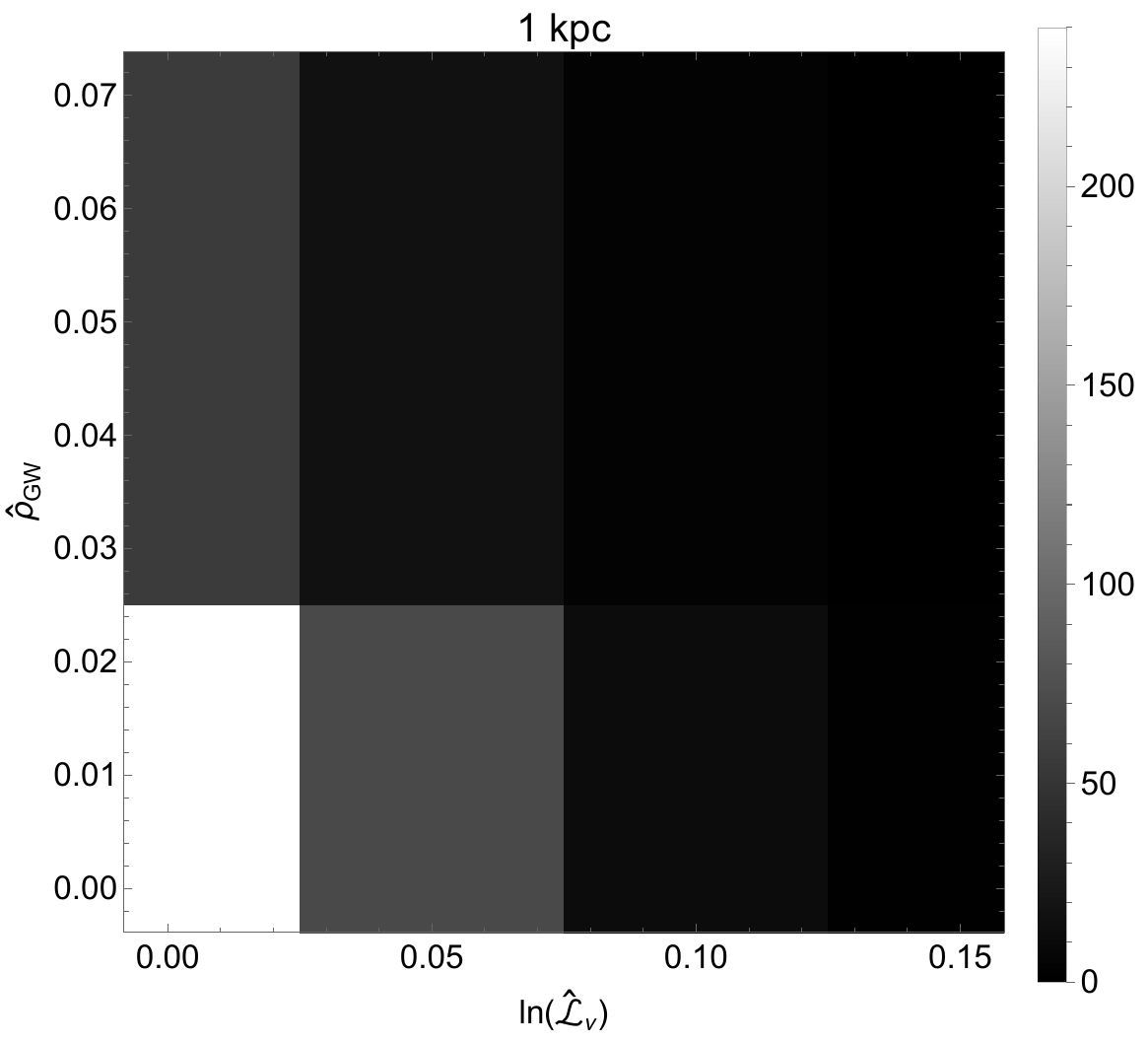}
		\includegraphics[width=0.35\textwidth]{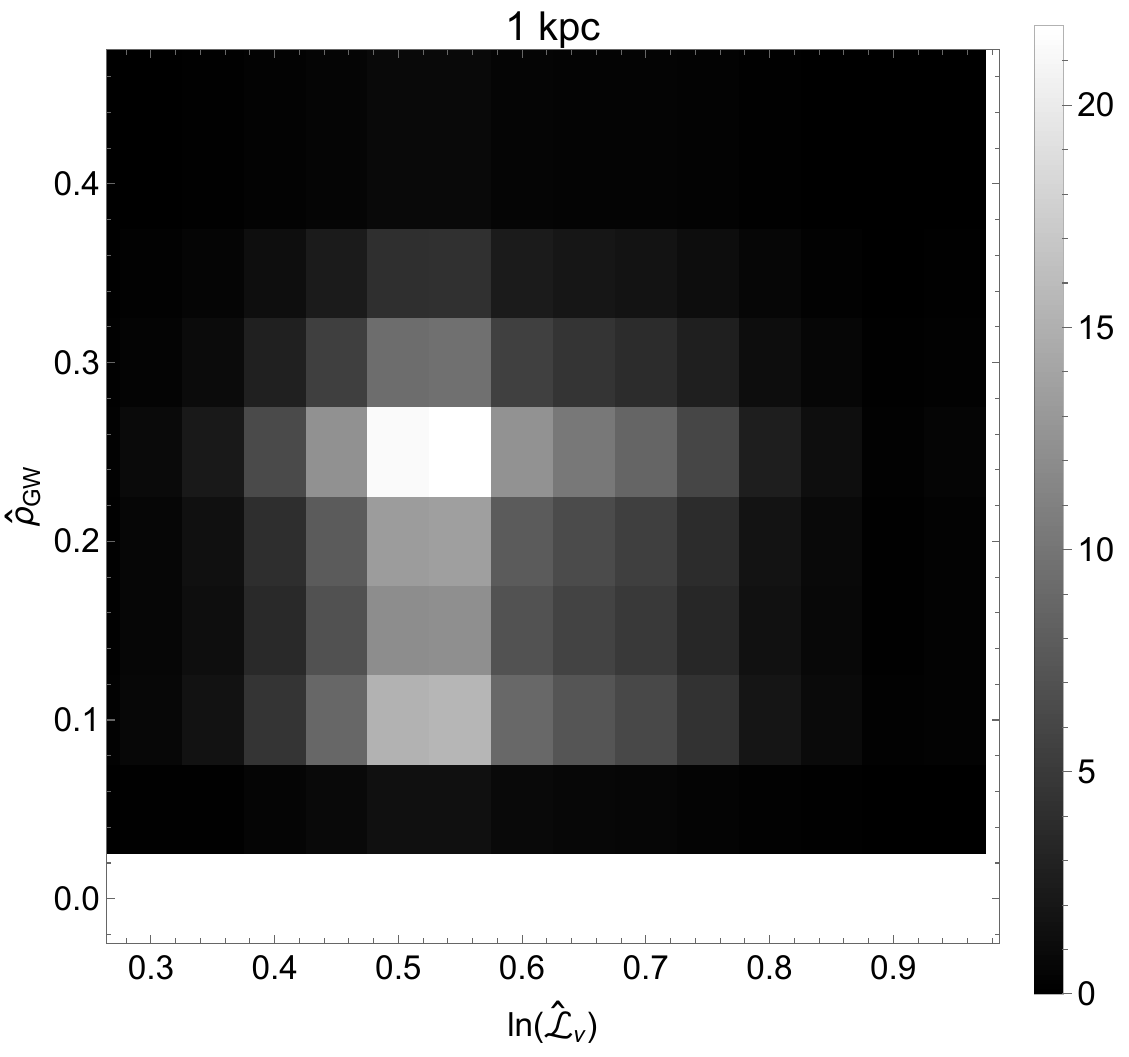}
		\includegraphics[width=0.35\textwidth]{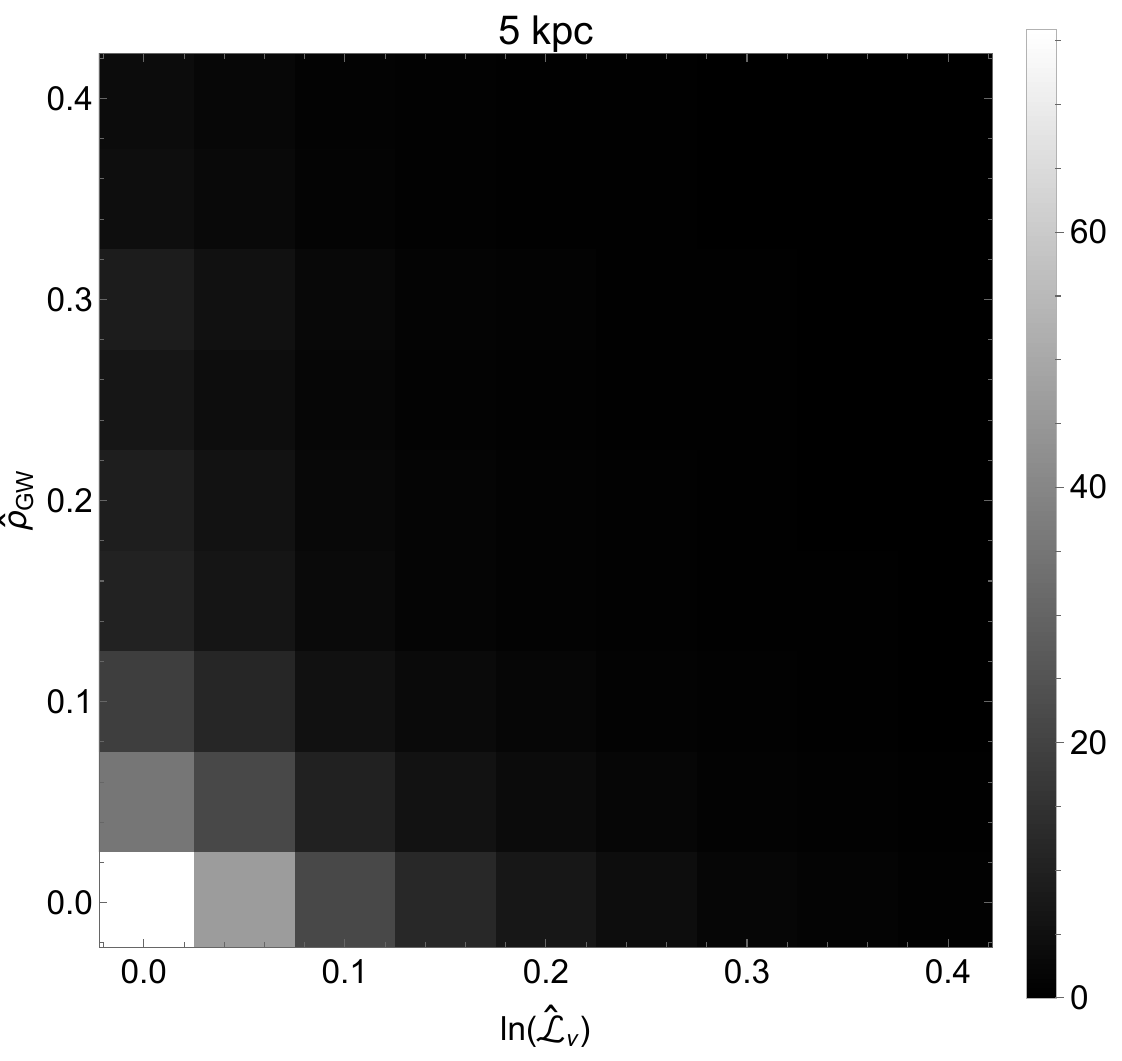}
		\includegraphics[width=0.35\textwidth]{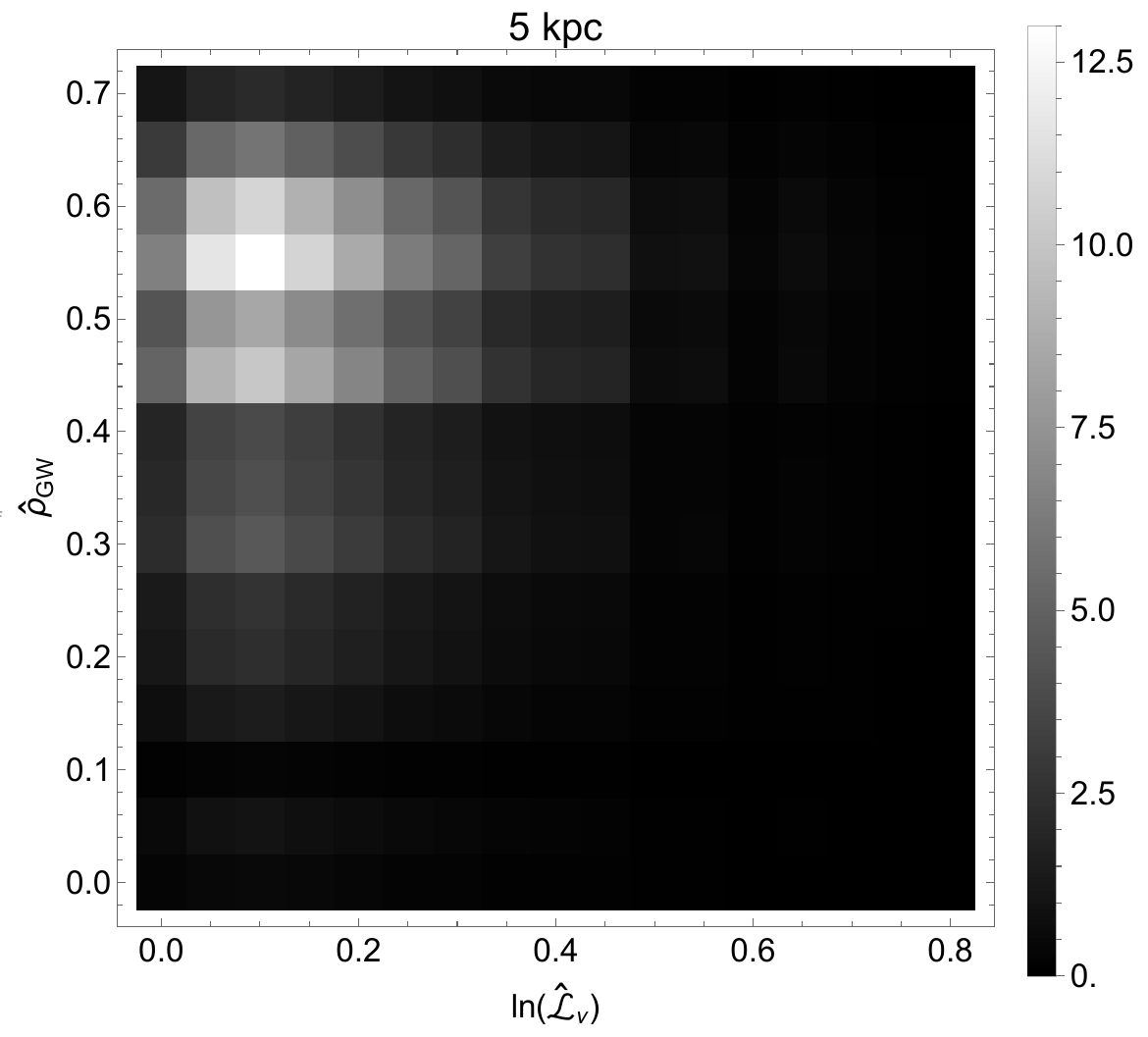}
		\includegraphics[width=0.35\textwidth]{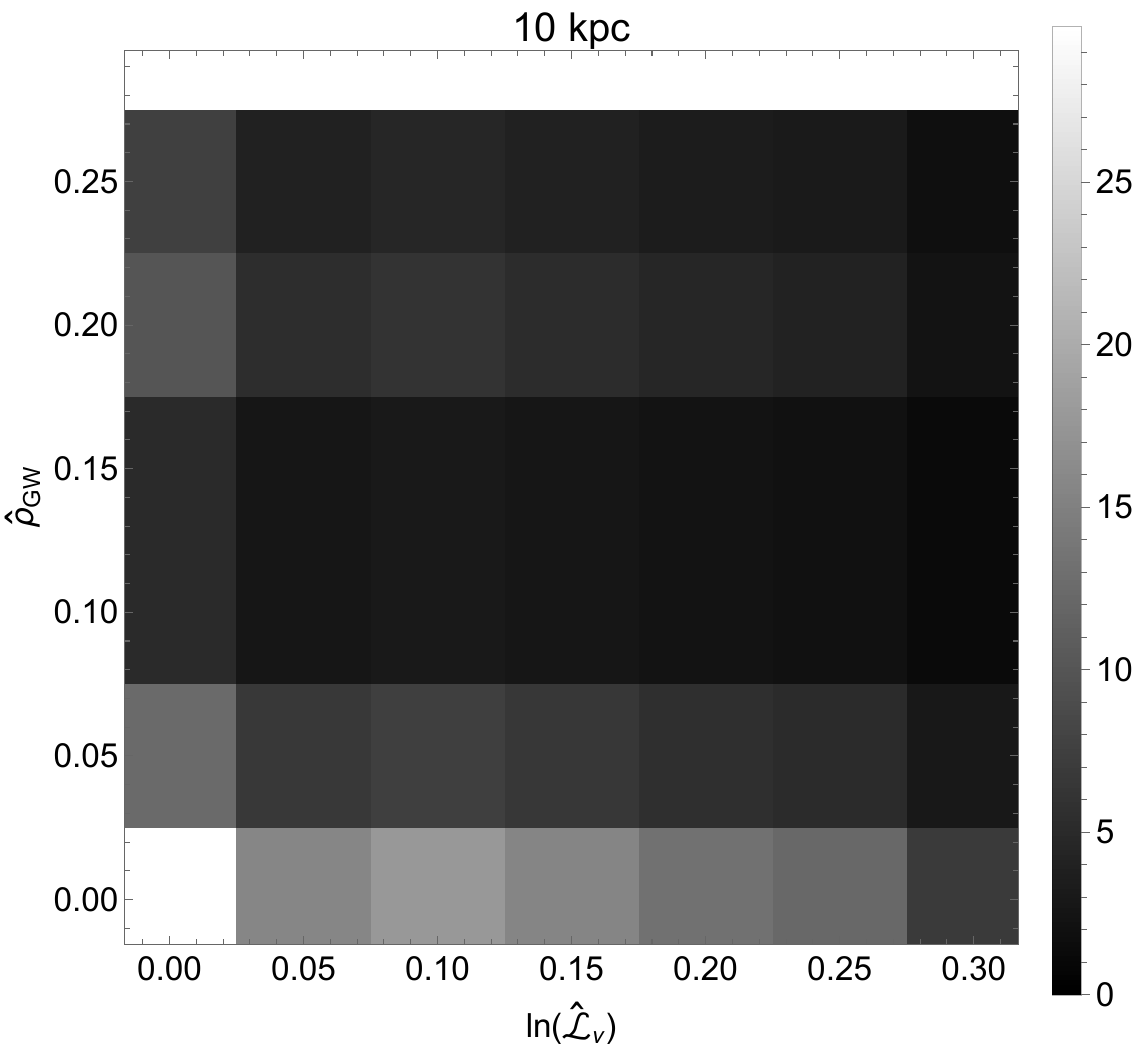}
		\includegraphics[width=0.35\textwidth]{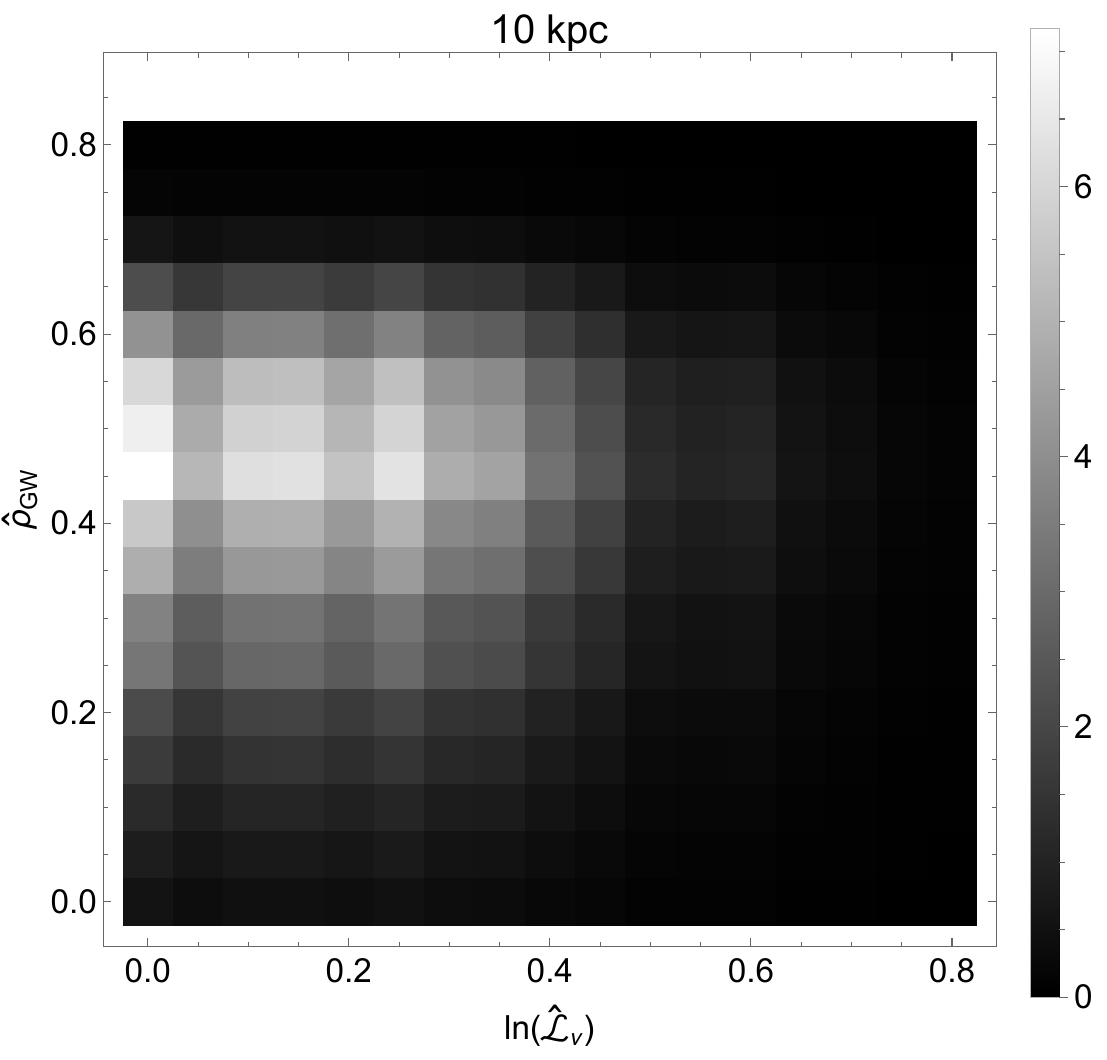}
	\end{center}
	\caption{2-dimensional probability distribution of the normalized likelihood metrics $\ln(\mathcal{L})$ and $\rho_\text{GW}$. Left column corresponds to the case where there is no SASI activity and the right column where there is; the first row corresponds to 1 kpc, the second to 5 kpc and the third to 10 kpc.}
	\label{Fig_2D_Prob_Distri}
\end{figure}

As mentioned in section \ref{Sect_SASImeter_Joint}, the joint analysis focuses solely on the identification probability of SASI activity via the neutrino and GW channel, the latter exclusively from the cWB XP O3 implementation, and so, results shown in this section are only valid for this and will not be specified as such further on; a discussion on a joint analysis using GW O4 data is left for section \ref{Sect_Conclusions}. As a first way to visualize this joint identification probability, in figure \ref{Fig_2D_Prob_Distri} a 2D map of the likelihood metrics $\ln(\mathcal{L})$ (x-axis) and $\rho_\text{GW}$ (y-axis) is shown. The left column corresponds to the likelihood metrics' values when the SASI activity was removed from the simulated data and the right column corresponds to values when there is SASI activity. The color map represents the joint likelihood value, with the clearer region signaling the most coincidental range of values between the two channels. As for the single-channel analysis, it is seen that the ``No-SASI'' plots (left column) exhibit small likelihood values across the three distances; however, as expected, a dependence on the distance is noticed: the likelihood distribution is more disperse and exhibits higher values as the distance increases. Particularly, it is noticed that the 10 kpc analysis shows a larger distribution and higher values for the neutrino likelihood metric than for the GW one, meaning that more noise is present in such channel. 

On the other hand, the SASI plots (right column) display a broader distribution with higher likelihood values. In the 1 kpc case, an almost balanced likelihood probability distribution is observed, meaning that the highest metric values, the clearest region, are concentrated near the middle point of the plot; however, it is observed that the neutrino channel exhibits larger likelihood values, meaning a slightly better confidence on this channel than in the GW one. However, this behavior drastically changes at 5 kpc, where now the GW channel exhibits even larger likelihood values than those for the neutrino channel which are, in turn, rather low; a similar scenario occurs for 10 kpc although not as drastically.

\begin{figure}[h!]
    \begin{center}
    	 \includegraphics[width=0.48\textwidth]{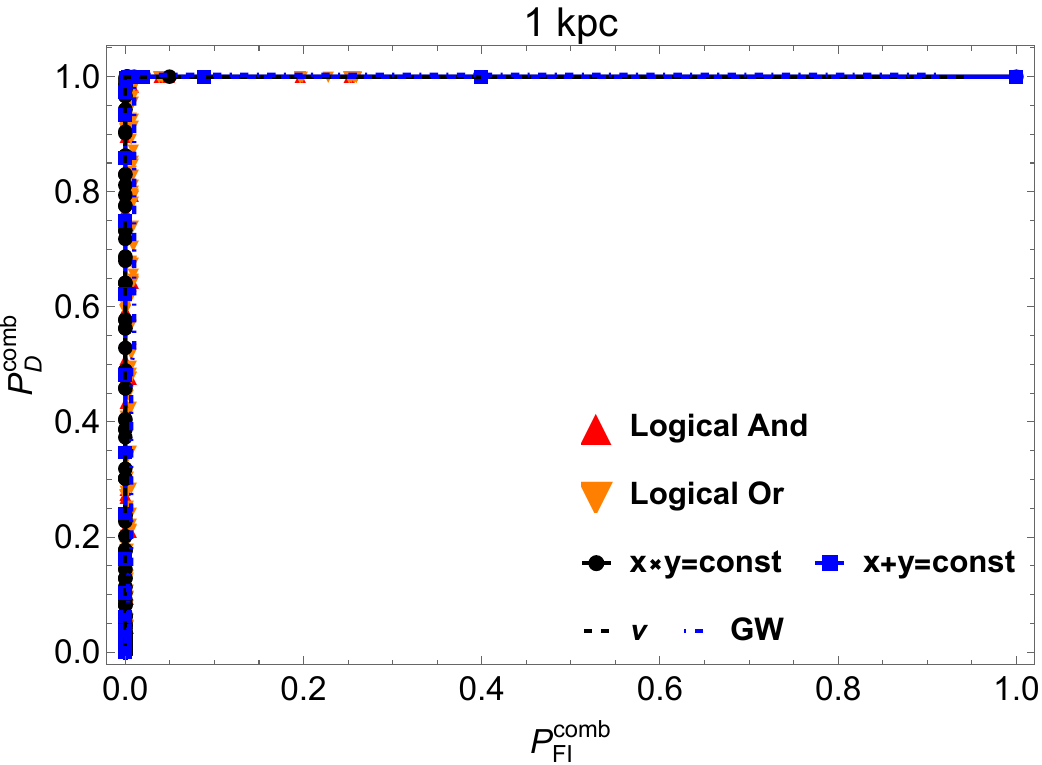}
        \includegraphics[width=0.48\textwidth]{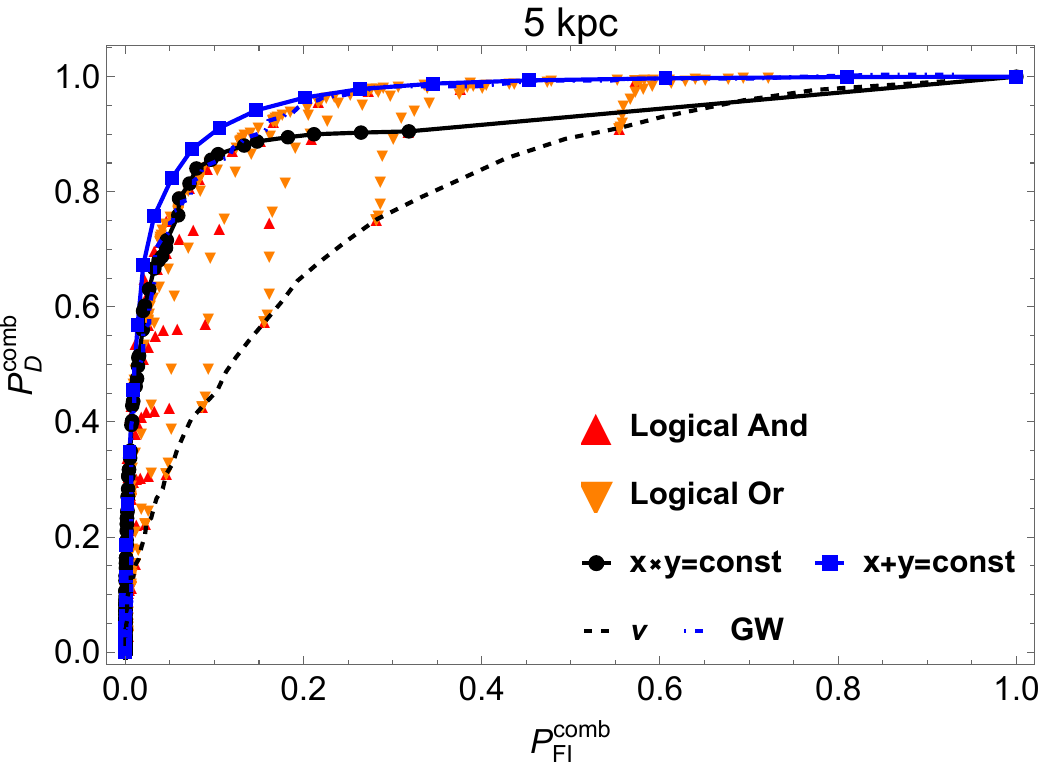}
        \includegraphics[width=0.48\textwidth]{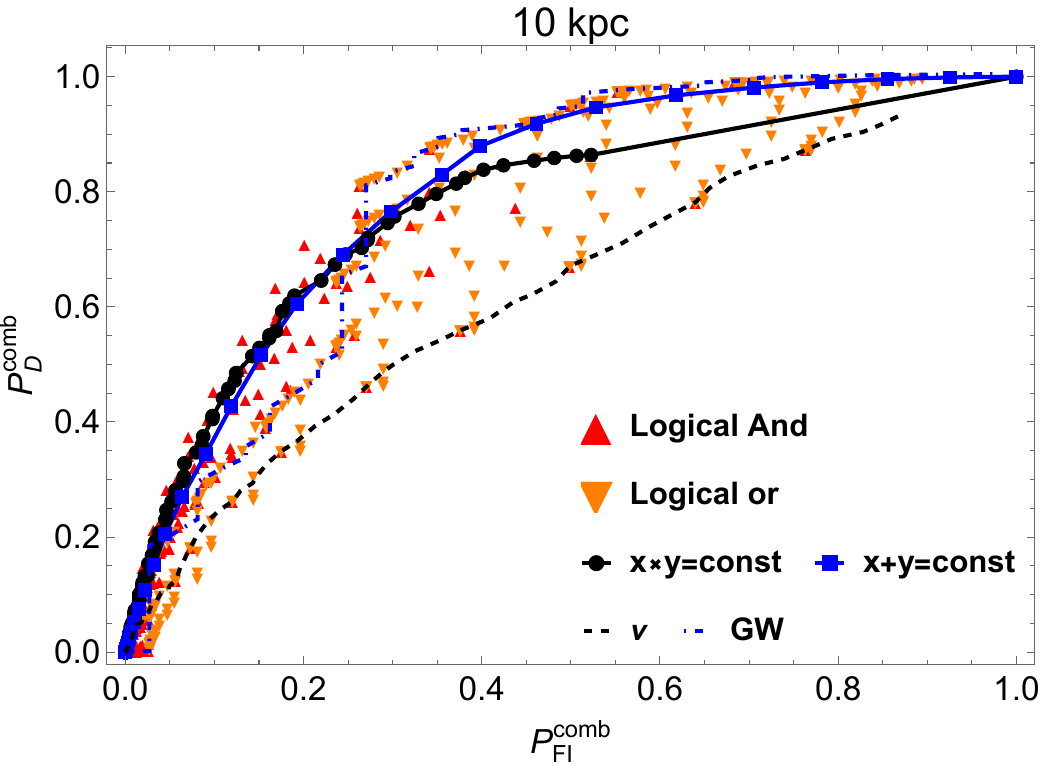}
    \end{center}
    \caption{Joint ROC curves for each of the distances and for each of the threshold selections described in section \ref{Sect_SASImeter_Joint}: \textit{logical and}, \textit{logical or}, $\rho + \mathcal{L} > \Lambda$ (labeled as $x+y = \text{const}$), and $\rho \times \mathcal{L} > \Lambda$ (labeled as $x \times y = \text{const}$).}
    \label{Fig_Joint_ROC}
\end{figure}

Delving into the previous statement, figure \ref{Fig_Joint_ROC} shows a key improvement with respect to the results in Lin2023 \cite{Zidu_SASImeter_Joint} that has its origin on the GW channel. This figure shows the ROC curves of the individual channels and the four combined thresholds defined in section \ref{Sect_SASImeter_Joint}. At 1 kpc, all ROC curves emulate the behavior of a perfect classifier. This happens because, as observed in the individual-channel ROC curves, figures \ref{Fig_ROC_GWs_O3} and \ref{fig:nuroc}, the $\nu-$ and GW-\texttt{SASImeter} have both an almost perfect detection efficiency; this translates into all combined ROC curves having, approximately, the same behavior. On the other hand, at 5 and 10 kpc, it is observed that the GW channel overperforms the neutrino detection efficiency, specially for the first; this translates into almost all joint ROC curves also overperforming the neutrino channel for almost all $P_\text{FI}$. In turn, these plots confirm one of the conclusions presented in Lin2023 \cite{Zidu_SASImeter_Joint}: the GW channel has a better detection efficiency at larger distances.

To directly show the improvements in the combined analysis due to the upgrades in the GW channel, table \ref{Table_Joint_ROC_Thresh} displays values of $P_\text{D}$ for different meaningful thresholds on $P_\text{FI}$ for this new analysis and that of Lin2023; this table particularly shows only the combined analysis for the $x+y=\mathrm{const}$ and $x\times y=\mathrm{const}$ ROC curves, so, for a complete comparison, reference is also made to figure 11 in Lin2023 \cite{Zidu_SASImeter_Joint}. For 1 kpc, it is shown that both ROC analysis display a perfect detection efficiency for all thresholds; however, the noticeable improvement introduced by the new implementations in the GW channel can be observed for the logical \textit{and} and \textit{or} analysis shown on the respective ROC curves. Those in Lin2023 clearly don't display the observed perfect behavior in figure \ref{Fig_Joint_ROC}, and that is due to the nature of the logic behind these specific ROC analysis. This shows that the improvements in the GW channel now position it on almost the same detection efficiency level as the neutrino channel, and this in turn make all combined detection efficiencies almost perfect.

Regarding the rest of the distances, table \ref{Table_Joint_ROC_Thresh} quantifies what can be clearly observed from the direct comparison between figure \ref{Fig_Joint_ROC} and figure 11 in Lin2023: the new analysis exhibits improved detection efficiencies across all ROC curves. Particularly, and as discussed in section \ref{Sect_Results_GW_O3}, the GW channel has better improvements at 5 kpc than at 10 kpc, and this is reflected in the combined analysis where it is observed that the difference between the $P_\text{D}$ values of the new and previous analysis are greater for 5 kpc than those for 10, specially for the lower $P_\text{FI}$ thresholds. However, even for 10 kpc, the new results exhibit larger detection probabilities than those of Lin2023.

\renewcommand{\arraystretch}{1.3}
\setlength{\tabcolsep}{7.6pt}
\begin{table}[h!]
\centering
\resizebox{0.75\textwidth}{!}{%
\begin{tabular}{ccccccc}
\hline
$P_\text{FI}$ & \multicolumn{6}{c}{$P_\text{D}$}                                                   \\
              & \multicolumn{6}{c}{$x + y$}                                                        \\
              & \multicolumn{2}{c}{1 kpc} & \multicolumn{2}{c}{5 kpc} & \multicolumn{2}{c}{10 kpc} \\ \cline{2-7} 
              & \textbf{XP O3}  & Lin2023 & \textbf{XP O3}  & Lin2023 & \textbf{XP O3}  & Lin2023  \\ \hline
0.01          & \textbf{1.00}   & 1.00    & \textbf{0.45}   & 0.30    & \textbf{0.06}   & 0.05     \\ \hline
0.05          & \textbf{1.00}   & 1.00    & \textbf{0.80}   & 0.56    & \textbf{0.21}   & 0.20     \\ \hline
0.10          & \textbf{1.00}   & 1.00    & \textbf{0.90}   & 0.70    & \textbf{0.37}   & 0.34     \\ \hline
0.25          & \textbf{1.00}   & 1.00    & \textbf{0.97}   & 0.87    & \textbf{0.70}   & 0.67     \\ \hline
0.50          & \textbf{1.00}   & 1.00    & \textbf{0.99}   & 0.96    & \textbf{0.93}   & 0.89     \\ \hline
              &                 &         &                 &         &                 &          \\
              &                 &         &                 &         &                 &          \\ \hline
$P_\text{FI}$ & \multicolumn{6}{c}{$P_\text{D}$}                                                   \\
              & \multicolumn{6}{c}{$x \times y$}                                                   \\
              & \multicolumn{2}{c}{1 kpc} & \multicolumn{2}{c}{5 kpc} & \multicolumn{2}{c}{10 kpc} \\ \cline{2-7} 
              & \textbf{XP O3}  & Lin2023 & \textbf{XP O3}  & Lin2023 & \textbf{XP O3}  & Lin2023  \\ \hline
0.01          & \textbf{1.00}   & 1.00    & \textbf{0.45}   & 0.30    & \textbf{0.07}   & 0.07     \\ \hline
0.05          & \textbf{1.00}   & 1.00    & \textbf{0.71}   & 0.56    & \textbf{0.26}   & 0.24     \\ \hline
0.10          & \textbf{1.00}   & 1.00    & \textbf{0.85}   & 0.71    & \textbf{0.41}   & 0.40     \\ \hline
0.25          & \textbf{1.00}   & 1.00    & \textbf{0.90}   & 0.86    & \textbf{0.69}   & 0.63     \\ \hline
0.50          & \textbf{1.00}   & 1.00    & \textbf{0.92}   & 0.90    & \textbf{0.86}   & 0.80     \\ \hline
\end{tabular}
}
\caption{Identification probability ($P_D$) for different false identification probabilities ($P_\text{FI}$), comparing the combined ROC criteria $x+y=\mathrm{const}$ and $x\times y=\mathrm{const}$. Results are shown for 1, 5, and 10 kpc, including both the new analysis and Lin2023.}
\label{Table_Joint_ROC_Thresh}
\end{table}

\section{Conclusions} \label{Sect_Conclusions}

The up-to-date results of the \texttt{SASImeter} pipeline, provided by candidate events produced by cWB XP, have been presented in this work, also including improvements in the SASI duration estimation. The SASI detectability has increased at all frequencies as well as the estimation of its parameters. For instance, the standard deviation of the central SASI frequency at 10 Kpc has dropped from $\sim 18.6 \; \text{Hz}$ in Lin2023 to only $\sim 5 \; \text{Hz}$ in O3 and $\sim 10 \; \text{Hz}$ in O4a, and even the smallest spread (at 1 Kpc) has fallen from $5.48 \; \text{Hz}$ to $4.82 \; \text{Hz}$ in O3 and $5.09 \; \text{Hz}$ in O4a, as seen from tables \ref{Table_GW_PE_O3} and \ref{Table_GW_PE_O4}. Similarly, the uncertainties in the time duration estimates decreased considerably; in Lin2023 \cite{Zidu_SASImeter_Joint}, the standard deviations represented $\sim 157 \%$, $\sim 112 \%$ and $\sim 134 \%$ of the estimated mean values for the time metric prescribed in equation \eqref{eq_time_dur_f1} at 1, 5 and 10 kpc respectively. Instead, for O3, the standard deviations represent $\sim 19 \%$, $\sim 46 \%$ and $\sim 47\%$ of the estimated mean values, while $\sim 22 \%$, $\sim 57 \%$ and $\sim 76 \%$ for O4a.

The combination of the neutrino and GW O3 messengers yields a strong detection efficiency compared to the results in Lin2023. As shown in figure \ref{Fig_Joint_ROC}, at 1 kpc every joint-criteria ROC is essentially ideal, since both channels individually already reach $P_D\approx1$ at very low false identification probabilities. At 5 and 10 kpc, joint detections generally outperform the neutrino channel, with only a few cases where they perform comparably or slightly worse. The numerical improvements are substantial: for instance, at 5 kpc and $P_{\rm FI}=0.1$, the combined $x+y=\text{const}$ threshold achieves $P_D\approx0.90$, compared to only $\sim0.70$ in the earlier work, as seen in table \ref{Table_Joint_ROC_Thresh}. These results reinforce and extend the conclusions of Lin2023 \cite{Zidu_SASImeter_Joint}: combining neutrino and GW O3 data increases the identification probability by allowing one messenger to compensate for the limitations of the other. 

The implementation of O4 data shows that the GW channel now matches the neutrino detection efficiency at 1 kpc and significantly surpasses it at larger distances, displaying an almost ideal identification performance. In this regime, and in virtue of what was previously mentioned, a joint-detection-efficiency analysis is not necessary, which motivates not repeating it with the latest data. However, it is necessary to stress that the parallel parameter estimation is highly relevant and valuable for a comprehensive astrophysical analysis. Consequently, the most important outcome of this updated study is that the GW channel, thanks to the most sensitive observational data to date and the XP version of the cWB detection pipeline, may be sufficiently reliable to identify SASI activity in a galactic core-collapse supernova on its own with high confidence. 

However, to further stress into the relevance of the combined analysis, it is important to highlight that the GWs and neutrinos emitted by the SASI are observer-dependent. Particularly, for the Kuroda2017 model, authors show \cite{Kuroda2017} that, for this particular simulation, the neutrino detection rate is significantly larger when the observer is along the axis of the SASI sloshing motion, while the GW emission is stronger toward the orthogonal direction of the same motion. Although these simulations are obtained for an observer along the positive z-axis, which is not a special direction relative to the SASI motion, it is relevant to stress that a hypothetical real detection could come from a direction such that the limit case where one channel is maximal and the other is minimal occurs; since the detection efficiency of the combined analysis can only be as worse as the worst-performing channel, then it becomes highly relevant, and necessary in practice, to evaluate both channels to confidently discriminate a SASI-born signal from a CCSN when there is confusion in both.

As mentioned in previous sections, the focus of this analysis relies on Kuroda2017 for comparative purposes with respect to previous \texttt{SASImeter} implementations, Lin2020 \cite{Zidu_SASImeter_Neutrino} and Lin2023 \cite{Zidu_SASImeter_Joint}. It is worth stressing that the pipeline's performance depends, in general, on the relative strength of the SASI activity to the rest of the GW signal (for a brief demonstration of an application of the \texttt{SASImeter} on a GW with no SASI activity, see Appendix \ref{App_1}), and in this specific simulation, the SASI activity is particularly strong. As for the relatively short duration of the simulation, it does not affect the performance of the GW-\texttt{SASImeter} as long as the whole SASI activity is included in what it has been defined as the SASI time-frequency region, this since the initial time of the HFF feature is only employed here to define such region; for a discussion of the interpretation of the slope and curvature of the HFF, the interested reader is refereed to \cite{Casallas2023, Casallas2025, Casallas_DCC}.

Furthermore, possible SASI frequency modulations are not expected to affect the identification performance of the GW-\texttt{SASImeter} as long as they do not drift outside the defined SASI region, this since the pipeline operates via the total energy located on it. Some simulations show that the GW SASI imprint may grow in frequency with time (spiral SASI) in late-shock revival scenarios, i.e., in long-lasting SASI, as seen in \cite{SASI_Spiral_Blondin, SASI_Spiral_Foglizzo, Vartanyan2023, Powell2021, Powell_DAWES, Burrows:2024pur}. If a spiral SASI frequency drifts above the chosen SASI frequency range, the identification performance would not be optimal but still useful since the pipeline avoids the production of false alarms by preventing leakage of HFF energy into the region. However, it is planned to investigate in future works an optimal identification method for spiral SASI; despite this, in terms of parameter estimations and drifting frequencies, this version of the \texttt{SASImeter} is still capable to probe the hypothesis that the frequency in the GW channel is double than in the neutrino one. To further discuss how the \texttt{SASImeter} methodology is valid for long-lasting signals, the same analysis was performed on a different GW model without SASI activity, Morozova2018, as discussed in Appendix \ref{App_1}. From the results on this model, it is observed that the \texttt{SASImeter} identification performance is that of a random classifier, which means that both the SASI and No-SASI cases register a similar presence of energy in the defined SASI region. As a consequence, requiring a small false identification probability automatically forces a small SASI detection probability.

From the neutrino perspective, the shortness of Kuroda2017 also represents an area of opportunity. As seen in figure \ref{Fig_Kuroda2017_Nu_Rate}, the SASI-induced modulation of neutrino signals ends abruptly at the tail, resulting in a non-detection of SASI for a $\nu$-\texttt{SASImeter} analysis starting at 150 ms and with durations $>~60~\mathrm{ms}$ in figure \ref{fig:nustartdur}. For longer SASI duration, the \texttt{SASImeter} can perform an analysis with finer frequency resolution, giving better performance for SASI frequency identification \cite{Zidu_SASImeter_Neutrino}.
 
Additionally, some conclusions can be drawn from the \texttt{SASImeter} estimations, specially in comparison with other analysis in the literature dedicated to the parameter estimation of the HFF. For example, in \cite{Casallas2023} authors apply a Neural-Network approach to estimate the slope of different GW models, including Kuroda2017; a direct comparison between those results and the ones provided here show that the SASImeter underestimates the slope by $\sim 1,000 \; \text{s}^{-2}$ with respect of that alternative estimation method across the three different galactic distances here prescribed. However, in \cite{HFF_Slope_Daniel}, authors show that a linear regression approach is sufficient enough to estimate the slope for HFF frequencies $\leq 1 \; \text{kHz}$, which is exactly the case for the Kuroda2017 model given the conditions of the simulation; however, a longer-lasting simulation could include a different evolution of the HFF that would require different methods for HFF-dedicated analyses. Furthermore, in this same work, authors correlate the HFF slope with the EoS of the progenitor; the estimations provided by the SASImeter are closer in range to the results for the nuclear equations of state SFHo and SFHx (see Table II in \cite{HFF_Slope_Daniel}), being the latter the true corresponding EoS for Kuroda2017.

Information about the PNS and the post-shock region are encoded both in the SASI time duration and its frequency. Authors, such as Foglizzo and collaborators, \cite{Foglizzo_SASI_Analytic, Foglizzo2014, Foglizzo2009, Foglizzo2007} and M\"uller \& Janka \cite{MullerJanka2014} have proposed analytic models that try to describe SASI properties, such as its oscillation period or its growth rate, based on key properties of the CCSN, such as the PNS mass, radius, and so compactness, the shock radius and the post-shock velocity; in this sense, the SASImeter can provide information which these models could be paired to. This type of analyses are left for future work.

Hence, all previous discussions show the relevance of the physical information provided by the \texttt{SASImeter} in the context of the multimessenger astrophysics of Core-Collapse Supernovae. Furthermore, these also motivate, and show, the necessary future work for the pipeline, specially in the GW channel, which will focus on addressing frequency growth and testing on more diverse models.

\appendix
\section{Appendix: \texttt{SASImeter} in a SASI-less model} \label{App_1}

\begin{figure}[h!]
    \centering
    \includegraphics[scale=0.35]{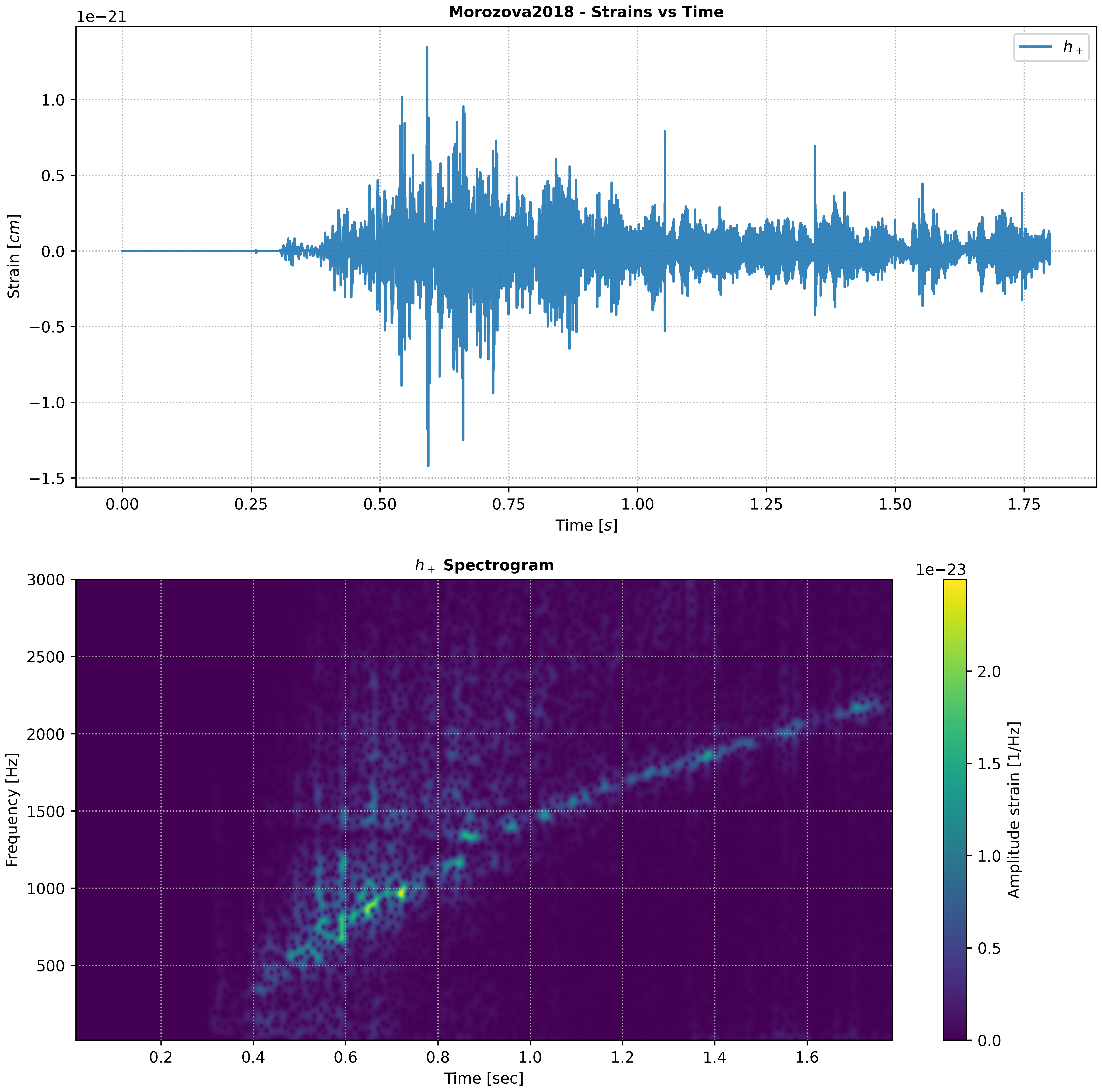}
    \caption{Morozova2018 waveform (at 10 kpc), a 2D Newtonian hydrodynamics with an approximate GR correction simulation from a $13 M_{\odot}$, non-rotating progenitor with the SFHo EoS. This particular GW signal has a long duration ($\sim 1,760 \; \text{ms}$) and has no evidence of SASI activity. \textit{Top row}: strain plot of polarization $h_+$. \textit{Bottom row}: spectrogram of the $h_+$ polarization.}
    \label{Fig_Morozova2018}
\end{figure}

\begin{figure}[h!]
    \begin{center}
        \includegraphics[width=0.45\linewidth]{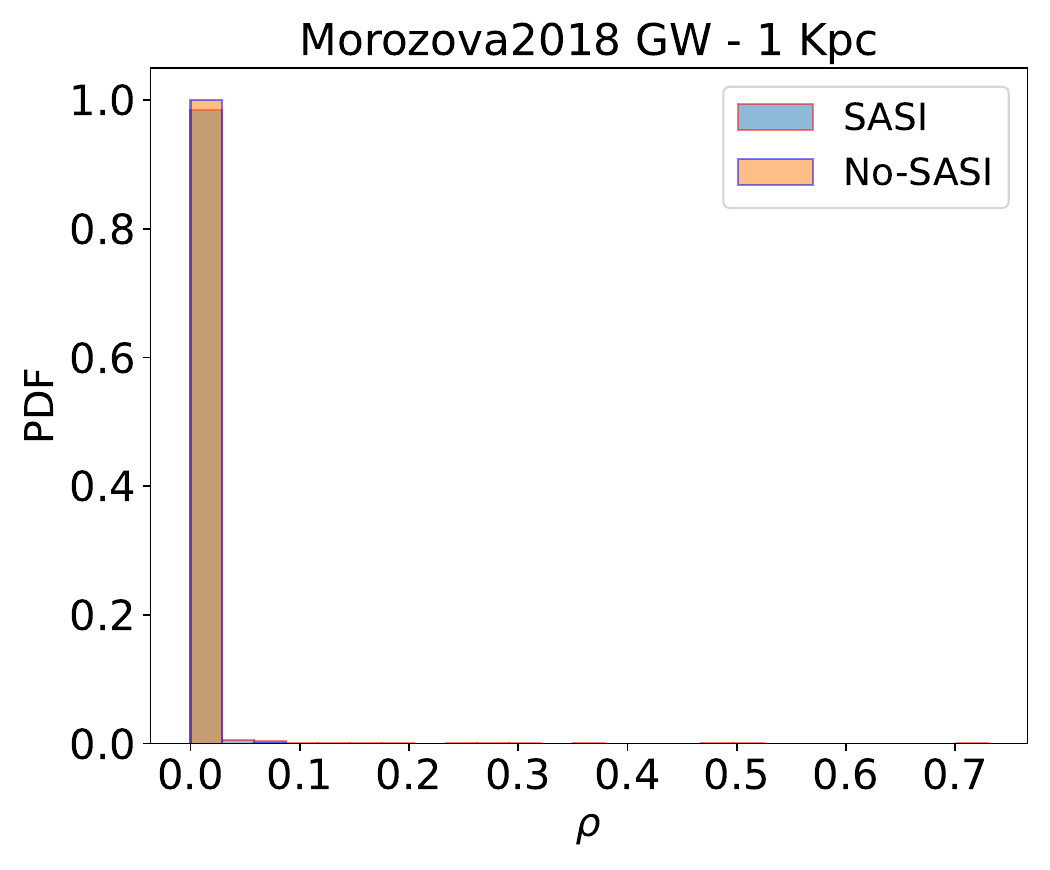}
        \includegraphics[width=0.45\linewidth]{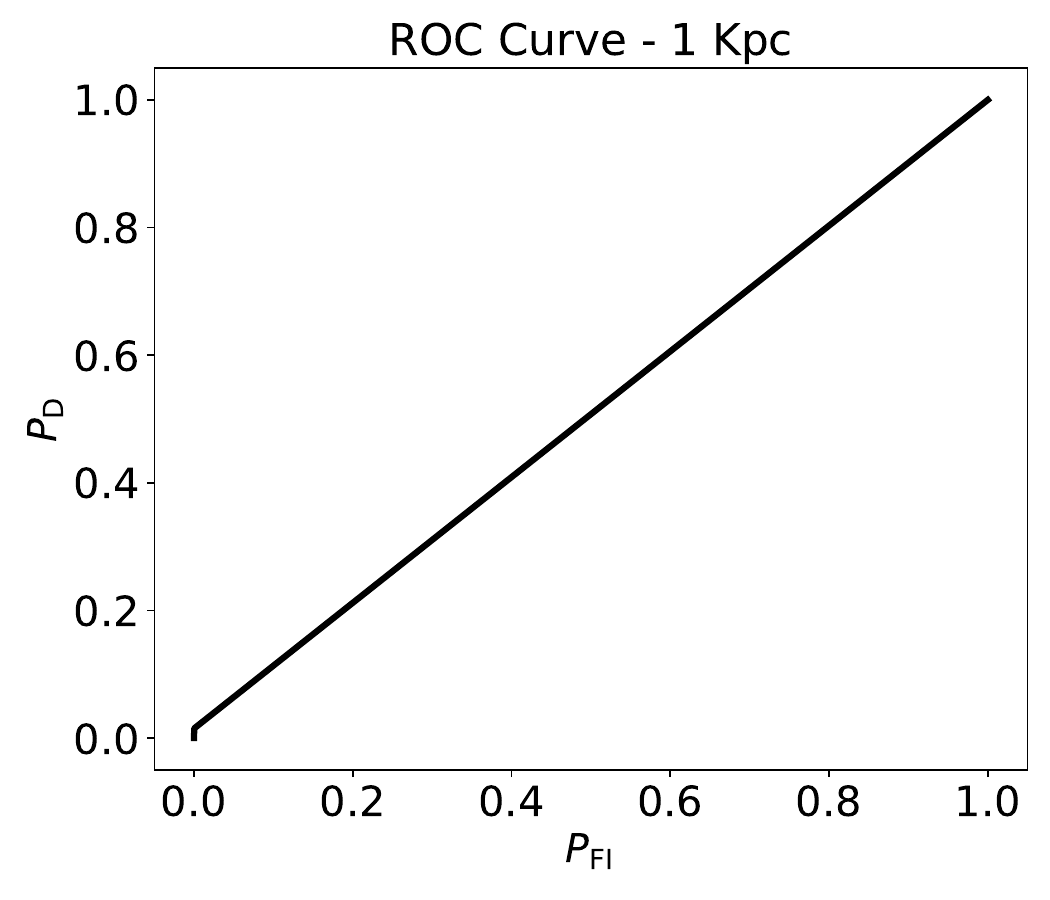}
    \end{center}
    \caption{Morozova2018 \texttt{SASImeter} identification results at 1 Kpc. Absence of SASI activity in the model leads to a bad identification performance as the pipeline is unable to distinguish between the SASI and No-SASI cases.}
    \label{Fig_Morozova2018_Results}
\end{figure}

To further demonstrate the \texttt{SASImeter} effectivity at identifying SASI activity and preventing false positive identifications, the pipeline was applied on a GW model, hereon referred to as Morozova2018, with features different from those of Kuroda2017. This simulated GW signal is obtained from a 13 solar masses, non-rotating progenitor with the SFHo EoS through a 2D Newtonian hydrodynamics with an approximate GR correction simulation \cite{Morozova2018}. As stated by its authors, this GW model shows no SASI activity, as also observed from figure \ref{Fig_Morozova2018}; furthermore, it is observed that this model is of a long duration, of approximately $1.76 \; \text{s}$, and that the HFF evolution reaches up to $\sim 2.25 \; \text{kHz}$. Then, in figure \ref{Fig_Morozova2018_Results}, the histograms for the $\rho_\text{GW}$ metric and the ROC curve at 1 kpc are shown. From the first, it is observed that both the SASI and No-SASI distributions are densely clustered near $\rho_\text{GW}=0$ and that they are extremely overlapped, which in turn translates into a ROC curve that exhibits a very poor identification performance, extremely close to that of a random classifier (a diagonal line). However, this is exactly what is expected from the \texttt{SASImeter} when there is no true SASI activity in a detection: both the SASI and No-SASI case register a similar, near zero presence of energy in the defined SASI region. This further proves that the methods employed by the \texttt{SASImeter} to determine the SASI time-frequency region are adequate and suppress spurious energy leakage, indeed avoiding false positives even when the GW signal is of a long duration and shows a more complex HFF evolution.

\section*{Acknowledgements}
Authors would like to thank the anonymous referees for their comments that led to improvements in the manuscript. The data underlying this work are available from the Gravitational Wave Open Science Center (GWOSC) at https://gwosc.org. V.S. is supported by a \textit{Beca Nacional de Posgrados} from the \textit{Secretaria de Ciencias, Humanidades, Tecnología e Innovación}. M.Z. is supported by the National Science Foundation Gravitational Physics Experimental and Data Analysis Program through awards PHY-2110555 and PHY-2405227. M.S. acknowledges Polish National Science Centre Grants No. UMO-2023/49/B/ST9/02777 and No. UMO-2024/03/1/ST9/00005, and the Polish National Agency for Academic Exchange within Polish Returns Programme Grant No. BPN/PPO/2023/1/00019.

\nocite{*}
\bibliographystyle{vancouver}
\bibliography{bibliography}

@article{SN_Multimess,
	author = {Fryer, Chris and Burns, Eric and Roming, Pete and Couch, Sean and Szczepa{\' n}czyk, Marek and Slane, Pat and Tamborra, Irene and Trappitsch, Reto},
	journal = {Bulletin of the AAS},
	number = {3},
    eid = {122},
	year = {2019},
	month = {may 31},
	url = {https://baas.aas.org/pub/2020n3i122},
	publisher = {},
	title = {Core-{Collapse} {Supernovae} and {Multi}-{Messenger} {Astronomy}},
	volume = {51},
}

@article{MDrago_Multimess,
  title = {Multimessenger observations of core-collapse supernovae: Exploiting the standing accretion shock instability},
  author = {Drago, Marco and Andresen, Haakon and Di Palma, Irene and Tamborra, Irene and Torres-Forn\'e, Alejandro},
  journal = {Phys. Rev. D},
  volume = {108},
  issue = {10},
  pages = {103036},
  numpages = {15},
  year = {2023},
  month = {Nov},
  publisher = {American Physical Society},
  doi = {10.1103/PhysRevD.108.103036},
  url = {https://link.aps.org/doi/10.1103/PhysRevD.108.103036}
}

@article{Michele_Tony_CCSNe,
  title = {Colloquium: Gravitational waves from neutrino-driven core collapse supernovae},
  author = {Mezzacappa, A. and Zanolin, M.},
  journal = {Rev. Mod. Phys.},
  pages = {--},
  year = {2025},
  month = {Sep},
  publisher = {American Physical Society},
  doi = {10.1103/pv6p-dtr2},
  url = {https://link.aps.org/doi/10.1103/pv6p-dtr2}
}

@article{Blondin_2006,
    doi = {10.1086/500817},
    url = {https://doi.org/10.1086/500817},
    year = {2006},
    month = {may},
    publisher = {},
    volume = {642},
    number = {1},
    pages = {401},
    author = {Blondin, John M. and Mezzacappa, Anthony},
    title = {The Spherical Accretion Shock Instability in the Linear Regime},
    journal = {The Astrophysical Journal}
}

@article{Walk_2020,
  title = {Neutrino emission characteristics of black hole formation in three-dimensional simulations of stellar collapse},
  author = {Walk, Laurie and Tamborra, Irene and Janka, Hans-Thomas and Summa, Alexander and Kresse, Daniel},
  journal = {Phys. Rev. D},
  volume = {101},
  issue = {12},
  pages = {123013},
  numpages = {24},
  year = {2020},
  month = {Jun},
  publisher = {American Physical Society},
  doi = {10.1103/PhysRevD.101.123013},
  url = {https://link.aps.org/doi/10.1103/PhysRevD.101.123013}
}

@article{EoS,
	author = {{Marek, A.} and {Janka, H.-T.} and {Müller, E.}},
	title = {Equation-of-state dependent features in shock-oscillation modulated
neutrino and gravitational-wave signals   from supernovae},
	DOI= "10.1051/0004-6361/200810883",
	url= "https://doi.org/10.1051/0004-6361/200810883",
	journal = {A\&A},
	year = 2009,
	volume = 496,
	number = 2,
	pages = "475-494",
}

@ARTICLE{Exp_Mechanism,
  title    = "Physical mechanism of core-collapse supernovae that neutrinos
              drive",
  author   = "Yamada, Shoichi and Nagakura, Hiroki and Akaho, Ryuichiro and
              Harada, Akira and Furusawa, Shun and Iwakami, Wakana and Okawa,
              Hirotada and Matsufuru, Hideo and Sumiyoshi, Kohsuke",
  journal  = "Proc Jpn Acad Ser B Phys Biol Sci",
  volume   =  100,
  number   =  3,
  pages    = "190--233",
  year     =  2024,
  address  = "Japan",
  language = "en"
}

@article{Zidu_SASImeter_Neutrino,
  title = {Detectability of standing accretion shock instabilities activity in supernova neutrino signals},
  author = {Lin, Zidu and Lunardini, Cecilia and Zanolin, Michele and Kotake, Kei and Richardson, Colter},
  journal = {Phys. Rev. D},
  volume = {101},
  issue = {12},
  pages = {123028},
  numpages = {15},
  year = {2020},
  month = {Jun},
  publisher = {American Physical Society},
  doi = {10.1103/PhysRevD.101.123028},
  url = {https://link.aps.org/doi/10.1103/PhysRevD.101.123028}
}

@article{Zidu_SASImeter_Joint,
  title = {Characterizing a supernova's standing accretion shock instability with neutrinos and gravitational waves},
  author = {Lin, Zidu and Rijal, Abhinav and Lunardini, Cecilia and Morales, Manuel D. and Zanolin, Michele},
  journal = {Phys. Rev. D},
  volume = {107},
  issue = {8},
  pages = {083017},
  numpages = {24},
  year = {2023},
  month = {Apr},
  publisher = {American Physical Society},
  doi = {10.1103/PhysRevD.107.083017},
  url = {https://link.aps.org/doi/10.1103/PhysRevD.107.083017}
}

@article{Kuroda2017,
doi = {10.3847/1538-4357/aa988d},
url = {https://doi.org/10.3847/1538-4357/aa988d},
year = {2017},
month = {dec},
publisher = {The American Astronomical Society},
volume = {851},
number = {1},
pages = {62},
author = {Kuroda, Takami and Kotake, Kei and Hayama, Kazuhiro and Takiwaki, Tomoya},
title = {Correlated Signatures of Gravitational-wave and Neutrino Emission in Three-dimensional General-relativistic Core-collapse Supernova Simulations},
journal = {The Astrophysical Journal}
}

@article{cwb,
  title = {Method for detection and reconstruction of gravitational wave transients with networks of advanced detectors},
  author = {Klimenko, S. and Vedovato, G. and Drago, M. and Salemi, F. and Tiwari, V. and Prodi, G. A. and Lazzaro, C. and Ackley, K. and Tiwari, S. and Da Silva, C. F. and Mitselmakher, G.},
  journal = {Phys. Rev. D},
  volume = {93},
  issue = {4},
  pages = {042004},
  numpages = {10},
  year = {2016},
  month = {Feb},
  publisher = {American Physical Society},
  doi = {10.1103/PhysRevD.93.042004},
  url = {https://link.aps.org/doi/10.1103/PhysRevD.93.042004}
}

@misc{wavescan,
      title={Wavescan: multiresolution regression of gravitational-wave data}, 
      author={Sergey Klimenko},
      year={2022},
      eprint={2201.01096},
      archivePrefix={arXiv},
      primaryClass={physics.data-an},
      url={https://arxiv.org/abs/2201.01096}, 
}

@ARTICLE{CCSNe_Theory,
  title    = "Core-collapse supernova explosion theory",
  author   = "Burrows, A and Vartanyan, D",
  journal  = "Nature",
  volume   =  589,
  number   =  7840,
  pages    = "29--39",
  month    =  jan,
  year     =  2021
}

@article{O3_data,
doi = {10.3847/1538-4365/acdc9f},
url = {https://doi.org/10.3847/1538-4365/acdc9f},
year = {2023},
month = {jul},
publisher = {The American Astronomical Society},
volume = {267},
number = {2},
pages = {29},
author = {Abbott, R. and Abe, H. and Acernese, F. and Ackley, K. and Adhicary, S. and Adhikari, N. and Adhikari, R. X. and Adkins, V. K. and Adya, V. B. and Affeldt, C. and Agarwal, D. and Agathos, M. and Aguiar, O. D. and Aiello, L. and Ain, A. and Ajith, P. and Akutsu, T. and Albanesi, S. and Alfaidi, R. A. and Al-Jodah, A. and Alléné, C. and Allocca, A. and Almualla, M. and Altin, P. A. and Amato, A. and Amez-Droz, L. and Amorosi, A. and Anand, S. and Ananyeva, A. and Andersen, R. and Anderson, S. B. and Anderson, W. G. and Andia, M. and Ando, M. and Andrade, T. and Andres, N. and Andrés-Carcasona, M. and Andrić, T. and Ansoldi, S. and Antelis, J. M. and Antier, S. and Aoumi, M. and Apostolatos, T. and Appavuravther, E. Z. and Appert, S. and Apple, S. K. and Arai, K. and Araya, A. and Araya, M. C. and Areeda, J. S. and Arène, M. and Aritomi, N. and Arnaud, N. and Arogeti, M. and Aronson, S. M. and Arun, K. G. and Asada, H. and Ashton, G. and Aso, Y. and Assiduo, M. and Assis de Souza Melo, S. and Aston, S. M. and Astone, P. and Aubin, F. and AultONeal, K. and Babak, S. and Badalyan, A. and Badaracco, F. and Badger, C. and Bae, S. and Bagnasco, S. and Bai, Y. and Baier, J. G. and Baiotti, L. and Baird, J. and Bajpai, R. and Baka, T. and Ball, M. and Ballardin, G. and Ballmer, S. W. and Baltus, G. and Banagiri, S. and Banerjee, B. and Bankar, D. and Baral, P. and Barayoga, J. C. and Barber, J. and Barish, B. C. and Barker, D. and Barneo, P. and Barone, F. and Barr, B. and Barsotti, L. and Barsuglia, M. and Barta, D. and Barthelmy, S. D. and Barton, M. A. and Bartos, I. and Basak, S. and Basalaev, A. and Bassiri, R. and Basti, A. and Bawaj, M. and Bayley, J. C. and Baylor, A. C. and Bazzan, M. and Bécsy, B. and Bedakihale, V. M. and Beirnaert, F. and Bejger, M. and Bell, A. S. and Benedetto, V. and Beniwal, D. and Benoit, W. and Bentley, J. D. and Yaala, M. Ben and Bera, S. and Berbel, M. and Bergamin, F. and Berger, B. K. and Bernuzzi, S. and Beroiz, M. and Berry, C. P. L. and Bersanetti, D. and Bertolini, A. and Betzwieser, J. and Beveridge, D. and Bevins, N. and Bhandare, R. and Bhandari, A. V. and Bhardwaj, U. and Bhatt, R. and Bhattacharjee, D. and Bhaumik, S. and Bianchi, A. and Bilenko, I. A. and Bilicki, M. and Billingsley, G. and Bini, S. and Birnholtz, O. and Biscans, S. and Bischi, M. and Biscoveanu, S. and Bisht, A. and Biswas, B. and Bitossi, M. and Bizouard, M.-A. and Blackburn, J. K. and Blair, C. D. and Blair, D. G. and Blair, R. M. and Bobba, F. and Bode, N. and Boër, M. and Bogaert, G. and Boileau, G. and Boldrini, M. and Bolingbroke, G. N. and Bonavena, L. D. and Bondarescu, R. and Bondu, F. and Bonilla, E. and Bonilla, G. S. and Bonnand, R. and Booker, P. and Bork, R. and Boschi, V. and Bose, N. and Bose, S. and Bossilkov, V. and Boudart, V. and Bouffanais, Y. and Bozzi, A. and Bradaschia, C. and Brady, P. R. and Braglia, M. and Branch, A. and Branchesi, M. and Brau, J. E. and Breschi, M. and Briant, T. and Brillet, A. and Brinkmann, M. and Brockill, P. and Brooks, A. F. and Brooks, J. and Brown, D. D. and Brunett, S. and Bruno, G. and Bruntz, R. and Bryant, J. and Bucci, F. and Buchanan, J. and Bulashenko, O. and Bulik, T. and Bulten, H. J. and Buonanno, A. and Burtnyk, K. and Buscicchio, R. and Buskulic, D. and Buy, C. and Byer, R. L. and Cabourn Davies, G. S. and Cabras, G. and Cabrita, R. and Cadonati, L. and Caesar, S. and Cagnoli, G. and Cahillane, C. and Calderón Bustillo, J. and Callaghan, J. D. and Callister, T. A. and Calloni, E. and Camp, J. B. and Canepa, M. and Santoro, G. Caneva and Cannavacciuolo, M. and Cannon, K. C. and Cao, H. and Cao, Z. and Capistran, L. A. and Capocasa, E. and Capote, E. and Carapella, G. and Carbognani, F. and Carlassara, M. and Carlin, J. B. and Carpinelli, M. and Carter, J. J. and Carullo, G. and Casanueva Diaz, J. and Casentini, C. and Castaldi, G. and Castro-Lucas, S. Y. and Caudill, S. and Cavaglià, M. and Cavalieri, R. and Cella, G. and Cerdá-Durán, P. and Cesarini, E. and Chaibi, W. and Chakalis, W. and Chalathadka Subrahmanya, S. and Champion, E. and Chan, C. and Chan, C. L. and Chandra, K. and Chang, I. P. and Chang, W. and Chanial, P. and Chao, S. and Chapman-Bird, C. and Charlton, E. L. and Charlton, P. and Chassande-Mottin, E. and Chastain, L. and Chatterjee, C. and Chatterjee, Debarati and Chatterjee, Deep and Chaturvedi, M. and Chaty, S. and Chatziioannou, K. and Chen, D. and Chen, H. and Chen, H. Y. and Chen, J. and Chen, K. H. and Chen, X. and Chen, Y.-R. and Chen, Y. and Cheng, H. and Chessa, P. and Cheung, H. Y. and Chia, H. Y. and Chiadini, F. and Chiang, C-I. and Chiang, C. and Chiarini, G. and Chiba, A. and Chiba, R. and Chierici, R. and Chincarini, A. and Chiofalo, M. L. and Chiummo, A. and Choudhary, S. and Christensen, N. and Chua, S. S. Y. and Chung, K. W. and Ciani, G. and Ciecielag, P. and Cieślar, M. and Cifaldi, M. and Ciobanu, A. A. and Ciolfi, R. and Clara, F. and Clark, J. A. and Clarke, T. A. and Clearwater, P. and Clesse, S. and Cleva, F. and Coccia, E. and Codazzo, E. and Cohadon, P.-F. and Colleoni, M. and Collette, C. G. and Colombo, A. and Colpi, M. and Compton, C. M. and Conti, L. and Cooper, S. J. and Corban, P. and Corbitt, T. R. and Cordero-Carrión, I. and Corezzi, S. and Cornish, N. J. and Corsi, A. and Cortese, S. and Coschizza, A. C. and Cottingham, R. and Coughlin, M. W. and Coulon, J.-P. and Countryman, S. T. and Coupechoux, J.-F. and Cousins, B. and Couvares, P. and Coward, D. M. and Cowart, M. J. and Cowburn, B. D. and Coyne, D. C. and Coyne, R. and Craig, K. and Creighton, J. D. E. and Creighton, T. D. and Criswell, A. W. and Crockett-Gray, J. C. G. and Croquette, M. and Crowder, S. G. and Cudell, J. R. and Cullen, T. J. and Cumming, A. and Cummings, R. and Cuoco, E. and Curyło, M. and Dabadie, P. and Dal Canton, T. and Dall’Osso, S. and Dálya, G. and D’Angelo, B. and Danilishin, S. and D’Antonio, S. and Danzmann, K. and Darroch, K. E. and Darsow-Fromm, C. and Dasgupta, A. and Datrier, L. E. H. and Datta, Sayantani and Dattilo, V. and Dave, I. and Davenport, A. and Davier, M. and Davis, D. and Davis, M. C. and Daw, E. J. and Dax, M. and DeBra, D. and Deenadayalan, M. and Degallaix, J. and De Laurentis, M. and Deléglise, S. and Del Favero, V. and De Lillo, F. and De Lillo, N. and Dell’Aquila, D. and Del Pozzo, W. and De Matteis, F. and D’Emilio, V. and Demos, N. and Dent, T. and Depasse, A. and De Pietri, R. and De Rosa, R. and De Rossi, C. and DeSalvo, R. and De Simone, R. and Dhurandhar, S. and Diab, R. and Diamond, P. Z. and Díaz, M. C. and Didio, N. A. and Dietrich, T. and Di Fiore, L. and Di Fronzo, C. and Di Giorgio, C. and Di Giovanni, F. and Di Giovanni, M. and Di Girolamo, T. and Diksha, D. and Di Lieto, A. and Di Michele, A. and Di Pace, S. and Di Palma, I. and Di Renzo, F. and Divyajyoti and Dmitriev, A. and Doctor, Z. and Dohmen, E. and Doleva, P. P. and Donahue, L. and D’Onofrio, L. and Donovan, F. and Dooley, K. L. and Dooney, T. and Doravari, S. and Dorosh, O. and Drago, M. and Driggers, J. C. and Drori, Y. and Ducoin, J.-G. and Dunn, L. and Dupletsa, U. and Durante, O. and D’Urso, D. and Duverne, P.-A. and Dwyer, S. E. and Eassa, C. and Easter, P. J. and Ebersold, M. and Eckhardt, T. and Eddolls, G. and Edelman, B. and Edo, T. B. and Edy, O. and Effler, A. and Eichholz, J. and Eisenmann, M. and Eisenstein, R. A. and Ejlli, A. and Engelby, E. and Engl, A. J. and Errico, L. and Essick, R. C. and Estellés, H. and Estevez, D. and Etzel, T. and Evans, C. and Evans, M. and Evans, T. M. and Evstafyeva, T. and Ewing, B. E. and Fabrizi, F. and Faedi, F. and Fafone, V. and Fair, H. and Fairhurst, S. and Fan, P. C. and Fan, X. and Farah, A. M. and Farr, B. and Farr, W. M. and Fauchon-Jones, E. J. and Favaro, G. and Favata, M. and Fays, M. and Feicht, J. and Fejer, M. M. and Fenyvesi, E. and Ferguson, D. L. and Fernandez-Galiana, A. and Ferrante, I. and Ferreira, T. A. and Fidecaro, F. and Figura, P. and Fiori, A. and Fiori, I. and Fishbach, M. and Fisher, R. P. and Fittipaldi, R. and Fiumara, V. and Flaminio, R. and Fleischer, S. M. and Fleming, L. S. and Floden, E. and Fong, H. K. and Font, J. A. and Fornal, B. and Forsyth, P. W. F. and Franke, A. and Frasca, S. and Frasconi, F. and Freed, J. P. and Frei, Z. and Freise, A. and Freitas, O. and Frey, R. and Fritschel, P. and Frolov, V. V. and Fronzé, G. G. and Fujimoto, Y. and Fukunaga, I. and Fulda, P. and Fyffe, M. and Gabbard, H. A. and Gabella, W. E. and Gadre, B. U. and Gaglani, K. and Gair, J. R. and Gais, J. and Galaudage, S. and Gallardo, S. and Gamba, R. and Ganapathy, D. and Ganguly, A. and Gao, D. and Gaonkar, S. G. and Garaventa, B. and Garcia-Bellido, J. and García-Núñez, C. and García-Quirós, C. and Gardner, K. A. and Gargiulo, J. and Garufi, F. and Gasbarra, C. and Gateley, B. and Gayathri, V. and Gemme, G. and Gennai, A. and George, J. and Gerberding, O. and Gergely, L. and Ghonge, S. and Ghosh, Abhirup and Ghosh, Archisman and Ghosh, Shaon and Ghosh, Shrobana and Ghosh, T. and Giacoppo, L. and Giaime, J. A. and Giardina, K. D. and Gibson, D. R. and Gier, C. and Giri, P. and Gissi, F. and Gkaitatzis, S. and Glanzer, J. and Gleckl, A. E. and Glotin, F. and Godfrey, J. and Godwin, P. and Goetz, E. and Goetz, R. and Golomb, J. and Goncharov, B. and González, G. and Gosselin, M. and Gouaty, R. and Gould, D. W. and Goyal, S. and Grace, B. and Grado, A. and Graham, V. and Granata, M. and Granata, V. and Gras, S. and Grassia, P. and Gray, C. and Gray, R. and Greco, G. and Green, A. C. and Green, R. and Green, S. and Green, S. R. and Gretarsson, A. M. and Gretarsson, E. M. and Griffith, D. and Griffiths, W. L. and Griggs, H. L. and Grignani, G. and Grimaldi, A. and Grote, H. and Gruson, A. S. and Guerra, D. and Guetta, D. and Guidi, G. M. and Guimaraes, A. R. and Gulati, H. K. and Gulminelli, F. and Gunny, A. M. and Guo, H. and Guo, Y. and Gupta, Anchal and Gupta, Anuradha and Gupta, Ish and Gupta, N. C. and Gupta, P. and Gupta, S. K. and Gurs, J. and Gushima, Y. and Gustafson, E. K. and Gutierrez, N. and Guzman, F. and Haegel, L. and Hain, G. and Haino, S. and Halim, O. and Hall, E. D. and Hamilton, E. Z. and Hammond, G. and Han, W.-B. and Haney, M. and Hanks, J. and Hanna, C. and Hannam, M. D. and Hannuksela, O. A. and Hansen, H. and Hanson, J. and Harada, R. and Harder, T. and Haris, K. and Harmark, T. and Harms, J. and Harry, G. M. and Harry, I. W. and Hartwig, D. and Haskell, B. and Haster, C.-J. and Hathaway, J. S. and Haughian, K. and Hayakawa, H. and Hayama, K. and Hayes, F. J. and Healy, J. and Heffernan, A. and Heidmann, A. and Heintze, M. C. and Heinze, J. and Heinzel, J. and Heitmann, H. and Hellman, F. and Hello, P. and Helmling-Cornell, A. F. and Hemming, G. and Hendry, M. and Heng, I. S. and Hennes, E. and Hennig, J.-S. and Hennig, M. and Henshaw, C. and Hernandez Vivanco, F. and Heurs, M. and Hewitt, A. L. and Higginbotham, S. and Hild, S. and Hill, P. and Himemoto, Y. and Hines, A. S. and Hirata, N. and Hirose, C. and Ho, J. and Hochheim, S. and Hofman, D. and Hohmann, J. N. and Holcomb, D. G. and Holland, N. A. and Holley-Bockelmann, K. and Hollows, I. J. and Holmes, Z. J. and Holt, K. and Holz, D. E. and Hong, Q. and Hornung, J. and Hoshino, S. and Hough, J. and Hourihane, S. and Howell, D. and Howell, E. J. and Hoy, C. G. and Hoyland, D. and Hsieh, B.-H. and Hsieh, H.-F. and Hsiung, C. and Hsu, H. and Hu, P. and Hu, Q. and Huang, H.-Y. and Huang, Y.-J. and Huang, Y. and Huang, Y. T. and Hübner, M. T. and Huddart, A. D. and Hughey, B. and Hui, D. C. Y. and Hui, V. and Husa, S. and Huttner, S. H. and Huxford, R. and Huynh-Dinh, T. and Hyland, J. and Iakovlev, A. and Iandolo, G. A. and Idzkowski, B. and Iess, A. and Inayoshi, K. and Inoue, Y. and Iorio, G. and Iosif, P. and Irwin, J. and Isi, M. and Ismail, M. A. and Itoh, Y. and Iyer, B. R. and JaberianHamedan, V. and Jacqmin, T. and Jacquet, P.-E. and Jadhav, S. J. and Jadhav, S. P. and Jain, D. and Jain, T. and James, A. L. and Jan, A. Z. and Jani, K. and Janiurek, L. and Janquart, J. and Janssens, K. and Janthalur, N. N. and Jaraba, S. and Jaranowski, P. and Jarov, S. and Jasal, P. and Jaume, R. and Javed, W. and Jenkins, A. C. and Jenner, K. and Jennings, A. and Jia, W. and Jiang, J. and Liu, Jian and Jin, H.-B. and Johansmeyer, K. and Johns, G. R. and Johnson, N. A. and Johnston, R. and Johny, N. and Jones, A. W. and Jones, D. H. and Jones, D. I. and Jones, P. and Jones, R. and Joshi, P. and Ju, L. and Jung, K. and Junker, J. and Juste, V. and Kajita, T. and Kalaghatgi, C. and Kalogera, V. and Kamai, B. and Kamiizumi, M. and Kanda, N. and Kandhasamy, S. and Kang, G. and Kanner, J. B. and Kapadia, S. J. and Kapasi, D. P. and Karat, S. and Karathanasis, C. and Karki, S. and Kasamatsu, D. and Kas-danouche, Y. A. and Kashyap, R. and Kasprzack, M. and Kastaun, W. and Kato, J. and Katsanevas, S. and Katsavounidis, E. and Katsuren, J. K. and Katzman, W. and Kaur, T. and Kawabe, K. and Kawazoe, K. and Kéfélian, F. and Keitel, D. and Kellard, I. and Kelley-Derzon, J. and Kennington, J. and Key, J. S. and Khadka, S. and Khalili, F. Y. and Khan, S. and Khanam, T. and Khazanov, E. A. and Khursheed, M. and Kijbunchoo, N. and Kim, C. and Kim, J. C. and Kim, K. and Kim, M. H. and Kim, P. and Kim, S. and Kim, W. S. and Kim, Y.-M. and Kimball, C. and Kimura, N. and Kinley-Hanlon, M. and Kirchhoff, R. and Kissel, J. S. and Kiyota, T. and Klimenko, S. and Klinger, T. and Knee, A. M. and Knust, N. and Kobayashi, Y. and Koch, P. and Koehlenbeck, S. M. and Koekoek, G. and Kohri, K. and Kokeyama, K. and Koley, S. and Koliadko, N. D. and Kolitsidou, P. and Kolstein, M. and Kondrashov, V. and Kong, A. K. H. and Kontos, A. and Korobko, M. and Kossak, R. V. and Kouvatsos, N. and Kovalam, M. and Koyama, N. and Kozak, D. B. and Kranzhoff, L. and Kranzhoff, S. L. and Kringel, V. and Krishnendu, N. V. and Królak, A. and Kuehn, G. and Kuijer, P. and Kukihara, M. and Kulkarni, S. and Kumar, A. and Kumar, Praveen and Kumar, Prayush and Kumar, Rahul and Kumar, Rakesh and Kume, J. and Kuns, K. and Kuroyanagi, S. and Kuwahara, S. and Kwak, K. and Lacaille, G. and Lagabbe, P. and Laghi, D. and Lakkis, M. H. and Lalande, E. and Lalleman, M. and Lamberts, A. and Landry, M. and Lane, B. B. and Lang, R. N. and Lange, J. and Lantz, B. and La Rana, A. and La Rosa, I. and Lartaux-Vollard, A. and Lasky, P. D. and Lawrence, J. and Laxen, M. and Lazzarini, A. and Lazzaro, C. and Leaci, P. and Leavey, S. and LeBohec, S. and Lecoeuche, Y. K. and Lee, E. and Lee, H. M. and Lee, H. W. and Lee, K. and Lee, R.-L. and Lee, R. and Lee, S. and Legred, I. N. and Lehmann, J. and Lehner, L. and Lemaître, A. and Lenti, M. and Leonardi, M. and Leonova, E. and Leroy, N. and Letendre, N. and Lethuillier, M. and Levesque, C. and Levin, Y. and Leyde, K. and Li, A. K. Y. and Li, K. L. and Li, T. G. F. and Li, X. and Lin, C.-Y. and Lin, E. T. and Lin, F-K. and Lin, F-L. and Lin, F. and Lin, H. L. and Lin, H. and Lin, L. C.-C. and Linde, F. and Linker, S. D. and Littenberg, T. B. and Liu, A. and Liu, G. C. and Llamas, F. and Lo, R. K. L. and Lo, T. and London, L. T. and Longo, A. and Lopez, D. and Lopez Portilla, M. and Lorenzini, M. and Loriette, V. and Lormand, M. and Losurdo, G. and Lott, T. P. and Lough, J. D. and Loughlin, H. A. and Lousto, C. O. and Lovelace, G. and Lowry, M. J. and Lück, H. and Lumaca, D. and Lundgren, A. P. and Lung, Y. and Lussier, A. W. and Lynam, J. E. and Ma, L. and Ma, S. and Ma’arif, M. and Macas, R. and MacInnis, M. and Macleod, D. M. and MacMillan, I. A. O. and Macquet, A. and Magaña Hernandez, I. and Magazzù, C. and Magee, R. M. and Maggiore, R. and Magnozzi, M. and Mahesh, M. and Mahesh, S. and Maini, M. and Majorana, E. and Makarem, C. N. and Maliakal, S. and Malik, A. and Man, N. and Mandic, V. and Mangano, V. and Mannix, B. and Mansell, G. L. and Mansingh, G. and Manske, M. and Mantovani, M. and Mapelli, M. and Marchesoni, F. and Pina, D. Marín and Marion, F. and Márka, S. and Márka, Z. and Markakis, C. and Markosyan, A. S. and Markowitz, A. and Maros, E. and Marquina, A. and Marsat, S. and Martelli, F. and Martin, I. W. and Martin, R. M. and Martinez, B. B. and Martinez, M. and Martinez, V. A. and Martinez, V. and Martinovic, K. and Martynov, D. V. and Marx, E. J. and Masalehdan, H. and Mason, K. and Masserot, A. and Reid, M. Masso and Mastrodicasa, M. and Mastrogiovanni, S. and Mateu-Lucena, M. and Matiushechkina, M. and Matsunaga, K. and Mavalvala, N. and McCarthy, R. and McClelland, D. E. and McClincy, P. K. and McCormick, S. and McCuller, L. and McGhee, G. I. and McGinn, J. and McIsaac, C. and McIver, J. and McLeod, A. and McRae, T. and McWilliams, S. T. and Meacher, D. and Mehmet, M. and Mehta, A. K. and Meijer, Q. and Melatos, A. and Mendell, G. and Menendez-Vazquez, A. and Menoni, C. S. and Mercer, R. A. and Mereni, L. and Merfeld, K. and Merilh, E. L. and Merritt, J. D. and Merzougui, M. and Messenger, C. and Messick, C. and Meyers, P. M. and Meylahn, F. and Mhaske, A. and Miani, A. and Miao, H. and Michaloliakos, I. and Michel, C. and Michimura, Y. and Middleton, H. and Mihaylov, D. P. and Miller, A. and Miller, A. L. and Miller, B. and Miller, S. and Millhouse, M. and Mills, J. C. and Milotti, E. and Minenkov, Y. and Mio, N. and Mir, Ll. M. and Miravet-Tenés, M. and Mishra, A. and Mishra, C. and Mishra, T. and Mistry, T. and Mitchell, A. L. and Mitra, S. and Mitrofanov, V. P. and Mitselmakher, G. and Mittleman, R. and Miyakawa, O. and Miyoki, S. and Mo, Geoffrey and Modafferi, L. M. and Moguel, E. and Mohapatra, S. R. P. and Mohite, S. R. and Molina-Ruiz, M. and Mondal, C. and Mondin, M. and Montani, M. and Moore, C. J. and Moragues, J. and Moraru, D. and Morawski, F. and More, A. and More, S. and Moreno, C. and Moreno, G. and Morisaki, S. and Moriwaki, Y. and Morras, G. and Moscatello, A. and Mours, B. and Mow-Lowry, C. M. and Mozzon, S. and Muciaccia, F. and Mukherjee, D. and Mukherjee, Soma and Mukherjee, Subroto and Mukherjee, Suvodip and Mukund, N. and Mullavey, A. and Munch, J. and Muñiz, E. A. and Murray, P. G. and Murray-Dean, J. and Muusse, S. and Nadji, S. L. and Nagar, A. and Nagar, T. and Nagarajan, N. and Nakamura, K. and Nakano, H. and Nakano, M. and Nakayama, Y. and Napolano, V. and Nardecchia, I. and Narikawa, T. and Narola, H. and Naticchioni, L. and Nayak, R. K. and Neil, B. F. and Neilson, J. and Nelson, A. and Nelson, T. J. N. and Nery, M. and Nesseris, S. and Neunzert, A. and Ng, K. Y. and Ng, S. W. S. and Nguyen, C. and Nguyen, P. and Nguyen, R. and Nguyen, T. and Nguyen Quynh, L. and Nichols, S. A. and Nieradka, G. and Nishino, Y. and Nishizawa, A. and Nissanke, S. and Nitoglia, E. and Niu, W. and Nocera, F. and Norman, M. and North, C. and Novak, J. and Nuño Siles, J. F. and Nurbek, G. and Nuttall, L. K. and Oberling, J. and O’Dell, J. and Oelker, E. and Oertel, M. and Oganesyan, G. and Oh, J. J. and Oh, K. and Oh, S. H. and O’Hanlon, T. and Ohashi, M. and Ohashi, T. and Ohkawa, M. and Ohme, F. and Ohta, H. and Oliveira, A. S. and Oliveri, R. and Oohara, K. and O’Reilly, B. and Ormiston, R. G. and Ormsby, N. D. and Orselli, M. and O’Shaughnessy, R. and O’Shea, E. and Oshima, Y. and Oshino, S. and Ossokine, S. and Osthelder, C. and Ottaway, D. J. and Overmier, H. and Pace, A. E. and Pagano, R. and Page, M. A. and Pai, A. and Pai, S. A. and Pal, S. and Palashov, O. and Pálfi, M. and Palomba, C. and Pan, K. C. and Panda, P. K. and Pang, P. T. H. and Pannarale, F. and Pant, B. C. and Panther, F. H. and Paoletti, F. and Paoli, A. and Paolone, A. and Papalexakis, E. E. and Pappas, G. and Parisi, A. and Park, J. and Parker, W. and Pascucci, D. and Pasqualetti, A. and Passaquieti, R. and Passuello, D. and Patel, M. and Pathak, M. and Patra, A. and Patricelli, B. and Patron, A. S. and Paul, S. and Payne, E. and Pearce, T. and Pedraza, M. and Pedurand, R. and Pegna, R. and Pegoraro, M. and Pele, A. and Arellano, F. E. Peña and Penn, S. and Perego, A. and Pereira, A. and Perez, C. J. and Périgois, C. and Perkins, C. C. and Perreca, A. and Perriès, S. and Perry, J. W. and Pesios, D. and Petermann, J. and Petrillo, C. and Pfeiffer, H. P. and Pham, H. and Pham, K. A. and Phukon, K. S. and Phurailatpam, H. and Piccinni, O. J. and Pichot, M. and Piendibene, M. and Piergiovanni, F. and Pierini, L. and Pierra, G. and Pierro, V. and Pillant, G. and Pillas, M. and Pilo, F. and Pinard, L. and Pineda-Bosque, C. and Pinto, I. M. and Piotrzkowski, B. J. and Piotrzkowski, K. and Pirello, M. and Pitkin, M. D. and Placidi, A. and Placidi, E. and Planas, M. L. and Plastino, W. and Poggiani, R. and Polini, E. and Pompili, L. and Pong, D. Y. T. and Ponrathnam, S. and Porcelli, E. and Portell, J. and Porter, E. K. and Posnansky, C. and Poulton, R. and Powell, Jade and Powell, Jonathan and Pracchia, M. and Pradier, T. and Prajapati, A. K. and Prasai, K. and Prasanna, R. and Pratten, G. and Principe, M. and Prodi, G. A. and Prokhorov, L. and Prosposito, P. and Prudenzi, L. and Puecher, A. and Pullin, J. and Punturo, M. and Puosi, F. and Puppo, P. and Pürrer, M. and Qi, H. and Quetschke, V. and Quinonez, P. J. and Quitzow-James, R. and Raab, F. J. and Raaijmakers, G. and Radulesco, N. and Raffai, P. and Rail, S. X. and Raja, S. and Rajan, C. and Ramirez, K. E. and Ramirez, T. D. and Ramos-Buades, A. and Rana, D. and Rana, J. and Randel, E. and Rangnekar, P. R. and Rapagnani, P. and Ray, A. and Raymond, V. and Raza, N. and Razzano, M. and Read, J. and Regimbau, T. and Rei, L. and Reid, S. and Reid, S. W. and Reitze, D. H. and Relton, P. and Renzini, A. and Rettegno, P. and Revenu, B. and Reza, A. and Rezac, M. and Rezaei, A. S. and Ricci, F. and Richards, D. and Richardson, J. W. and Rijal, A. and Riles, K. and Riley, H. K. and Rinaldi, S. and Robertson, C. and Robertson, N. A. and Robinet, F. and Rocchi, A. and Rodriguez, S. and Rolland, L. and Rollins, J. G. and Romanelli, M. and Romano, R. and Romel, C. L. and Romero, A. and Romero-Shaw, I. M. and Romie, J. H. and Ronchini, S. and Roocke, T. J. and Rosa, L. and Rosauer, T. J. and Rose, C. A. and Rosińska, D. and Ross, M. P. and Rossello, M. and Roussel, A. and Rowan, S. and Rowlinson, S. J. and Roy, S. and Royzman, A. and Rozza, D. and Ruggi, P. and Ruiz Morales, E. and Ruiz-Rocha, K. and Ryan, K. and Sachdev, S. and Sadecki, T. and Sadiq, J. and Saffarieh, P. and Saha, S. S. and Saha, S. and Saito, Y. and Sakai, K. and Sakellariadou, M. and Sako, T. and Sakon, S. and Salafia, O. S. and Salces-Carcoba, F. and Salconi, L. and Saleem, M. and Salemi, F. and Sallé, M. and Samajdar, A. and Sanchez, E. J. and Sanchez, J. H. and Sanchez, L. E. and Sanchis-Gual, N. and Sanders, J. R. and Sanuy, A. and Saravanan, T. R. and Sarin, N. and Sasli, A. and Sassi, P. and Sassolas, B. and Satari, H. and Sauter, O. and Savage, R. L. and Savant, V. and Sawada, T. and Sawant, H. L. and Sayah, S. and Schaetzl, D. and Scheel, M. and Scherf, S. J. and Scheuer, J. and Schiworski, M. G. and Schmidt, P. and Schmidt, S. and Schmitz, S. J. and Schnabel, R. and Schneewind, M. and Schofield, R. M. S. and Schönbeck, A. and Schuler, H. and Schulte, B. W. and Schutz, B. F. and Schwartz, E. and Scott, J. and Scott, S. M. and Seetharamu, T. C. and Seglar-Arroyo, M. and Sekiguchi, Y. and Sellers, D. and Sengupta, A. S. and Sentenac, D. and Seo, E. G. and Sequino, V. and Sergeev, A. and Servignat, G. and Setyawati, Y. and Shaffer, T. and Shahriar, M. S. and Shaikh, M. A. and Shams, B. and Shao, L. and Sharma, P. and Chaudhary, S. Sharma and Shawhan, P. and Shcheblanov, N. S. and Sheela, A. and Shen, B. and Shepard, K. G. and Sheridan, E. and Shikano, Y. and Shikauchi, M. and Shimizu, H. and Shimode, K. and Shinkai, H. and Shoemaker, D. H. and Shoemaker, D. M. and ShyamSundar, S. and Sider, A. and Siegel, H. and Sieniawska, M. and Sigg, D. and Silenzi, L. and Singer, L. P. and Singh, D. and Singh, M. K. and Singh, N. and Singha, A. and Sintes, A. M. and Sipala, V. and Skliris, V. and Slagmolen, B. J. J. and Slaven-Blair, T. J. and Smetana, J. and Smith, J. R. and Smith, L. and Smith, R. J. E. and Soldateschi, J. and Somala, S. N. and Somiya, K. and Soni, K. and Soni, S. and Sordini, V. and Sorrentino, F. and Sorrentino, N. and Sotani, H. and Soulard, R. and Souradeep, T. and Sowell, E. and Spagnuolo, V. and Spencer, A. P. and Spera, M. and Spinicelli, P. and Srivastava, A. K. and Srivastava, V. and Stachie, C. and Stachurski, F. and Steer, D. A. and Steinlechner, J. and Steinlechner, S. and Stergioulas, N. and StPierre, M. and Strang, L. C. and Stratta, G. and Strong, M. D. and Strunk, A. and Sturani, R. and Stuver, A. L. and Suchenek, M. and Sudhagar, S. and Sueltmann, N. and Sugiyama, T. and Suh, H. G. and Sullivan, A. G. and Summerscales, T. Z. and Sun, L. and Sunil, S. and Sur, A. and Suresh, J. and Sutton, P. J. and Suzuki, Takamasa and Suzuki, Takanori and Swinkels, B. L. and Syx, A. and Szczepańczyk, M. J. and Szewczyk, P. and Tacca, M. and Tagoshi, H. and Tait, S. C. and Takahashi, H. and Takahashi, R. and Takamori, A. and Takano, S. and Takeda, H. and Takeda, M. and Talbot, C. J. and Talbot, C. and Tamaki, M. and Tamanini, N. and Tanabe, D. and Tanaka, K. and Tanaka, T. and Tanasijczuk, A. J. and Tanioka, S. and Tanner, D. B. and Tao, D. and Tao, L. and Tapia, R. D. and San Martín, E. N. Tapia and Tarafder, R. and Taranto, C. and Taruya, A. and Tasson, J. D. and Teloi, M. and Tenorio, R. and Terhune, J. E. S. and Terkowski, L. and Themann, H. and Thirugnanasambandam, M. P. and Thomas, L. M. and Thomas, M. and Thomas, P. and Thomas, S. and Thompson, J. E. and Thondapu, S. R. and Thorne, K. A. and Thrane, E. and Tiwari, Shubhanshu and Tiwari, Srishti and Tiwari, V. and Toivonen, A. M. and Tolley, A. E. and Tomaru, T. and Tomita, K. and Tomura, T. and Tonelli, M. and Torres-Forné, A. and Torrie, C. I. and e Melo, I. Tosta and Tournefier, E. and Trapananti, A. and Travasso, F. and Traylor, G. and Trenado, J. and Trevor, M. and Tringali, M. C. and Tripathee, A. and Troiano, L. and Trovato, A. and Trozzo, L. and Trudeau, R. J. and Tsang, K. W. and Tsang, T. and Tse, M. and Tso, R. and Tsuchida, S. and Tsukada, L. and Tsutsui, T. and Turbang, K. and Turconi, M. and Turski, C. and Tuyenbayev, D. and Ubach, H. and Ubhi, A. S. and Uchikata, N. and Uchiyama, T. and Udall, R. P. and Uehara, T. and Ueno, K. and Unnikrishnan, C. S. and Ushiba, T. and Utina, A. and Vahlbruch, H. and Vaidya, N. and Vajente, G. and Vajpeyi, A. and Valdes, G. and Valentini, M. and Vallero, S. and Valsan, V. and van Bakel, N. and van Beuzekom, M. and van Dael, M. and van den Brand, J. F. J. and Van Den Broeck, C. and Vander-Hyde, D. C. and van der Sluys, M. and Van de Walle, A. and van Dongen, J. and van Haevermaet, H. and van Heijningen, J. V. and Vanosky, J. and Putten, M. H. P. M. van and van Ranst, Z. and van Remortel, N. and Vardaro, M. and Vargas, A. F. and Varma, V. and Vasúth, M. and Vecchio, A. and Vedovato, G. and Veitch, J. and Veitch, P. J. and Venneberg, J. and Venugopalan, G. and Verdier, P. and Verkindt, D. and Verma, P. and Verma, Y. and Vermeulen, S. M. and Veske, D. and Vetrano, F. and Viceré, A. and Vidyant, S. and Viets, A. D. and Vijaykumar, A. and Villa-Ortega, V. and Vina, M. and Vincent, E. T. and Vinet, J.-Y. and Viret, S. and Virtuoso, A. and Vitale, S. and Vocca, H. and Voigt, D. and von Reis, E. R. G. and von Wrangel, J. S. A. and Vorvick, C. and Vyatchanin, S. P. and Wade, L. E. and Wade, M. and Wagner, K. J. and Walet, R. C. and Walker, M. and Wallace, G. S. and Wallace, L. and Wang, H. and Wang, J. Z. and Wang, W. H. and Ward, R. L. and Warner, J. and Was, M. and Washimi, T. and Washington, N. Y. and Watada, K. and Watarai, D. and Watchi, J. and Wayt, K. E. and Weaver, B. and Weaving, C. R. and Webster, S. A. and Weinert, M. and Weinstein, A. J. and Weiss, R. and Weller, C. M. and Weller, R. A. and Wellmann, F. and Wen, L. and Weßels, P. and Wette, K. and Whelan, J. T. and White, D. D. and Whiting, B. F. and Whittle, C. and Wilk, O. S. and Wilken, D. and Willetts, K. and Williams, D. and Williams, M. J. and Williamson, A. R. and Willis, J. L. and Willke, B. and Wipf, C. C. and Woan, G. and Woehler, J. and Wofford, J. K. and Wong, D. and Wong, H. T. and Wong, I. C. F. and Wright, M. and Wu, C. and Wu, D. S. and Wu, H. and Wysocki, D. M. and Xiao, L. and Xu, V. A. and Yadav, N. and Yamada, T. and Yamamoto, H. and Yamamoto, K. and Yamamoto, M. and Yamamoto, T. and Yamamoto, T. S. and Yamashita, K. and Yamazaki, R. and Yang, F. W. and Yang, K. Z. and Yang, Y.-C. and Yap, M. J. and Yeeles, D. W. and Yelikar, A. B. and Yeung, T. Y. and Yokoyama, J. and Yokozawa, T. and Yoo, J. and Yu, Hang and Yu, Haocun and Yuzurihara, H. and Zadrożny, A. and Zannelli, A. J. and Zanolin, M. and Zeeshan, M. and Zeidler, S. and Zelenova, T. and Zendri, J.-P. and Zevin, M. and Zhang, J. and Zhang, L. and Zhang, R. and Zhang, T. and Zhang, Y. and Zhao, C. and Zhao, Yue and Zhao, Yuhang and Zheng, Y. and Zhong, H. and Zhou, R. and Zhu, X. J. and Zhu, Z.-H. and Zimmerman, A. B. and Zucker, M. E. and Zweizig, J. and (The LIGO Scientific Collaboration, the Virgo Collaboration, and the KAGRA Collaboration)},
title = {Open Data from the Third Observing Run of LIGO, Virgo, KAGRA, and GEO},
journal = {The Astrophysical Journal Supplement Series}
}

@article{Casallas2023,
  title = {Characterizing the temporal evolution of the high-frequency gravitational wave emission for a core collapse supernova with laser interferometric data: A neural network approach},
  author = {Casallas-Lagos, Alejandro and Antelis, Javier M. and Moreno, Claudia and Zanolin, Michele and Mezzacappa, Anthony and Szczepa\ifmmode \acute{n}\else \'{n}\fi{}czyk, Marek J.},
  journal = {Phys. Rev. D},
  volume = {108},
  issue = {8},
  pages = {084027},
  numpages = {17},
  year = {2023},
  month = {Oct},
  publisher = {American Physical Society},
  doi = {10.1103/PhysRevD.108.084027},
  url = {https://link.aps.org/doi/10.1103/PhysRevD.108.084027}
}

@Article{Casallas2025,
AUTHOR = {Casallas-Lagos, Alejandro and Antelis, Javier M. and Moreno, Claudia and Franco-Hernández, Ramiro},
TITLE = {Estimation of the High-Frequency Feature Slope in Gravitational Wave Signals from Core Collapse Supernovae Using Machine Learning},
JOURNAL = {Applied Sciences},
VOLUME = {15},
YEAR = {2025},
NUMBER = {1},
ARTICLE-NUMBER = {65},
URL = {https://www.mdpi.com/2076-3417/15/1/65},
ISSN = {2076-3417},
DOI = {10.3390/app15010065}
}

@article{Tamborra_2013,
  title = {Neutrino Signature of Supernova Hydrodynamical Instabilities in Three Dimensions},
  author = {Tamborra, Irene and Hanke, Florian and M\"uller, Bernhard and Janka, Hans-Thomas and Raffelt, Georg},
  journal = {Phys. Rev. Lett.},
  volume = {111},
  issue = {12},
  pages = {121104},
  numpages = {5},
  year = {2013},
  month = {Sep},
  publisher = {American Physical Society},
  doi = {10.1103/PhysRevLett.111.121104},
  url = {https://link.aps.org/doi/10.1103/PhysRevLett.111.121104}
}

@article{Lund_2010,
  title = {Fast time variations of supernova neutrino fluxes and their detectability},
  author = {Lund, Tina and Marek, Andreas and Lunardini, Cecilia and Janka, Hans-Thomas and Raffelt, Georg},
  journal = {Phys. Rev. D},
  volume = {82},
  issue = {6},
  pages = {063007},
  numpages = {13},
  year = {2010},
  month = {Sep},
  publisher = {American Physical Society},
  doi = {10.1103/PhysRevD.82.063007},
  url = {https://link.aps.org/doi/10.1103/PhysRevD.82.063007}
}

@article{Blondin_2003,
doi = {10.1086/345812},
url = {https://doi.org/10.1086/345812},
year = {2003},
month = {feb},
publisher = {},
volume = {584},
number = {2},
pages = {971},
author = {Blondin, John M. and Mezzacappa, Anthony and DeMarino, Christine},
title = {Stability of Standing Accretion Shocks, with an Eye toward Core-Collapse Supernovae},
journal = {The Astrophysical Journal}
}

@article{AllSky_O4a,
  title = {All-sky search for short gravitational-wave bursts in the first part of the fourth LIGO-Virgo-KAGRA observing run},
  author = {Abac, A. G. and Abouelfettouh, I. and Acernese, F. and Ackley, K. and Adamcewicz, C. and Adhicary, S. and Adhikari, D. and Adhikari, N. and Adhikari, R. X. and Adkins, V. K. and Afroz, S. and Agapito, A. and Agarwal, D. and Agathos, M. and Aggarwal, N. and Aggarwal, S. and Aguiar, O. D. and Ahrend, I.-L. and Aiello, L. and Ain, A. and Ajith, P. and Akutsu, T. and Albanesi, S. and Ali, W. and Al-Kershi, S. and All\'en\'e, C. and Allocca, A. and Al-Shammari, S. and Altin, P. A. and Alvarez-Lopez, S. and Amar, W. and Amarasinghe, O. and Amato, A. and Amicucci, F. and Amra, C. and Ananyeva, A. and Anderson, S. B. and Anderson, W. G. and Andia, M. and Ando, M. and Andr\'es-Carcasona, M. and Andri\ifmmode \acute{c}\else \'{c}\fi{}, T. and Anglin, J. and Ansoldi, S. and Antelis, J. M. and Antier, S. and Aoumi, M. and Appavuravther, E. Z. and Appert, S. and Apple, S. K. and Arai, K. and Araya, A. and Araya, M. C. and Sedda, M. Arca and Areeda, J. S. and Aritomi, N. and Armato, F. and Armstrong, S. and Arnaud, N. and Arogeti, M. and Aronson, S. M. and Ashton, G. and Aso, Y. and Asprea, L. and Assiduo, M. and Melo, S. Assis de Souza and Aston, S. M. and Astone, P. and Attadio, F. and Aubin, F. and AultONeal, K. and Avallone, G. and Avila, E. A. and Babak, S. and Badger, C. and Bae, S. and Bagnasco, S. and Baiotti, L. and Bajpai, R. and Baka, T. and Baker, A. M. and Baker, K. A. and Baldi, G. and Baldicchi, N. and Ball, M. and Ballardin, G. and Ballmer, S. W. and Banagiri, S. and Banerjee, B. and Bankar, D. and Baptiste, T. M. and Baral, P. and Baratti, M. and Barayoga, J. C. and Barish, B. C. and Barker, D. and Barman, N. and Barneo, P. and Barone, F. and Barr, B. and Barsotti, L. and Barsuglia, M. and Barta, D. and Bartoletti, A. M. and Barton, M. A. and Bartos, I. and Basalaev, A. and Bassiri, R. and Basti, A. and Bawaj, M. and Baxi, P. and Bayley, J. C. and Baylor, A. C. and Baynard, P. A. and Bazzan, M. and Bedakihale, V. M. and Beirnaert, F. and Bejger, M. and Belardinelli, D. and Bell, A. S. and Bellie, D. S. and Bellizzi, L. and Benoit, W. and Bentara, I. and Bentley, J. D. and Yaala, M. Ben and Bera, S. and Bergamin, F. and Berger, B. K. and Bernuzzi, S. and Beroiz, M. and Bersanetti, D. and Bertheas, T. and Bertolini, A. and Betzwieser, J. and Beveridge, D. and Bevilacqua, G. and Bevins, N. and Bhandare, R. and Bhatt, R. and Bhattacharjee, D. and Bhattacharyya, S. and Bhaumik, S. and Biancalana, V. and Bianchi, A. and Bilenko, I. A. and Billingsley, G. and Binetti, A. and Bini, S. and Binu, C. and Biot, S. and Birnholtz, O. and Biscoveanu, S. and Bisht, A. and Bitossi, M. and Bizouard, M.-A. and Blaber, S. and Blackburn, J. K. and Blagg, L. A. and Blair, C. D. and Blair, D. G. and Bode, N. and Boettner, N. and Boileau, G. and Boldrini, M. and Bolingbroke, G. N. and Bolliand, A. and Bonavena, L. D. and Bondarescu, R. and Bondu, F. and Bonilla, E. and Bonilla, M. S. and Bonino, A. and Bonnand, R. and Borchers, A. and Borhanian, S. and Boschi, V. and Bose, S. and Bossilkov, V. and Bothra, Y. and Boudon, A. and Bourg, L. and Boyle, M. and Bozzi, A. and Bradaschia, C. and Brady, P. R. and Branch, A. and Branchesi, M. and Braun, I. and Briant, T. and Brillet, A. and Brinkmann, M. and Brockill, P. and Brockmueller, E. and Brooks, A. F. and Brown, B. C. and Brown, D. D. and Brozzetti, M. L. and Brunett, S. and Bruno, G. and Bruntz, R. and Bryant, J. and Bu, Y. and Bucci, F. and Buchanan, J. and Bulashenko, O. and Bulik, T. and Bulten, H. J. and Buonanno, A. and Burtnyk, K. and Buscicchio, R. and Buskulic, D. and Buy, C. and Byer, R. L. and Davies, G. S. Cabourn and Cabrita, R. and C\'aceres-Barbosa, V. and Cadonati, L. and Cagnoli, G. and Cahillane, C. and Calafat, A. and Callister, T. A. and Calloni, E. and Callos, S. R. and Santoro, G. Caneva and Cannon, K. C. and Cao, H. and Capistran, L. A. and Capocasa, E. and Capote, E. and Capurri, G. and Carapella, G. and Carbognani, F. and Carlassara, M. and Carlin, J. B. and Carlson, T. K. and Carney, M. F. and Carpinelli, M. and Carrillo, G. and Carter, J. J. and Carullo, G. and Casallas-Lagos, A. and Diaz, J. Casanueva and Casentini, C. and Castro-Lucas, S. Y. and Caudill, S. and Cavagli\`a, M. and Cavalieri, R. and Ceja, A. and Cella, G. and Cerd\'a-Dur\'an, P. and Cesarini, E. and Chabbra, N. and Chaibi, W. and Chakraborty, A. and Chakraborty, P. and Chakraborty, S. and Subrahmanya, S. Chalathadka and Chan, J. C. L. and Chan, M. and Chang, K. and Chao, S. and Charlton, P. and Chassande-Mottin, E. and Chatterjee, C. and Chatterjee, Debarati and Chatterjee, Deep and Chaturvedi, M. and Chaty, S. and Chatziioannou, K. and Chen, A. and Chen, A. H.-Y. and Chen, D. and Chen, H. and Chen, H. Y. and Chen, S. and Chen, Yanbei and Chen, Yitian and Cheng, H. P. and Chessa, P. and Cheung, H. T. and Cheung, S. Y. and Chiadini, F. and Chiarini, G. and Chiba, A. and Chincarini, A. and Chiofalo, M. L. and Chiummo, A. and Chou, C. and Choudhary, S. and Christensen, N. and Chua, S. S. Y. and Ciani, G. and Ciecielag, P. and Cie\ifmmode \acute{s}\else \'{s}\fi{}lar, M. and Cifaldi, M. and Cirok, B. and Clara, F. and Clark, J. A. and Clarke, T. A. and Clearwater, P. and Clesse, S. and Cleva, F. and Coccia, E. and Codazzo, E. and Cohadon, P.-F. and Colace, S. and Colangeli, E. and Colleoni, M. and Collette, C. G. and Collins, J. and Colloms, S. and Colombo, A. and Compton, C. M. and Connolly, G. and Conti, L. and Corbitt, T. R. and Cordero-Carri\'on, I. and Corezzi, S. and Cornish, N. J. and Coronado, I. and Corsi, A. and Cottingham, R. and Coughlin, M. W. and Couineaux, A. and Couvares, P. and Coward, D. M. and Coyne, R. and Cozzumbo, A. and Creighton, J. D. E. and Creighton, T. D. and Cremonese, P. and Crook, S. and Crouch, R. and Csizmazia, J. and Cudell, J. R. and Cullen, T. J. and Cumming, A. and Cuoco, E. and Cusinato, M. and Da Concei\ifmmode \mbox{\c{c}}\else \c{c}\fi{}\~ao, L. V. and Canton, T. Dal and Pra, S. Dal and D\'alya, G. and D'Angelo, B. and Danilishin, S. and D'Antonio, S. and Danzmann, K. and Darroch, K. E. and Dartez, L. P. and Das, R. and Dasgupta, A. and Dattilo, V. and Daumas, A. and Davari, N. and Dave, I. and Davenport, A. and Davier, M. and Davies, T. F. and Davis, D. and Davis, L. and Davis, M. C. and Davis, P. and Daw, E. J. and Dax, M. and De Bolle, J. and Deenadayalan, M. and Degallaix, J. and De Laurentis, M. and De Lillo, F. and Della Torre, S. and Del Pozzo, W. and Demagny, A. and De Marco, F. and Demasi, G. and De Matteis, F. and Demos, N. and Dent, T. and Depasse, A. and DePergola, N. and De Pietri, R. and De Rosa, R. and De Rossi, C. and Desai, M. and DeSalvo, R. and DeSimone, A. and De Simone, R. and Dhani, A. and Diab, R. and D\'{\i}az, M. C. and Di Cesare, M. and Dideron, G. and Dietrich, T. and Di Fiore, L. and Di Fronzo, C. and Di Giovanni, M. and Di Girolamo, T. and Diksha, D. and Ding, J. and Di Pace, S. and Di Palma, I. and Di Piero, D. and Di Renzo, F. and Divyajyoti and Dmitriev, A. and Docherty, J. P. and Doctor, Z. and Doerksen, N. and Dohmen, E. and Doke, A. and De Souza, A. Domiciano and D'Onofrio, L. and Donovan, F. and Dooley, K. L. and Dooney, T. and Doravari, S. and Dorosh, O. and Doyle, W. J. D. and Drago, M. and Driggers, J. C. and Dunn, L. and Dupletsa, U. and Duverne, P.-A. and D'Urso, D. and Roy, P. Dutta and Duval, H. and Dwyer, S. E. and Eassa, C. and Ebersold, M. and Eckhardt, T. and Eddolls, G. and Effler, A. and Eichholz, J. and Einsle, H. and Eisenmann, M. and Emma, M. and Endo, K. and Enficiaud, R. and Errico, L. and Espinosa, R. and Esposito, M. and Essick, R. C. and Estell\'es, H. and Etzel, T. and Evans, M. and Evstafyeva, T. and Ewing, B. E. and Ezquiaga, J. M. and Fabrizi, F. and Fafone, V. and Fairhurst, S. and Farah, A. M. and Farr, B. and Farr, W. M. and Favaro, G. and Favata, M. and Fays, M. and Fazio, M. and Feicht, J. and Fejer, M. M. and Felicetti, R. and Fenyvesi, E. and Fernandes, J. and Fernandes, T. and Fernando, D. and Ferraiuolo, S. and Ferreira, T. A. and Fidecaro, F. and Figura, P. and Fiori, A. and Fiori, I. and Fishbach, M. and Fisher, R. P. and Fittipaldi, R. and Fiumara, V. and Flaminio, R. and Fleischer, S. M. and Fleming, L. S. and Floden, E. and Fong, H. and Font, J. A. and Fontinele-Nunes, F. and Foo, C. and Fornal, B. and Franceschetti, K. and Frappez, F. and Frasca, S. and Frasconi, F. and Freed, J. P. and Frei, Z. and Freise, A. and Freitas, O. and Frey, R. and Frischhertz, W. and Fritschel, P. and Frolov, V. V. and Fronz\'e, G. G. and Fuentes-Garcia, M. and Fujii, S. and Fujimori, T. and Fulda, P. and Fyffe, M. and Gadre, B. and Gair, J. R. and Galaudage, S. and Galdi, V. and Gamba, R. and Gamboa, A. and Gamoji, S. and Ganapathy, D. and Ganguly, A. and Garaventa, B. and Garc\'{\i}a-Bellido, J. and Garc\'{\i}a-Quir\'os, C. and Gardner, J. W. and Gardner, K. A. and Garg, S. and Gargiulo, J. and Garrido, X. and Garron, A. and Garufi, F. and Garver, P. A. and Gasbarra, C. and Gateley, B. and Gautier, F. and Gayathri, V. and Gayer, T. and Gemme, G. and Gennai, A. and Gennari, V. and George, J. and George, R. and Gerberding, O. and Gergely, L. and Ghosh, Archisman and Ghosh, Sayantan and Ghosh, Shaon and Ghosh, Shrobana and Ghosh, Suprovo and Ghosh, Tathagata and Giaime, J. A. and Giardina, K. D. and Gibson, D. R. and Gier, C. and Gkaitatzis, S. and Glanzer, J. and Glotin, F. and Godfrey, J. and Godley, R. V. and Godwin, P. and Goettel, A. S. and Goetz, E. and Golomb, J. and Lopez, S. Gomez and Goncharov, B. and Gonz\'alez, G. and Goodarzi, P. and Goode, S. and Goodwin-Jones, A. W. and Gosselin, M. and Gouaty, R. and Gould, D. W. and Govorkova, K. and Grado, A. and Graham, V. and Granados, A. E. and Granata, M. and Granata, V. and Gras, S. and Grassia, P. and Graves, J. and Gray, C. and Gray, R. and Greco, G. and Green, A. C. and Green, L. and Green, S. M. and Green, S. R. and Greenberg, C. and Gretarsson, A. M. and Griffin, H. K. and Griffith, D. and Griggs, H. L. and Grignani, G. and Grimaud, C. and Grote, H. and Grunewald, S. and Guerra, D. and Guetta, D. and Guidi, G. M. and Guimaraes, A. R. and Gulati, H. K. and Gulminelli, F. and Guo, H. and Guo, W. and Guo, Y. and Gupta, Anuradha and Gupta, I. and Gupta, N. C. and Gupta, S. K. and Gupta, V. and Gupte, N. and Gurs, J. and Gutierrez, N. and Guttman, N. and Guzman, F. and Haba, D. and Haberland, M. and Haino, S. and Hall, E. D. and Hamilton, E. Z. and Hammond, G. and Haney, M. and Hanks, J. and Hanna, C. and Hannam, M. D. and Hannuksela, O. A. and Hanselman, A. G. and Hansen, H. and Hanson, J. and Hanumasagar, S. and Harada, R. and Hardison, A. R. and Harikumar, S. and Haris, K. and Harley-Trochimczyk, I. and Harmark, T. and Harms, J. and Harry, G. M. and Harry, I. W. and Hart, J. and Haskell, B. and Haster, C. J. and Haughian, K. and Hayakawa, H. and Hayama, K. and Heintze, M. C. and Heinze, J. and Heinzel, J. and Heitmann, H. and Hellman, F. and Helmling-Cornell, A. F. and Hemming, G. and Henderson-Sapir, O. and Hendry, M. and Heng, I. S. and Hennig, M. H. and Henshaw, C. and Heurs, M. and Hewitt, A. L. and Heynen, J. and Heyns, J. and Higginbotham, S. and Hild, S. and Hill, S. and Himemoto, Y. and Hirata, N. and Hirose, C. and Hofman, D. and Hogan, B. E. and Holland, N. A. and Hollows, I. J. and Holz, D. E. and Honet, L. and Horton-Bailey, D. J. and Hough, J. and Hourihane, S. and Howard, N. T. and Howell, E. J. and Hoy, C. G. and Hrishikesh, C. A. and Hsi, P. and Hsieh, H.-F. and Hsieh, H.-Y. and Hsiung, C. and Hsu, S.-H. and Hsu, W.-F. and Hu, Q. and Huang, H. Y. and Huang, Y. and Huang, Y. T. and Huddart, A. D. and Hughey, B. and Hui, V. and Husa, S. and Huxford, R. and Iampieri, L. and Iandolo, G. A. and Ianni, M. and Iannone, G. and Iascau, J. and Ide, K. and Iden, R. and Ierardi, A. and Ikeda, S. and Imafuku, H. and Inoue, Y. and Iorio, G. and Iosif, P. and Iqbal, M. H. and Irwin, J. and Ishikawa, R. and Isi, M. and Isleif, K. S. and Itoh, Y. and Iwaya, M. and Iyer, B. R. and Jacquet, C. and Jacquet, P.-E. and Jacquot, T. and Jadhav, S. J. and Jadhav, S. P. and Jain, M. and Jain, T. and James, A. L. and Jani, K. and Janquart, J. and Janthalur, N. N. and Jaraba, S. and Jaranowski, P. and Jaume, R. and Javed, W. and Jennings, A. and Jensen, M. and Jia, W. and Jiang, J. and Jin, H.-B. and Johns, G. R. and Johnson, N. A. and Johnston, M. C. and Johnston, R. and Johny, N. and Jones, D. H. and Jones, D. I. and Jones, R. and Jose, H. E. and Joshi, P. and Joshi, S. K. and Joubert, G. and Ju, J. and Ju, L. and Jung, K. and Junker, J. and Juste, V. and Kabagoz, H. B. and Kajita, T. and Kaku, I. and Kalogera, V. and Kalomenopoulos, M. and Kamiizumi, M. and Kanda, N. and Kandhasamy, S. and Kang, G. and Kannachel, N. C. and Kanner, J. B. and KantiMahanty, S. A. and Kapadia, S. J. and Kapasi, D. P. and Karthikeyan, M. and Kasprzack, M. and Kato, H. and Kato, T. and Katsavounidis, E. and Katzman, W. and Kaushik, R. and Kawabe, K. and Kawamoto, R. and Keitel, D. and Kemperman, L. J. and Kennington, J. and Kerkow, F. A. and Kesharwani, R. and Key, J. S. and Khadela, R. and Khadka, S. and Khadkikar, S. S. and Khalili, F. Y. and Khan, F. and Khanam, T. and Khursheed, M. and Khusid, N. M. and Kiendrebeogo, W. and Kijbunchoo, N. and Kim, C. and Kim, J. C. and Kim, K. and Kim, M. H. and Kim, S. and Kim, Y.-M. and Kimball, C. and Kimes, K. and Kinnear, M. and Kissel, J. S. and Klimenko, S. and Knee, A. M. and Knox, E. J. and Knust, N. and Kobayashi, K. and Koehlenbeck, S. M. and Koekoek, G. and Kohri, K. and Kokeyama, K. and Koley, S. and Kolitsidou, P. and Koloniari, A. E. and Komori, K. and Kong, A. K. H. and Kontos, A. and Koponen, L. M. and Korobko, M. and Kou, X. and Koushik, A. and Kouvatsos, N. and Kovalam, M. and Koyama, T. and Kozak, D. B. and Kranzhoff, S. L. and Kringel, V. and Krishnendu, N. V. and Kroker, S. and Kr\'olak, A. and Kruska, K. and Kubisz, J. and Kuehn, G. and Kulkarni, S. and Ramamohan, A. Kulur and Kumar, Achal and Kumar, Anil and Kumar, Praveen and Kumar, Prayush and Kumar, Rahul and Kumar, Rakesh and Kume, J. and Kuns, K. and Kuntimaddi, N. and Kuroyanagi, S. and Kuwahara, S. and Kwak, K. and Kwan, K. and Kwon, S. and Lacaille, G. and Laghi, D. and Laity, A. H. and Lalande, E. and Lalleman, M. and Lalremruati, P. C. and Landry, M. and Lane, B. B. and Lang, R. N. and Lange, J. and Langgin, R. and Lantz, B. and La Rosa, I. and Larsen, J. and Lartaux-Vollard, A. and Lasky, P. D. and Lawrence, J. and Laxen, M. and Lazarte, C. and Lazzarini, A. and Lazzaro, C. and Leaci, P. and Leali, L. and Lecoeuche, Y. K. and Lee, H. M. and Lee, H. W. and Lee, J. and Lee, K. and Lee, R.-K. and Lee, R. and Lee, Sungho and Lee, Sunjae and Lee, Y. and Legred, I. N. and Lehmann, J. and Lehner, L. and Le Jean, M. and Lema\^{\i}tre, A. and Lenti, M. and Leonardi, M. and Lequime, M. and Leroy, N. and Lesovsky, M. and Letendre, N. and Lethuillier, M. and Levin, Y. and Leyde, K. and Li, A. K. Y. and Li, K. L. and Li, T. G. F. and Li, X. and Li, Y. and Li, Z. and Lihos, A. and Lin, E. T. and Lin, F. and Lin, L. C.-C. and Lin, Y.-C. and Lindsay, C. and Linker, S. D. and Liu, A. and Liu, G. C. and Liu, Jian and Villarreal, F. Llamas and Llobera-Querol, J. and Lo, R. K. L. and Locquet, J.-P. and Loggins, S. C. G. and Loizou, M. R. and London, L. T. and Longo, A. and Lopez, D. and Portilla, M. Lopez and Lorenzini, M. and Lorenzo-Medina, A. and Loriette, V. and Lormand, M. and Losurdo, G. and Lotti, E. and Lott, T. P. and Lough, J. D. and Loughlin, H. A. and Lousto, C. O. and Low, N. and Lu, N. and Lucchesi, L. and L\"uck, H. and Lumaca, D. and Lundgren, A. P. and Lussier, A. W. and Macas, R. and MacInnis, M. and Macleod, D. M. and MacMillan, I. A. O. and Macquet, A. and Maeda, K. and Maenaut, S. and Magare, S. S. and Magee, R. M. and Maggio, E. and Maggiore, R. and Magnozzi, M. and Mahesh, M. and Maini, M. and Majhi, S. and Majorana, E. and Makarem, C. N. and Malakar, D. and Malaquias-Reis, J. A. and Mali, U. and Maliakal, S. and Malik, A. and Mallick, L. and Malz, A.-K. and Man, N. and Mancarella, M. and Mandic, V. and Mangano, V. and Mannix, B. and Mansell, G. L. and Manske, M. and Mantovani, M. and Mapelli, M. and Marinelli, C. and Marion, F. and Markosyan, A. S. and Markowitz, A. and Maros, E. and Marsat, S. and Martelli, F. and Martin, I. W. and Martin, R. M. and Martinez, B. B. and Martinez, D. A. and Martinez, M. and Martinez, V. and Martini, A. and Martins, J. C. and Martynov, D. V. and Marx, E. J. and Massaro, L. and Masserot, A. and Masso-Reid, M. and Mastrogiovanni, S. and Matcovich, T. and Matiushechkina, M. and Maurin, L. and Mavalvala, N. and Maxwell, N. and McCarrol, G. and McCarthy, R. and McClelland, D. E. and McCormick, S. and McCuller, L. and McEachin, S. and McElhenny, C. and McGhee, G. I. and McGinn, J. and McGowan, K. B. M. and McIver, J. and McLeod, A. and McMahon, I. and McRae, T. and McTeague, R. and Meacher, D. and Meagher, B. N. and Mechum, R. and Meijer, Q. and Melatos, A. and Menoni, C. S. and Mera, F. and Mercer, R. A. and Mereni, L. and Merfeld, K. and Merilh, E. L. and M\'erou, J. R. and Merritt, J. D. and Merzougui, M. and Messick, C. and Mestichelli, B. and Meyer-Conde, M. and Meylahn, F. and Mhaske, A. and Miani, A. and Miao, H. and Michel, C. and Michimura, Y. and Middleton, H. and Mihaylov, D. P. and Miller, S. J. and Millhouse, M. and Milotti, E. and Milotti, V. and Minenkov, Y. and Minihan, E. M. and Mir, Ll. M. and Mirasola, L. and Miravet-Ten\'es, M. and Miritescu, C.-A. and Mishra, A. and Mishra, C. and Mishra, T. and Mitchell, A. L. and Mitchell, J. G. and Mitra, S. and Mitrofanov, V. P. and Mitsuhashi, K. and Mittleman, R. and Miyakawa, O. and Miyoki, S. and Miyoko, A. and Mo, G. and Mobilia, L. and Mohapatra, S. R. P. and Mohite, S. R. and Molina-Ruiz, M. and Mondin, M. and Montani, M. and Moore, C. J. and Moraru, D. and More, A. and More, S. and Moreno, C. and Moreno, E. A. and Moreno, G. and Serra, A. Moreso and Morisaki, S. and Moriwaki, Y. and Morras, G. and Moscatello, A. and Mould, M. and Mours, B. and Mow-Lowry, C. M. and Muccillo, L. and Muciaccia, F. and Mukherjee, D. and Mukherjee, Samanwaya and Mukherjee, Soma and Mukherjee, Subroto and Mukherjee, Suvodip and Mukund, N. and Mullavey, A. and Mullock, H. and Mundi, J. and Mungioli, C. L. and Murakoshi, M. and Murray, P. G. and Nabari, D. and Nadji, S. L. and Nagar, A. and Nagarajan, N. and Nakagaki, K. and Nakamura, K. and Nakano, H. and Nakano, M. and Nanadoumgar-Lacroze, D. and Nandi, D. and Napolano, V. and Narayan, P. and Nardecchia, I. and Narikawa, T. and Narola, H. and Naticchioni, L. and Nayak, R. K. and Negri, L. and Nela, A. and Nelle, C. and Nelson, A. and Nelson, T. J. N. and Nery, M. and Neunzert, A. and Ng, S. and Quynh, L. Nguyen and Nichols, S. A. and Nielsen, A. B. and Nishino, Y. and Nishizawa, A. and Nissanke, S. and Niu, W. and Nocera, F. and Noller, J. and Norman, M. and North, C. and Novak, J. and Nowicki, R. and Siles, J. F. Nu\~no and Nuttall, L. K. and Obayashi, K. and Oberling, J. and O'Dell, J. and Oelker, E. and Oertel, M. and Oganesyan, G. and O'Hanlon, T. and Ohashi, M. and Ohme, F. and Oliveri, R. and Omer, R. and O'Neal, B. and Onishi, M. and Oohara, K. and O'Reilly, B. and Orselli, M. and O'Shaughnessy, R. and O'Shea, S. and Oshino, S. and Osthelder, C. and Ota, I. and Ottaway, D. J. and Ouzriat, A. and Overmier, H. and Owen, B. J. and Ozaki, R. and Pace, A. E. and Pagano, R. and Page, M. A. and Pai, A. and Paiella, L. and Pal, A. and Pal, S. and Palaia, M. A. and P\'alfi, M. and Palma, P. P. and Palomba, C. and Palud, P. and Pan, H. and Pan, J. and Pan, K. C. and Panda, P. K. and Pandey, Shiksha and Pandey, Swadha and Pang, P. T. H. and Pannarale, F. and Pannone, K. A. and Pant, B. C. and Panther, F. H. and Panzeri, M. and Paoletti, F. and Paolone, A. and Papadopoulos, A. and Papalexakis, E. E. and Papalini, L. and Papigkiotis, G. and Paquis, A. and Parisi, A. and Park, B.-J. and Park, J. and Parker, W. and Pascale, G. and Pascucci, D. and Pasqualetti, A. and Passaquieti, R. and Passenger, L. and Passuello, D. and Patane, O. and Patel, A. V. and Pathak, D. and Patra, A. and Patricelli, B. and Patterson, B. G. and Paul, K. and Paul, S. and Payne, E. and Pearce, T. and Pedraza, M. and Pele, A. and Arellano, F. E. Pe\~na and Peng, X. and Peng, Y. and Penn, S. and Penuliar, M. D. and Perego, A. and Pereira, Z. and P\'erigois, C. and Perna, G. and Perreca, A. and Perret, J. and Perri\`es, S. and Perry, J. W. and Pesios, D. and Peters, S. and Petracca, S. and Petrillo, C. and Pfeiffer, H. P. and Pham, H. and Pham, K. A. and Phukon, K. S. and Phurailatpam, H. and Piarulli, M. and Piccari, L. and Piccinni, O. J. and Pichot, M. and Piendibene, M. and Piergiovanni, F. and Pierini, L. and Pierra, G. and Pierro, V. and Pietrzak, M. and Pillas, M. and Pilo, F. and Pinard, L. and Pinto, I. M. and Pinto, M. and Piotrzkowski, B. J. and Pirello, M. and Pitkin, M. D. and Placidi, A. and Placidi, E. and Planas, M. L. and Plastino, W. and Plunkett, C. and Poggiani, R. and Polini, E. and Pomper, J. and Pompili, L. and Poon, J. and Porcelli, E. and Porter, E. K. and Posnansky, C. and Poulton, R. and Powell, J. and Prabhu, G. S. and Pracchia, M. and Pradhan, B. K. and Pradier, T. and Prajapati, A. K. and Prasai, K. and Prasanna, R. and Prasia, P. and Pratten, G. and Principe, G. and Prodi, G. A. and Prosperi, P. and Prosposito, P. and Providence, A. C. and Puecher, A. and Pullin, J. and Puppo, P. and P\"urrer, M. and Qi, H. and Qin, J. and Qu\'em\'ener, G. and Quetschke, V. and Quinonez, P. J. and Qutob, N. and Rading, R. and Rainho, I. and Raja, S. and Rajan, C. and Rajbhandari, B. and Ramirez, K. E. and Vidal, F. A. Ramis and Arevalo, M. Ramos and Ramos-Buades, A. and Ranjan, S. and Ransom, K. and Rapagnani, P. and Ratto, B. and Ravichandran, A. and Ray, A. and Raymond, V. and Razzano, M. and Read, J. and Regimbau, T. and Reid, S. and Reissel, C. and Reitze, D. H. and Renzini, A. I. and Revenu, B. and Pe\~na, A. Revilla and Reyes, R. and Ricca, L. and Ricci, F. and Ricci, M. and Ricciardone, A. and Rice, J. and Richardson, J. W. and Richardson, M. L. and Rijal, A. and Riles, K. and Riley, H. K. and Rinaldi, S. and Rittmeyer, J. and Robertson, C. and Robinet, F. and Robinson, M. and Rocchi, A. and Rolland, L. and Rollins, J. G. and Romano, A. E. and Romano, R. and Romero, A. and Romero-Shaw, I. M. and Romie, J. H. and Ronchini, S. and Roocke, T. J. and Rosa, L. and Rosauer, T. J. and Rose, C. A. and Rosi\ifmmode \acute{n}\else \'{n}\fi{}ska, D. and Ross, M. P. and Rossello-Sastre, M. and Rowan, S. and Roy, S. K. and Roy, S. and Rozza, D. and Ruggi, P. and Ruhama, N. and Morales, E. Ruiz and Ruiz-Rocha, K. and Sachdev, S. and Sadecki, T. and Saffarieh, P. and Safi-Harb, S. and Sah, M. R. and Saha, S. and Sainrat, T. and Menon, S. Sajith and Sakai, K. and Sakai, Y. and Sakellariadou, M. and Sakon, S. and Salafia, O. S. and Salces-Carcoba, F. and Salconi, L. and Saleem, M. and Salemi, F. and Sall\'e, M. and Salunkhe, S. U. and Salvador, S. and Salvarese, A. and Samajdar, A. and Sanchez, A. and Sanchez, E. J. and Sanchez, L. E. and Sanchis-Gual, N. and Sanders, J. R. and S\"anger, E. M. and Santoliquido, F. and Sarandrea, F. and Saravanan, T. R. and Sarin, N. and Sarkar, P. and Sasli, A. and Sassi, P. and Sassolas, B. and Sato, R. and Sato, S. and Sato, Yukino and Sato, Yu and Sauter, O. and Savage, R. L. and Sawada, T. and Sawant, H. L. and Sayah, S. and Scacco, V. and Schaetzl, D. and Scheel, M. and Schiebelbein, A. and Schiworski, M. G. and Schmidt, P. and Schmidt, S. and Schnabel, R. and Schneewind, M. and Schofield, R. M. S. and Schouteden, K. and Schulte, B. W. and Schutz, B. F. and Schwartz, E. and Scialpi, M. and Scott, J. and Scott, S. M. and Sedas, R. M. and Seetharamu, T. C. and Seglar-Arroyo, M. and Sekiguchi, Y. and Sellers, D. and Sembo, N. and Sengupta, A. S. and Seo, E. G. and Seo, J. W. and Sequino, V. and Serra, M. and Sevrin, A. and Shaffer, T. and Shah, U. S. and Shaikh, M. A. and Shao, L. and Sharma, A. K. and Sharma, Preeti and Sharma, Prianka and Sharma, Ritwik and Chaudhary, S. Sharma and Shawhan, P. and Shcheblanov, N. S. and Sheridan, E. and Shi, Z.-H. and Shikauchi, M. and Shimomura, R. and Shinkai, H. and Shirke, S. and Shoemaker, D. H. and Shoemaker, D. M. and Short, R. W. and ShyamSundar, S. and Sider, A. and Siegel, H. and Sigg, D. and Silenzi, L. and Silvestri, L. and Simmonds, M. and Singer, L. P. and Singh, Amitesh and Singh, Anika and Singh, D. and Singh, N. and Singh, S. and Sintes, A. M. and Sipala, V. and Skliris, V. and Slagmolen, B. J. J. and Slater, D. A. and Slaven-Blair, T. J. and Smetana, J. and Smith, J. R. and Smith, L. and Smith, R. J. E. and Smith, W. J. and Filho, S. Soares de Albuquerque and Soares-Santos, M. and Somiya, K. and Song, I. and Soni, S. and Sordini, V. and Sorrentino, F. and Sotani, H. and Spada, F. and Spagnuolo, V. and Spencer, A. P. and Spinicelli, P. and Srivastava, A. K. and Stachurski, F. and Stark, C. J. and Steer, D. A. and Steinle, N. and Steinlechner, J. and Steinlechner, S. and Stergioulas, N. and Stevens, P. and StPierre, M. and Strong, M. D. and Strunk, A. and Stuver, A. L. and Suchenek, M. and Sudhagar, S. and Sudo, Y. and Sueltmann, N. and Suleiman, L. and Sullivan, K. D. and Sun, J. and Sun, L. and Sunil, S. and Suresh, J. and Sutton, B. J. and Sutton, P. J. and Suzuki, K. and Suzuki, M. and Swinkels, B. L. and Syx, A. and Szczepa\ifmmode \acute{n}\else \'{n}\fi{}czyk, M. J. and Szewczyk, P. and Tacca, M. and Tagoshi, H. and Takada, K. and Takahashi, H. and Takahashi, R. and Takamori, A. and Takano, S. and Takeda, H. and Takeshita, K. and Schmiegelow, I. Takimoto and Takou-Ayaoh, M. and Talbot, C. and Tamaki, M. and Tamanini, N. and Tanabe, D. and Tanaka, K. and Tanaka, S. J. and Tanioka, S. and Tanner, D. B. and Tanner, W. and Tao, L. and Tapia, R. D. and Mart\'{\i}n, E. N. Tapia San and Taranto, C. and Taruya, A. and Tasson, J. D. and Tau, J. G. and Tellez, D. and Tenorio, R. and Themann, H. and Theodoropoulos, A. and Thirugnanasambandam, M. P. and Thomas, L. M. and Thomas, M. and Thomas, P. and Thompson, J. E. and Thondapu, S. R. and Thorne, K. A. and Thrane, E. and Tissino, J. and Tiwari, A. and Tiwari, Pawan and Tiwari, Praveer and Tiwari, S. and Tiwari, V. and Todd, M. R. and Toffano, M. and Toivonen, A. M. and Toland, K. and Tolley, A. E. and Tomaru, T. and Tommasini, V. and Tomura, T. and Tong, H. and Tong-Yu, C. and Torres-Forn\'e, A. and Torrie, C. I. and e Melo, I. Tosta and Tournefier, E. and Nery, M. Trad and Tran, K. and Trapananti, A. and Travaglini, R. and Travasso, F. and Traylor, G. and Trevor, M. and Tringali, M. C. and Tripathee, A. and Troian, G. and Trovato, A. and Trozzo, L. and Trudeau, R. J. and Tsang, T. and Tsuchida, S. and Tsukada, L. and Turbang, K. and Turconi, M. and Turski, C. and Ubach, H. and Uchikata, N. and Uchiyama, T. and Udall, R. P. and Uehara, T. and Ueno, K. and Undheim, V. and Uronen, L. E. and Ushiba, T. and Vacatello, M. and Vahlbruch, H. and Vaidya, N. and Vajente, G. and Vajpeyi, A. and Valencia, J. and Valentini, M. and Vallejo-Pe\~na, S. A. and Vallero, S. and Valsan, V. and van Dael, M. and Van den Bossche, E. and van den Brand, J. F. J. and Van Den Broeck, C. and van der Sluys, M. and Van de Walle, A. and van Dongen, J. and Vandra, K. and VanDyke, M. and van Haevermaet, H. and van Heijningen, J. V. and Van Hove, P. and Vanier, J. and VanKeuren, M. and Vanosky, J. and van Remortel, N. and Vardaro, M. and Vargas, A. F. and Varma, V. and Vazquez, A. N. and Vecchio, A. and Vedovato, G. and Veitch, J. and Veitch, P. J. and Venikoudis, S. and Venterea, R. C. and Verdier, P. and Vereecken, M. and Verkindt, D. and Verma, B. and Verma, Y. and Vermeulen, S. M. and Vetrano, F. and Veutro, A. and Vicer\'e, A. and Vidyant, S. and Viets, A. D. and Vijaykumar, A. and Vilkha, A. and Espinosa, N. Villanueva and Villa-Ortega, V. and Vincent, E. T. and Vinet, J.-Y. and Viret, S. and Vitale, S. and Vocca, H. and Voigt, D. and von Reis, E. R. G. and von Wrangel, J. S. A. and Vossius, W. E. and Vujeva, L. and Vyatchanin, S. P. and Wack, J. and Wade, L. E. and Wade, M. and Wagner, K. J. and Wallace, L. and Wang, E. J. and Wang, H. and Wang, J. Z. and Wang, W. H. and Wang, Y. F. and Waratkar, G. and Warner, J. and Was, M. and Washimi, T. and Washington, N. Y. and Watarai, D. and Weaver, B. and Webster, S. A. and Weickhardt, N. L. and Weinert, M. and Weinstein, A. J. and Weiss, R. and Wen, L. and Wette, K. and Whelan, J. T. and Whiting, B. F. and Whittle, C. and Wickens, E. G. and Wilken, D. and Wilkin, A. T. and Williams, B. M. and Williams, D. and Williams, M. J. and Williams, N. S. and Willis, J. L. and Willke, B. and Wils, M. and Wilson, L. and Winborn, C. W. and Winterflood, J. and Wipf, C. C. and Woan, G. and Woehler, J. and Wolfe, N. E. and Wong, H. T. and Wong, I. C. F. and Wong, K. and Wouters, T. and Wright, J. L. and Wright, M. and Wu, B. and Wu, C. and Wu, D. S. and Wu, H. and Wu, K. and Wu, Q. and Wu, Y. and Wu, Z. and Wuchner, E. and Wysocki, D. M. and Xu, V. A. and Xu, Y. and Yadav, N. and Yamamoto, H. and Yamamoto, K. and Yamamoto, T. S. and Yamamoto, T. and Yamazaki, R. and Yan, T. and Yang, K. Z. and Yang, Y. and Yarbrough, Z. and Yebana, J. and Yeh, S.-W. and Yelikar, A. B. and Yin, X. and Yokoyama, J. and Yokozawa, T. and Yuan, S. and Yuzurihara, H. and Zanolin, M. and Zeeshan, M. and Zelenova, T. and Zendri, J.-P. and Zeoli, M. and Zerrad, M. and Zevin, M. and Zhang, L. and Zhang, N. and Zhang, R. and Zhang, T. and Zhao, C. and Zhao, Yue and Zhao, Yuhang and Zhao, Z.-C. and Zheng, Y. and Zhong, H. and Zhou, H. and Zhu, H. O. and Zhu, Z.-H. and Zimmerman, A. B. and Zimmermann, L. and Zucker, M. E. and Zweizig, J.},
  collaboration = {The LIGO Scientific Collaboration, the Virgo Collaboration, and the KAGRA Collaboration},
  journal = {Phys. Rev. D},
  volume = {112},
  issue = {10},
  pages = {102005},
  numpages = {29},
  year = {2025},
  month = {Nov},
  publisher = {American Physical Society},
  doi = {10.1103/wjdz-jdby},
  url = {https://link.aps.org/doi/10.1103/wjdz-jdby}
}

@article{SASI_Spiral_Foglizzo,
  title = {Standing accretion shock instability in the collapse of a rotating stellar core},
  author = {Walk, Laurie and Foglizzo, Thierry and Tamborra, Irene},
  journal = {Phys. Rev. D},
  volume = {107},
  issue = {6},
  pages = {063014},
  numpages = {15},
  year = {2023},
  month = {Mar},
  publisher = {American Physical Society},
  doi = {10.1103/PhysRevD.107.063014},
  url = {https://link.aps.org/doi/10.1103/PhysRevD.107.063014}
}

@Article{SASI_Spiral_Blondin,
author={Blondin, John M.
and Mezzacappa, Anthony},
title={Pulsar spins from an instability in the accretion shock of supernovae},
journal={Nature},
year={2007},
month={Jan},
day={01},
volume={445},
number={7123},
pages={58-60},
issn={1476-4687},
doi={10.1038/nature05428},
url={https://doi.org/10.1038/nature05428}
}

@article{ Foglizzo_SASI_Analytic,
	author = {{Foglizzo, T.}},
	title = {Analytic insight into the physics of the standing accretion shock instability - I. Shock instability in a non-rotating stellar core},
	DOI= "10.1051/0004-6361/202452300",
	url= "https://doi.org/10.1051/0004-6361/202452300",
	journal = {A\&A},
	year = 2024,
	volume = 692,
	pages = "A196",
}

@Article{GW_Detect_Perspectives,
AUTHOR = {Szczepańczyk, Marek and Zanolin, Michele},
TITLE = {Gravitational Waves from a Core-Collapse Supernova: Perspectives with Detectors in the Late 2020s and Early 2030s},
JOURNAL = {Galaxies},
VOLUME = {10},
YEAR = {2022},
NUMBER = {3},
ARTICLE-NUMBER = {70},
URL = {https://www.mdpi.com/2075-4434/10/3/70},
ISSN = {2075-4434},
DOI = {10.3390/galaxies10030070}
}

@misc{cWB_2G,
      title={Optimizing searches for gravitational wave bursts using coherent WaveBurst 2G}, 
      author={Alessandro Martini and Andrea Miani and Marco Drago and Claudia Lazzaro and Francesco Salemi and Sophie Bini and Osvaldo Freitas and Edoardo Milotti and Giacomo Principe and Shubhanshu Tiwari and Agata Trovato and Gabriele Vedovato and Yumeng Xu and Giovanni Andrea Prodi},
      year={2025},
      eprint={2510.21411},
      archivePrefix={arXiv},
      primaryClass={gr-qc},
      url={https://arxiv.org/abs/2510.21411}, 
}

@article{O4_Performance,
  title = {Advanced LIGO detector performance in the fourth observing run},
  author = {Capote, E. and Jia, W. and Aritomi, N. and Nakano, M. and Xu, V. and Abbott, R. and Abouelfettouh, I. and Adhikari, R. X. and Ananyeva, A. and Appert, S. and Apple, S. K. and Arai, K. and Aston, S. M. and Ball, M. and Ballmer, S. W. and Barker, D. and Barsotti, L. and Berger, B. K. and Betzwieser, J. and Bhattacharjee, D. and Billingsley, G. and Biscans, S. and Blair, C. D. and Bode, N. and Bonilla, E. and Bossilkov, V. and Branch, A. and Brooks, A. F. and Brown, D. D. and Bryant, J. and Cahillane, C. and Cao, H. and Clara, F. and Collins, J. and Compton, C. M. and Cottingham, R. and Coyne, D. C. and Crouch, R. and Csizmazia, J. and Cumming, A. and Dartez, L. P. and Davis, D. and Demos, N. and Dohmen, E. and Driggers, J. C. and Dwyer, S. E. and Effler, A. and Ejlli, A. and Etzel, T. and Evans, M. and Feicht, J. and Frey, R. and Frischhertz, W. and Fritschel, P. and Frolov, V. V. and Fuentes-Garcia, M. and Fulda, P. and Fyffe, M. and Ganapathy, D. and Gateley, B. and Gayer, T. and Giaime, J. A. and Giardina, K. D. and Glanzer, J. and Goetz, E. and Goetz, R. and Goodwin-Jones, A. W. and Gras, S. and Gray, C. and Griffith, D. and Grote, H. and Guidry, T. and Gurs, J. and Hall, E. D. and Hanks, J. and Hanson, J. and Heintze, M. C. and Helmling-Cornell, A. F. and Holland, N. A. and Hoyland, D. and Huang, H. Y. and Inoue, Y. and James, A. L. and Jamies, A. and Jennings, A. and Jones, D. H. and Kabagoz, H. B. and Karat, S. and Karki, S. and Kasprzack, M. and Kawabe, K. and Kijbunchoo, N. and King, P. J. and Kissel, J. S. and Komori, K. and Kontos, A. and Kumar, Rahul and Kuns, K. and Landry, M. and Lantz, B. and Laxen, M. and Lee, K. and Lesovsky, M. and Villarreal, F. Llamas and Lormand, M. and Loughlin, H. A. and Macas, R. and MacInnis, M. and Makarem, C. N. and Mannix, B. and Mansell, G. L. and Martin, R. M. and Mason, K. and Matichard, F. and Mavalvala, N. and Maxwell, N. and McCarrol, G. and McCarthy, R. and McClelland, D. E. and McCormick, S. and McRae, T. and Mera, F. and Merilh, E. L. and Meylahn, F. and Mittleman, R. and Moraru, D. and Moreno, G. and Mullavey, A. and Nelson, T. J. N. and Neunzert, A. and Notte, J. and Oberling, J. and O'Hanlon, T. and Osthelder, C. and Ottaway, D. J. and Overmier, H. and Parker, W. and Patane, O. and Pele, A. and Pham, H. and Pirello, M. and Pullin, J. and Quetschke, V. and Ramirez, K. E. and Ransom, K. and Reyes, J. and Richardson, J. W. and Robinson, M. and Rollins, J. G. and Romel, C. L. and Romie, J. H. and Ross, M. P. and Ryan, K. and Sadecki, T. and Sanchez, A. and Sanchez, E. J. and Sanchez, L. E. and Savage, R. L. and Schaetzl, D. and Schiworski, M. G. and Schnabel, R. and Schofield, R. M. S. and Schwartz, E. and Sellers, D. and Shaffer, T. and Short, R. W. and Sigg, D. and Slagmolen, B. J. J. and Soike, C. and Soni, S. and Srivastava, V. and Sun, L. and Tanner, D. B. and Thomas, M. and Thomas, P. and Thorne, K. A. and Todd, M. R. and Torrie, C. I. and Traylor, G. and Ubhi, A. S. and Vajente, G. and Vanosky, J. and Vecchio, A. and Veitch, P. J. and Vibhute, A. M. and von Reis, E. R. G. and Warner, J. and Weaver, B. and Weiss, R. and Whittle, C. and Willke, B. and Wipf, C. C. and Wright, J. L. and Yamamoto, H. and Zhang, L. and Zucker, M. E.},
  journal = {Phys. Rev. D},
  volume = {111},
  issue = {6},
  pages = {062002},
  numpages = {30},
  year = {2025},
  month = {Mar},
  publisher = {American Physical Society},
  doi = {10.1103/PhysRevD.111.062002},
  url = {https://link.aps.org/doi/10.1103/PhysRevD.111.062002}
}

@article{Opt_O3_SN_Search,
  title = {Optically targeted search for gravitational waves emitted by core-collapse supernovae during the third observing run of Advanced LIGO and Advanced Virgo},
  author = {Szczepa\ifmmode \acute{n}\else \'{n}\fi{}czyk, Marek J. and Zheng, Yanyan and Antelis, Javier M. and Benjamin, Michael and Bizouard, Marie-Anne and Casallas-Lagos, Alejandro and Cerd\'a-Dur\'an, Pablo and Davis, Derek and Gondek-Rosi\ifmmode \acute{n}\else \'{n}\fi{}ska, Dorota and Klimenko, Sergey and Moreno, Claudia and Obergaulinger, Martin and Powell, Jade and Ramirez, Dymetris and Ratto, Brad and Richardson, Colter and Rijal, Abhinav and Stuver, Amber L. and Szewczyk, Pawe\l{} and Vedovato, Gabriele and Zanolin, Michele and Bartos, Imre and Bhaumik, Shubhagata and Bulik, Tomasz and Drago, Marco and Font, Jos\'e A. and De Colle, Fabio and Garc\'{\i}a-Bellido, Juan and Gayathri, V. and Hughey, Brennan and Mitselmakher, Guenakh and Mishra, Tanmaya and Mukherjee, Soma and Nguyen, Quynh Lan and Chan, Man Leong and Di Palma, Irene and Piotrzkowski, Brandon J. and Singh, Neha},
  journal = {Phys. Rev. D},
  volume = {110},
  issue = {4},
  pages = {042007},
  numpages = {20},
  year = {2024},
  month = {Aug},
  publisher = {American Physical Society},
  doi = {10.1103/PhysRevD.110.042007},
  url = {https://link.aps.org/doi/10.1103/PhysRevD.110.042007}
}

@article{HuangT_Veutro,
  title = {Extracting SASI signatures from gravitational waves of core-collapse supernovae using the Hilbert-Huang transform},
  author = {Veutro, A. and Di Palma, I. and Zegarelli, A.},
  journal = {Phys. Rev. D},
  volume = {112},
  issue = {10},
  pages = {103020},
  numpages = {9},
  year = {2025},
  month = {Nov},
  publisher = {American Physical Society},
  doi = {10.1103/tdl7-vys1},
  url = {https://link.aps.org/doi/10.1103/tdl7-vys1}
}

@article{LIGO_Adv,
doi = {10.1088/0264-9381/32/7/074001},
url = {https://doi.org/10.1088/0264-9381/32/7/074001},
year = {2015},
month = {mar},
publisher = {IOP Publishing},
volume = {32},
number = {7},
pages = {074001},
author = {The LIGO Scientific Collaboration and Aasi, J and Abbott, B P and Abbott, R and Abbott, T and Abernathy, M R and Ackley, K and Adams, C and Adams, T and Addesso, P and Adhikari, R X and Adya, V and Affeldt, C and Aggarwal, N and Aguiar, O D and Ain, A and Ajith, P and Alemic, A and Allen, B and Amariutei, D and Anderson, S B and Anderson, W G and Arai, K and Araya, M C and Arceneaux, C and Areeda, J S and Ashton, G and Ast, S and Aston, S M and Aufmuth, P and Aulbert, C and Aylott, B E and Babak, S and Baker, P T and Ballmer, S W and Barayoga, J C and Barbet, M and Barclay, S and Barish, B C and Barker, D and Barr, B and Barsotti, L and Bartlett, J and Barton, M A and Bartos, I and Bassiri, R and Batch, J C and Baune, C and Behnke, B and Bell, A S and Bell, C and Benacquista, M and Bergman, J and Bergmann, G and Berry, C P L and Betzwieser, J and Bhagwat, S and Bhandare, R and Bilenko, I A and Billingsley, G and Birch, J and Biscans, S and Biwer, C and Blackburn, J K and Blackburn, L and Blair, C D and Blair, D and Bock, O and Bodiya, T P and Bojtos, P and Bond, C and Bork, R and Born, M and Bose, Sukanta and Brady, P R and Braginsky, V B and Brau, J E and Bridges, D O and Brinkmann, M and Brooks, A F and Brown, D A and Brown, D D and Brown, N M and Buchman, S and Buikema, A and Buonanno, A and Cadonati, L and Calderón Bustillo, J and Camp, J B and Cannon, K C and Cao, J and Capano, C D and Caride, S and Caudill, S and Cavaglià, M and Cepeda, C and Chakraborty, R and Chalermsongsak, T and Chamberlin, S J and Chao, S and Charlton, P and Chen, Y and Cho, H S and Cho, M and Chow, J H and Christensen, N and Chu, Q and Chung, S and Ciani, G and Clara, F and Clark, J A and Collette, C and Cominsky, L and Constancio, M and Cook, D and Corbitt, T R and Cornish, N and Corsi, A and Costa, C A and Coughlin, M W and Countryman, S and Couvares, P and Coward, D M and Cowart, M J and Coyne, D C and Coyne, R and Craig, K and Creighton, J D E and Creighton, T D and Cripe, J and Crowder, S G and Cumming, A and Cunningham, L and Cutler, C and Dahl, K and Dal Canton, T and Damjanic, M and Danilishin, S L and Danzmann, K and Dartez, L and Dave, I and Daveloza, H and Davies, G S and Daw, E J and DeBra, D and Del Pozzo, W and Denker, T and Dent, T and Dergachev, V and DeRosa, R T and DeSalvo, R and Dhurandhar, S and D´ıaz, M and Di Palma, I and Dojcinoski, G and Dominguez, E and Donovan, F and Dooley, K L and Doravari, S and Douglas, R and Downes, T P and Driggers, J C and Du, Z and Dwyer, S and Eberle, T and Edo, T and Edwards, M and Edwards, M and Effler, A and Eggenstein, H.-B and Ehrens, P and Eichholz, J and Eikenberry, S S and Essick, R and Etzel, T and Evans, M and Evans, T and Factourovich, M and Fairhurst, S and Fan, X and Fang, Q and Farr, B and Farr, W M and Favata, M and Fays, M and Fehrmann, H and Fejer, M M and Feldbaum, D and Ferreira, E C and Fisher, R P and Frei, Z and Freise, A and Frey, R and Fricke, T T and Fritschel, P and Frolov, V V and Fuentes-Tapia, S and Fulda, P and Fyffe, M and Gair, J R and Gaonkar, S and Gehrels, N and Gergely´, L Á and Giaime, J A and Giardina, K D and Gleason, J and Goetz, E and Goetz, R and Gondan, L and González, G and Gordon, N and Gorodetsky, M L and Gossan, S and Goßler, S and Gräf, C and Graff, P B and Grant, A and Gras, S and Gray, C and Greenhalgh, R J S and Gretarsson, A M and Grote, H and Grunewald, S and Guido, C J and Guo, X and Gushwa, K and Gustafson, E K and Gustafson, R and Hacker, J and Hall, E D and Hammond, G and Hanke, M and Hanks, J and Hanna, C and Hannam, M D and Hanson, J and Hardwick, T and Harry, G M and Harry, I W and Hart, M and Hartman, M T and Haster, C-J and Haughian, K and Hee, S and Heintze, M and Heinzel, G and Hendry, M and Heng, I S and Heptonstall, A W and Heurs, M and Hewitson, M and Hild, S and Hoak, D and Hodge, K A and Hollitt, S E and Holt, K and Hopkins, P and Hosken, D J and Hough, J and Houston, E and Howell, E J and Hu, Y M and Huerta, E and Hughey, B and Husa, S and Huttner, S H and Huynh, M and Huynh-Dinh, T and Idrisy, A and Indik, N and Ingram, D R and Inta, R and Islas, G and Isler, J C and Isogai, T and Iyer, B R and Izumi, K and Jacobson, M and Jang, H and Jawahar, S and Ji, Y and Jiménez-Forteza, F and Johnson, W W and Jones, D I and Jones, R and Ju, L and Haris, K and Kalogera, V and Kandhasamy, S and Kang, G and Kanner, J B and Katsavounidis, E and Katzman, W and Kaufer, H and Kaufer, S and Kaur, T and Kawabe, K and Kawazoe, F and Keiser, G M and Keitel, D and Kelley, D B and Kells, W and Keppel, D G and Key, J S and Khalaidovski, A and Khalili, F Y and Khazanov, E A and Kim, C and Kim, K and Kim, N G and Kim, N and Kim, Y.-M and King, E J and King, P J and Kinzel, D L and Kissel, J S and Klimenko, S and Kline, J and Koehlenbeck, S and Kokeyama, K and Kondrashov, V and Korobko, M and Korth, W Z and Kozak, D B and Kringel, V and Krishnan, B and Krueger, C and Kuehn, G and Kumar, A and Kumar, P and Kuo, L and Landry, M and Lantz, B and Larson, S and Lasky, P D and Lazzarini, A and Lazzaro, C and Le, J and Leaci, P and Leavey, S and Lebigot, E O and Lee, C H and Lee, H K and Lee, H M and Leong, J R and Levin, Y and Levine, B and Lewis, J and Li, T G F and Libbrecht, K and Libson, A and Lin, A C and Littenberg, T B and Lockerbie, N A and Lockett, V and Logue, J and Lombardi, A L and Lormand, M and Lough, J and Lubinski, M J and Lück, H and Lundgren, A P and Lynch, R and Ma, Y and Macarthur, J and MacDonald, T and Machenschalk, B and MacInnis, M and Macleod, D M and Magaña-Sandoval, F and Magee, R and Mageswaran, M and Maglione, C and Mailand, K and Mandel, I and Mandic, V and Mangano, V and Mansell, G L and Márka, S and Márka, Z and Markosyan, A and Maros, E and Martin, I W and Martin, R M and Martynov, D and Marx, J N and Mason, K and Massinger, T J and Matichard, F and Matone, L and Mavalvala, N and Mazumder, N and Mazzolo, G and McCarthy, R and McClelland, D E and McCormick, S and McGuire, S C and McIntyre, G and McIver, J and McLin, K and McWilliams, S and Meadors, G D and Meinders, M and Melatos, A and Mendell, G and Mercer, R A and Meshkov, S and Messenger, C and Meyers, P M and Miao, H and Middleton, H and Mikhailov, E E and Miller, A and Miller, J and Millhouse, M and Ming, J and Mirshekari, S and Mishra, C and Mitra, S and Mitrofanov, V P and Mitselmakher, G and Mittleman, R and Moe, B and Mohanty, S D and Mohapatra, S R P and Moore, B and Moraru, D and Moreno, G and Morriss, S R and Mossavi, K and Mow-Lowry, C M and Mueller, C L and Mueller, G and Mukherjee, S and Mullavey, A and Munch, J and Murphy, D and Murray, P G and Mytidis, A and Nash, T and Nayak, R K and Necula, V and Nedkova, K and Newton, G and Nguyen, T and Nielsen, A B and Nissanke, S and Nitz, A H and Nolting, D and Normandin, M E N and Nuttall, L K and Ochsner, E and O’Dell, J and Oelker, E and Ogin, G H and Oh, J J and Oh, S H and Ohme, F and Oppermann, P and Oram, R and O’Reilly, B and Ortega, W and O’Shaughnessy, R and Osthelder, C and Ott, C D and Ottaway, D J and Ottens, R S and Overmier, H and Owen, B J and Padilla, C and Pai, A and Pai, S and Palashov, O and Pal-Singh, A and Pan, H and Pankow, C and Pannarale, F and Pant, B C and Papa, M A and Paris, H and Patrick, Z and Pedraza, M and Pekowsky, L and Pele, A and Penn, S and Perreca, A and Phelps, M and Pierro, V and Pinto, I M and Pitkin, M and Poeld, J and Post, A and Poteomkin, A and Powell, J and Prasad, J and Predoi, V and Premachandra, S and Prestegard, T and Price, L R and Principe, M and Privitera, S and Prix, R and Prokhorov, L and Puncken, O and Pürrer, M and Qin, J and Quetschke, V and Quintero, E and Quiroga, G and Quitzow-James, R and Raab, F J and Rabeling, D S and Radkins, H and Raffai, P and Raja, S and Rajalakshmi, G and Rakhmanov, M and Ramirez, K and Raymond, V and Reed, C M and Reid, S and Reitze, D H and Reula, O and Riles, K and Robertson, N A and Robie, R and Rollins, J G and Roma, V and Romano, J D and Romanov, G and Romie, J H and Rowan, S and Rüdiger, A and Ryan, K and Sachdev, S and Sadecki, T and Sadeghian, L and Saleem, M and Salemi, F and Sammut, L and Sandberg, V and Sanders, J R and Sannibale, V and Santiago-Prieto, I and Sathyaprakash, B S and Saulson, P R and Savage, R and Sawadsky, A and Scheuer, J and Schilling, R and Schmidt, P and Schnabel, R and Schofield, R M S and Schreiber, E and Schuette, D and Schutz, B F and Scott, J and Scott, S M and Sellers, D and Sengupta, A S and Sergeev, A and Serna, G and Sevigny, A and Shaddock, D A and Shahriar, M S and Shaltev, M and Shao, Z and Shapiro, B and Shawhan, P and Shoemaker, D H and Sidery, T L and Siemens, X and Sigg, D and Silva, A D and Simakov, D and Singer, A and Singer, L and Singh, R and Sintes, A M and Slagmolen, B J J and Smith, J R and Smith, M R and Smith, R J E and Smith-Lefebvre, N D and Son, E J and Sorazu, B and Souradeep, T and Staley, A and Stebbins, J and Steinke, M and Steinlechner, J and Steinlechner, S and Steinmeyer, D and Stephens, B C and Steplewski, S and Stevenson, S and Stone, R and Strain, K A and Strigin, S and Sturani, R and Stuver, A L and Summerscales, T Z and Sutton, P J and Szczepanczyk, M and Szeifert, G and Talukder, D and Tanner, D B and Tápai, M and Tarabrin, S P and Taracchini, A and Taylor, R and Tellez, G and Theeg, T and Thirugnanasambandam, M P and Thomas, M and Thomas, P and Thorne, K A and Thorne, K S and Thrane, E and Tiwari, V and Tomlinson, C and Torres, C V and Torrie, C I and Traylor, G and Tse, M and Tshilumba, D and Ugolini, D and Unnikrishnan, C S and Urban, A L and Usman, S A and Vahlbruch, H and Vajente, G and Valdes, G and Vallisneri, M and van Veggel, A A and Vass, S and Vaulin, R and Vecchio, A and Veitch, J and Veitch, P J and Venkateswara, K and Vincent-Finley, R and Vitale, S and Vo, T and Vorvick, C and Vousden, W D and Vyatchanin, S P and Wade, A R and Wade, L and Wade, M and Walker, M and Wallace, L and Walsh, S and Wang, H and Wang, M and Wang, X and Ward, R L and Warner, J and Was, M and Weaver, B and Weinert, M and Weinstein, A J and Weiss, R and Welborn, T and Wen, L and Wessels, P and Westphal, T and Wette, K and Whelan, J T and Whitcomb, S E and White, D J and Whiting, B F and Wilkinson, C and Williams, L and Williams, R and Williamson, A R and Willis, J L and Willke, B and Wimmer, M and Winkler, W and Wipf, C C and Wittel, H and Woan, G and Worden, J and Xie, S and Yablon, J and Yakushin, I and Yam, W and Yamamoto, H and Yancey, C C and Yang, Q and Zanolin, M and Zhang, Fan and Zhang, L and Zhang, M and Zhang, Y and Zhao, C and Zhou, M and Zhu, X J and Zucker, M E and Zuraw, S and Zweizig, J},
title = {Advanced LIGO},
journal = {Classical and Quantum Gravity}
}

@article{Virgo_Adv,
doi = {10.1088/0264-9381/32/2/024001},
url = {https://doi.org/10.1088/0264-9381/32/2/024001},
year = {2014},
month = {dec},
publisher = {IOP Publishing},
volume = {32},
number = {2},
pages = {024001},
author = {Acernese, F and Agathos, M and Agatsuma, K and Aisa, D and Allemandou, N and Allocca, A and Amarni, J and Astone, P and Balestri, G and Ballardin, G and Barone, F and Baronick, J-P and Barsuglia, M and Basti, A and Basti, F and Bauer, Th S and Bavigadda, V and Bejger, M and Beker, M G and Belczynski, C and Bersanetti, D and Bertolini, A and Bitossi, M and Bizouard, M A and Bloemen, S and Blom, M and Boer, M and Bogaert, G and Bondi, D and Bondu, F and Bonelli, L and Bonnand, R and Boschi, V and Bosi, L and Bouedo, T and Bradaschia, C and Branchesi, M and Briant, T and Brillet, A and Brisson, V and Bulik, T and Bulten, H J and Buskulic, D and Buy, C and Cagnoli, G and Calloni, E and Campeggi, C and Canuel, B and Carbognani, F and Cavalier, F and Cavalieri, R and Cella, G and Cesarini, E and Mottin, E Chassande- and Chincarini, A and Chiummo, A and Chua, S and Cleva, F and Coccia, E and Cohadon, P-F and Colla, A and Colombini, M and Conte, A and Coulon, J-P and Cuoco, E and Dalmaz, A and D’Antonio, S and Dattilo, V and Davier, M and Day, R and Debreczeni, G and Degallaix, J and Deléglise, S and Pozzo, W Del and Dereli, H and Rosa, R De and Fiore, L Di and Lieto, A Di and Virgilio, A Di and Doets, M and Dolique, V and Drago, M and Ducrot, M and Endrőczi, G and Fafone, V and Farinon, S and Ferrante, I and Ferrini, F and Fidecaro, F and Fiori, I and Flaminio, R and Fournier, J-D and Franco, S and Frasca, S and Frasconi, F and Gammaitoni, L and Garufi, F and Gaspard, M and Gatto, A and Gemme, G and Gendre, B and Genin, E and Gennai, A and Ghosh, S and Giacobone, L and Giazotto, A and Gouaty, R and Granata, M and Greco, G and Groot, P and Guidi, G M and Harms, J and Heidmann, A and Heitmann, H and Hello, P and Hemming, G and Hennes, E and Hofman, D and Jaranowski, P and Jonker, R J G and Kasprzack, M and Kéfélian, F and Kowalska, I and Kraan, M and Królak, A and Kutynia, A and Lazzaro, C and Leonardi, M and Leroy, N and Letendre, N and Li, T G F and Lieunard, B and Lorenzini, M and Loriette, V and Losurdo, G and Magazzù, C and Majorana, E and Maksimovic, I and Malvezzi, V and Man, N and Mangano, V and Mantovani, M and Marchesoni, F and Marion, F and Marque, J and Martelli, F and Martellini, L and Masserot, A and Meacher, D and Meidam, J and Mezzani, F and Michel, C and Milano, L and Minenkov, Y and Moggi, A and Mohan, M and Montani, M and Morgado, N and Mours, B and Mul, F and Nagy, M F and Nardecchia, I and Naticchioni, L and Nelemans, G and Neri, I and Neri, M and Nocera, F and Pacaud, E and Palomba, C and Paoletti, F and Paoli, A and Pasqualetti, A and Passaquieti, R and Passuello, D and Perciballi, M and Petit, S and Pichot, M and Piergiovanni, F and Pillant, G and Piluso, A and Pinard, L and Poggiani, R and Prijatelj, M and Prodi, G A and Punturo, M and Puppo, P and Rabeling, D S and Rácz, I and Rapagnani, P and Razzano, M and Re, V and Regimbau, T and Ricci, F and Robinet, F and Rocchi, A and Rolland, L and Romano, R and Rosińska, D and Ruggi, P and Saracco, E and Sassolas, B and Schimmel, F and Sentenac, D and Sequino, V and Shah, S and Siellez, K and Straniero, N and Swinkels, B and Tacca, M and Tonelli, M and Travasso, F and Turconi, M and Vajente, G and van Bakel, N and van Beuzekom, M and van den Brand, J F J and Van Den Broeck, C and van der Sluys, M V and van Heijningen, J and Vasúth, M and Vedovato, G and Veitch, J and Verkindt, D and Vetrano, F and Viceré, A and Vinet, J-Y and Visser, G and Vocca, H and Ward, R and Was, M and Wei, L-W and Yvert, M and żny, A Zadro and Zendri, J-P},
title = {Advanced Virgo: a second-generation interferometric gravitational wave detector},
journal = {Classical and Quantum Gravity}
}

@article{Freq_SASI_Analysis,
  title = {Dedicated-frequency analysis of gravitational-wave bursts from core-collapse supernovae with minimal assumptions},
  author = {Lee, Yi Shuen C. and Szczepa\ifmmode \acute{n}\else \'{n}\fi{}czyk, Marek J. and Mishra, Tanmaya and Millhouse, Margaret and Melatos, Andrew},
  journal = {Phys. Rev. D},
  volume = {112},
  issue = {8},
  pages = {082006},
  numpages = {20},
  year = {2025},
  month = {Oct},
  publisher = {American Physical Society},
  doi = {10.1103/kg3l-dtxc},
  url = {https://link.aps.org/doi/10.1103/kg3l-dtxc}
}

@article{HFF_Slope_Daniel,
  title = {Dependence of the reconstructed core-collapse supernova gravitational wave high-frequency feature on the nuclear equation of state in real interferometric data},
  author = {Murphy, R. Daniel and Casallas-Lagos, Alejandro and Mezzacappa, Anthony and Zanolin, Michele and Landfield, Ryan E. and Lentz, Eric J. and Marronetti, Pedro and Antelis, Javier M. and Moreno, Claudia},
  journal = {Phys. Rev. D},
  volume = {110},
  issue = {8},
  pages = {083006},
  numpages = {21},
  year = {2024},
  month = {Oct},
  publisher = {American Physical Society},
  doi = {10.1103/PhysRevD.110.083006},
  url = {https://link.aps.org/doi/10.1103/PhysRevD.110.083006}
}

@article{HuangT_Takeda,
  title = {Application of the Hilbert-Huang transform for analyzing standing-accretion-shock-instability induced gravitational waves in a core-collapse supernova},
  author = {Takeda, M. and Hiranuma, Y. and Kanda, N. and Kotake, K. and Kuroda, T. and Negishi, R. and Oohara, K. and Sakai, K. and Sakai, Y. and Sawada, T. and Takahashi, H. and Tsuchida, S. and Watanabe, Y. and Yokozawa, T.},
  journal = {Phys. Rev. D},
  volume = {104},
  issue = {8},
  pages = {084063},
  numpages = {12},
  year = {2021},
  month = {Oct},
  publisher = {American Physical Society},
  doi = {10.1103/PhysRevD.104.084063},
  url = {https://link.aps.org/doi/10.1103/PhysRevD.104.084063}
}

@article{Andresen2017,
    author = {Andresen, H. and Müller, B. and Müller, E. and Janka, H.-Th.},
    title = {Gravitational wave signals from 3D neutrino hydrodynamics simulations of core-collapse supernovae},
    journal = {Monthly Notices of the Royal Astronomical Society},
    volume = {468},
    number = {2},
    pages = {2032-2051},
    year = {2017},
    month = {03},
    issn = {0035-8711},
    doi = {10.1093/mnras/stx618},
    url = {https://doi.org/10.1093/mnras/stx618},
    eprint = {https://academic.oup.com/mnras/article-pdf/468/2/2032/11210813/stx618.pdf},
}

@article{Kawahara2018,
doi = {10.3847/1538-4357/aae57b},
url = {https://doi.org/10.3847/1538-4357/aae57b},
year = {2018},
month = {nov},
publisher = {The American Astronomical Society},
volume = {867},
number = {2},
pages = {126},
author = {Kawahara, Hajime and Kuroda, Takami and Takiwaki, Tomoya and Hayama, Kazuhiro and Kotake, Kei},
title = {A Linear and Quadratic Time–Frequency Analysis of Gravitational Waves from Core-collapse Supernovae},
journal = {The Astrophysical Journal}
}

@article{Mezzacappa2020,
  title = {Gravitational-wave signal of a core-collapse supernova explosion of a $15\text{ }{M}_{\ensuremath{\bigodot}}$ star},
  author = {Mezzacappa, Anthony and Marronetti, Pedro and Landfield, Ryan E. and Lentz, Eric J. and Yakunin, Konstantin N. and Bruenn, Stephen W. and Hix, W. Raphael and Messer, O. E. Bronson and Endeve, Eirik and Blondin, John M. and Harris, J. Austin},
  journal = {Phys. Rev. D},
  volume = {102},
  issue = {2},
  pages = {023027},
  numpages = {20},
  year = {2020},
  month = {Jul},
  publisher = {American Physical Society},
  doi = {10.1103/PhysRevD.102.023027},
  url = {https://link.aps.org/doi/10.1103/PhysRevD.102.023027}
}

@article{Powell2025,
doi = {10.1088/1361-6382/ae0dea},
url = {https://doi.org/10.1088/1361-6382/ae0dea},
year = {2025},
month = {oct},
publisher = {IOP Publishing},
volume = {42},
number = {21},
pages = {215002},
author = {Powell, Jade and Müller, Bernhard},
title = {Gravitational waves from core-collapse supernovae with no electromagnetic counterparts},
journal = {Classical and Quantum Gravity}
}

@article{Murphy2009,
doi = {10.1088/0004-637X/707/2/1173},
url = {https://doi.org/10.1088/0004-637X/707/2/1173},
year = {2009},
month = {dec},
publisher = {The American Astronomical Society},
volume = {707},
number = {2},
pages = {1173},
author = {Murphy, Jeremiah W. and Ott, Christian D. and Burrows, Adam},
title = {A MODEL FOR GRAVITATIONAL WAVE EMISSION FROM NEUTRINO-DRIVEN CORE-COLLAPSE SUPERNOVAE},
journal = {The Astrophysical Journal}
}

@techreport{Casallas_DCC,
  author       = {Casallas-Lagos, Alejandro and Szczepa{\' n}czyk, Marek and Zanolin, Michele and Mezzacappa, Anthony},
  title        = {Estimating the Curvature of the CCSN GW High Frequency Feature in Interferometric Noise},
  institution  = {LIGO Scientific Collaboration},
  number       = {LIGO-G2600497},
  year         = {2026},
  url          = {https://dcc.ligo.org/LIGO-G2600497},
  note         = {Accessed: 2026-04-03}
}

@article{Foglizzo2007,
doi = {10.1086/509612},
url = {https://doi.org/10.1086/509612},
year = {2007},
month = {jan},
publisher = {},
volume = {654},
number = {2},
pages = {1006},
author = {Foglizzo, T. and Galletti, P. and Scheck, L. and Janka, H.-Th.},
title = {Instability of a Stalled Accretion Shock: Evidence for the Advective-Acoustic Cycle},
journal = {The Astrophysical Journal}
}

@article{Powell2021,
title = "The final core collapse of pulsational pair instability supernovae",
author = "Jade Powell and Bernhard M{\"u}ller and Alexander Heger",
year = "2021",
month = may,
day = "1",
doi = "10.1093/mnras/stab614",
language = "English",
volume = "503",
pages = "2108--2122",
journal = "Monthly Notices of the Royal Astronomical Society",
issn = "0035-8711",
publisher = "Oxford University Press",
number = "2",

}

@article{Powell_DAWES,
    author = "Powell, Jade and Lasky, Paul D.",
    title = "{The DAWES review NN: Gravitational-wave burst astrophysics}",
    eprint = "2410.12105",
    archivePrefix = "arXiv",
    primaryClass = "astro-ph.HE",
    doi = "10.1017/pasa.2025.10",
    journal = "Publ. Astron. Soc. Austral.",
    volume = "42",
    pages = "e030",
    year = "2025"
}

@article{Foglizzo2014,
    author = {Fernández, Rodrigo and Müller, Bernhard and Foglizzo, Thierry and Janka, Hans-Thomas},
    title = {Characterizing SASI- and convection-dominated core-collapse supernova explosions in two dimensions},
    journal = {Monthly Notices of the Royal Astronomical Society},
    volume = {440},
    number = {3},
    pages = {2763-2780},
    year = {2014},
    month = {05},
    issn = {0035-8711},
    doi = {10.1093/mnras/stu408},
    url = {https://doi.org/10.1093/mnras/stu408},
    eprint = {https://academic.oup.com/mnras/article-pdf/440/3/2763/23991662/stu408.pdf},
}

@article{MullerJanka2014,
doi = {10.1088/0004-637X/788/1/82},
url = {https://doi.org/10.1088/0004-637X/788/1/82},
year = {2014},
month = {may},
publisher = {The American Astronomical Society},
volume = {788},
number = {1},
pages = {82},
author = {Müller, Bernhard and Janka, Hans-Thomas},
title = {A NEW MULTI-DIMENSIONAL GENERAL RELATIVISTIC NEUTRINO HYDRODYNAMICS CODE FOR CORE-COLLAPSE SUPERNOVAE. IV. THE NEUTRINO SIGNAL},
journal = {The Astrophysical Journal}
}

@article{Foglizzo2009,
doi = {10.1088/0004-637X/694/2/820},
url = {https://doi.org/10.1088/0004-637X/694/2/820},
year = {2009},
month = {mar},
publisher = {The American Astronomical Society},
volume = {694},
number = {2},
pages = {820},
author = {Foglizzo, T.},
title = {A SIMPLE TOY MODEL OF THE ADVECTIVE–ACOUSTIC INSTABILITY. I. PERTURBATIVE APPROACH},
journal = {The Astrophysical Journal}
}

\end{document}